\documentclass[11pt]{article}
\pdfoutput=1
\usepackage{appendix}
\usepackage{tikz}
\usetikzlibrary{matrix,arrows,decorations.pathmorphing,decorations.pathreplacing}
\usepackage{jheppub}
\usepackage{amsmath,amssymb,euscript,array,mathrsfs,appendix,ctable,marvosym}
\usepackage{arydshln}
\usepackage{todonotes}
\usepackage{graphicx}
\usepackage[normalem]{ulem}
\usepackage{cleveref}
\usepackage[new]{old-arrows}

\usepackage{extarrows}

\usepackage[skins,theorems]{tcolorbox}
\tcbset{highlight math style={enhanced,
  colframe=red,colback=white,arc=0pt,boxrule=1pt}}

\input{tcilatex}

\numberwithin{equation}{section}

\title{Contact 4d Chern-Simons theory: Generalities}
\author{David M. Schmidtt\footnote{david@df.ufscar.br}} 

\affiliation{Physics Department, S\~ao Carlos Federal University, \\
Rodovia Washington Lu\'is, km 235, S\~ao Carlos - SP, Brazil} 

\abstract{We refine and generalize the results of \cite{Yo}, where evidence in favor of applying the non-Abelian localization method to handle the 4d Chern-Simons theory path integral formulation was presented. We show, via duality manipulations and invoking some symplectic geometry results, both inspired by the Beasley-Witten work \cite{NA loc CS}, that the path integral of a regularized version of the 4d Chern-Simons theory, formally takes the canonical symplectic form required by the method of non-Abelian localization. The new theory is defined on a deformed quotient space and interpolates between the conventional 3d Chern-Simons theory on a Seifert manifold M \cite{NA loc CS}, trivially embedded into $\mathbb{R}\times \text{M}$, and the Costello-Yamazaki \cite{CY} 4d Chern-Simons theory defined on the same 4d manifold. It is also shown that the regularized theory is consistent, following an idea of Beasley \cite{Wilson NA loc}, with the insertion of coadjoint orbit defects of the 1d Chern-Simons theory type. This approach opens the possibility for using exact path integral methods to explore the quantum integrable structure of certain 2d integrable sigma models of the non-ultralocal type, which are widely known to be somehow immune to the use of more traditional quantization methods, like the algebraic Bethe ansatz.
\begin{flushleft}
\end{flushleft}
}

\begin{document}

\maketitle


\section{Introduction}\label{1}

It is now a consensus the importance of the description to integrable lattice models and 2d integrable field theories (IFT's) provided by the 4d Chern-Simons (CS) theory introduced in \cite{C1,C2} and further studied in \cite{CWY1,CWY2,CY}. In recent years, this new approach has attracted a great deal of attention due to its potential in offering new insights into the quantum integrable structure of these systems and their general properties. One notable characteristic of this approach lies in its power for describing successfully, at the quantum level, lattice models and IFT's of the so-called ultralocal type. However, despite of this success it shows some difficulties\footnote{A heuristic and a technical reason explaining why, can be found in $\S 2$ of \cite{Witten2016}.} in describing the more elusive non-ultralocal IFT's \cite{Maillet}. The latter being a peculiar kind of 2d IFT's theories escaping the grasp of some well-established quantization methods, like the powerful algebraic Bethe ansatz.

Due to the relevance of non-ultralocal IFT's to string theory and the AdS/CFT correspondence, a natural question that raises is if it is possible to generalize the original 4d CS theory formulation of Costello and Yamazaki (CY) \cite{CY} in a way to include such theories in a sensible way. An initial idea for doing this was first proposed in \cite{Yo}, based on the seminal work \cite{NA loc CS} devoted to a reformulation of the conventional 3d CS theory on a Seifert manifold in a way amenable to the use of the method of non-Abelian localization \cite{Witten revisited}. As initially concluded in \cite{Yo}, a generalization of the original 4d CS theory based on the ideias of \cite{NA loc CS} was viable, but in terms of an action functional that displays two major characteristics, apparently incompatible with the ones already present in the CY 4d CS theory formulation:

I) It manifests a gauge symmetry behavior that is not present in the original theory, in the sense of lack of covariance of some quantities and the need for imposing constraints on the gauge group parameters that are unnatural from the 4d CS theory perspective in order to recover it. Indeed, the gauge group elements were required to depend on an specific combination of 2d coordinates (light-cone coordinates), which is not a canonical condition existing in the CY 4d CS theory.

II) It does not display the same boundary equations of motion (eom) existing in the original theory that are so important for constructing systematically the 2d IFT's action functionals living on the pole-defect surface and their associated Lax connections. Recall that the 4d CS theory is extremely sensitive to choices of and changes in the boundary eom. A direct consequence of this was that the generalized 4d CS theory was not able to reproduce successfully some of the 2d IFT's described correctly, by the CY 4d CS theory.

At first sight, these two apparent drawbacks seem to be inevitable. However, as we shall see, a closer examination to the 4d CS theory generalization procedure of \cite{Yo}, shows there is a consistent way to bypass both and it is the main purpose of this paper to `rectify' I) and II) by constructing a slightly different generalization that is compatible with the properties of the original 4d CS theory. This will be done by showing\footnote{Actually, we deduce a more general expression including (co)-adjoint orbit insertions, Cf. \eqref{main result}.} the following \textit{formal} equivalence of path integrals
\begin{equation}
\int\nolimits_{\mathcal{A}} \mathcal{D}\mathbb{A} \, \text{exp}\left[\frac{i}{\hbar} S\big(\mathbb{A}\big)_{\text{reg}}\right]= \mathcal{N}\int\nolimits_{\overline{\mathcal{A}}}\text{exp} \left[ i\hat{\Omega}-\frac{1}{2\epsilon}\big(\mu,\mu\big)  \right ], \label{MR}
\end{equation}
where
\begin{equation}
S(\mathbb{A})_{\text{reg}}=ic\dint\nolimits_{\mathbb{M}}\omega_{\zeta} \wedge CS\left( 
\mathbb{A}\right)\label{regg}
\end{equation}
and
\begin{equation}
\big(\mu,\mu\big)=-\frac{i}{c}S(\mathbb{A})_{\text{reg}}-\dint\nolimits_{\mathbb{M}}\gamma_{\text{top}} \text{Tr}\left( \Phi_{\text{on}} ^{2}%
\right). \label{contactt}
\end{equation}
In the next paragraphs we will roughly describe the ingredients on both sides of \eqref{MR} in order to summarize their meaning.

\textit{Left hand side.} Starting with the regularized 4d CS theory action functional \eqref{regg}, we have that $\mathbb{M}=\mathbb{R}\times \text{M}$, with M being a Seifert manifold\footnote{See the reference \cite{Orlik}, for further details.}, which is essentially a non-trivial circle bundle over a Riemann surface $C$, possibly with orbifold points. $\omega_{\zeta}$ is a 1-form on $\mathbb{M}$, depending on a real deformation parameter $\zeta$, with the property that in the limit $\zeta \rightarrow 0$, it reduces to a 1-form $\omega$ on M defined in terms of the usual 4d CS theory twist form $\omega_{C}$ on $C$. $CS\left( 
\mathbb{A}\right)$ is the well-known CS 3-form for a connection $\mathbb{A}\in \mathcal{A}$, where $\mathcal{A}$ is the space of gauge connections on $\mathbb{M}$ subordinated to a particular set of boundary eom solutions to be determined below. $c$ is a real constant. 

What is relevant to be mentioned, at this stage, about the left hand side (lhs) of \eqref{MR} is that, in the limit $\zeta\rightarrow 0$, the action \eqref{regg} reproduces successfully the same 2d IFT's described by the original 4d CS theory formulation. This is because, in this limit, \eqref{regg} reduces to
\begin{equation}
S(\mathbb{A})_{4d\text{-CS}}=ic\dint\nolimits_{\mathbb{M}}\omega \wedge CS\left( 
\mathbb{A}\right), \label{111}
\end{equation}
which is precisely the CY 4d CS theory but now defined now $\mathbb{R}\times \text{M}$. Below, we shall consider a sample of some of the most emblematic integrable string sigma models of the principal chiral model (PCM) type, all being non-ultralocal, in order to exemplify this relation. Hence, the lhs in \eqref{MR} is naturally linked to a path integral description of these 2d IFT's and this is because after gauge fixing and imposing and appropriate set of solutions to the boundary eom, the IFT's are thus recovered from \eqref{111}.

\textit{Right hand side.} Starting with the quadratic action \eqref{contactt}, we have that $\gamma_{\text{top}}$ is a metric-independent volume form on $\mathbb{M}$ constructed out of a contact form $\alpha$ on M and the 1-form $d\tau$ on $\mathbb{R}$, and $\Phi_{\text{on}}$, is an adjoint scalar field depending on the field strength of $\mathbb{A}$ and the differential forms $\omega_{\zeta}$, $\kappa$ and $\gamma_{\text{top}}$, with $\kappa$ defined in terms of $\alpha$ and $d\tau$. In addition, $\overline{\mathcal{A}}\subset \mathcal{A}$ is a functional symplectic manifold endowed with a real symplectic form $i\hat{\Omega}$ that is acted, in a Hamiltonian way, by a symmetry group whose associated moment map is $\mu$. The term $\mathcal{N}$ is the result of some formal functional integrations and $\epsilon$ is a real coupling constant depending on $c$ and $\hbar$. 

What is important to be emphasized, at this point, about the right hand side (rhs) of \eqref{MR}, is that it is of the canonical form required by the method of non-Abelian localization of symplectic integrals \cite{Witten revisited}. Hence, in principle, the rhs of \eqref{MR} can be computed exactly and from the relation of the lhs to some 2d IFT's in the limit $\zeta \rightarrow 0$, we can consider the possibility of having found a novel way to probe the quantum integrable structure of these non-ultralocal IFT's, using a known equivariant localization technique. However, despite of the fact that \eqref{MR} is of the required canonical form, the underlying symplectic manifolds entering the localization formula are, as we shall see, of the pseudo-K\"ahler type which is a direct consequence of the complex nature of the Lie groups involved in the 4d CS theory formulation. This is to be contrasted with the known results \cite{NA loc CS, Wilson NA loc} involving real compact Lie groups, where the relevant manifolds are all K\"ahler. 

Let us now briefly explain why the usual 4d CS theory must be regularized in the first place. Simply put, the regularization process consists in making the following replacements 
\begin{equation}
\big(\Sigma \times C,\, \omega_{C}    \big)\longrightarrow \big(\mathbb{R} \times \text{M},\, \omega_{\zeta}   \big)\label{222}
\end{equation}
in the conventional 4d CS theory plus a certain shift symmetry to be identified later on. This ensures that the volume form $\gamma_{\text{top}}$ defined on $\mathbb{R} \times \text{M}$, which now takes the explicit form
\begin{equation}
\gamma_{\text{top}} \sim n\zeta d\tau \wedge \alpha \wedge \underline{\pi}^{\ast}(\sigma_{C}), \label{333}
\end{equation} 
never vanishes provided the circle bundle is non-trivial\footnote{Notice that $n$ enters explicitly on the rhs of \eqref{MR}, via $\gamma_{\text{top}}$ in \eqref{contactt}. }, i.e. $n \neq 0$, and the deforming parameter is such that $\zeta \neq 0$. Above, $\sigma_{C}$ is a symplectic form on the base manifold $C$. In turn, the existence of \eqref{333} is essential for defining the inner product used to construct the quadratic expression on the lhs of the equality \eqref{contactt} and also for making the whole construction mathematically consistent. In summary, the result \eqref{MR} is only possible because of the substitution \eqref{222}, hence its importance in defining \eqref{regg}. All this will be explained in detail in the remainder of the text, of course.   

This paper is motivated by ideas introduced in the seminal works \cite{NA loc CS} by Beasley and Witten and \cite{Wilson NA loc} by Beasley and it is organized as follows: In $\S \eqref{2}$, we employ a classical duality approach to introduce in an straightforward manner the so-called contact 4d CS theory, which is basically given by the rhs of the expression \eqref{contactt} above. This section presents in an intuitive way the modifications to the usual 4d CS theory that are necessary for implementing the ideas of \cite{NA loc CS}. It covers the original CY 4d CS theory, the regularized 4d CS theory, Cf. \eqref{regg}, the extended 4d CS theory and finally, the contact 4d CS theory. We also introduce, at this stage, complex (co)-adjoint orbit defects of the 1d CS theory type and briefly discuss the properties of all these actions functionals under the action of a formal gauge group. In $\S \eqref{3}$, we discuss reality conditions in detail and take the opportunity to modify the discussion reported in \cite{Yo}, in order to present it in a more appropriate and complete way. In $\S \eqref{4}$ we revisit and adapt to our construction, some of the results of \cite{unifying} in order to show how the action \eqref{regg}, in the limit $\zeta \rightarrow 0$, reproduces a sample of some well-known non-ultralocal IFT's, all of them associated to $C=\mathbb{CP}^{1}$. Among them, we explicitly recover the principal chiral model (PCM) with Wess-Zumino (WZ) term, the Homogeneous Yang-Baxter sigma model, the $\lambda$-deformed PCM and the Yang-Baxter sigma model or $\eta$-deformed PCM. Boundary eom are also carefully studied in this section and we show why the action \eqref{regg} turns out to be compatible with the boundary eom that are present in the original 4d CS theory formulation. The compatibility being deeply and nicely related to the properties of the contact form to be defined below. In $\S \eqref{5}$, a more rigorous approach to the contact 4d CS theory is employed. It is based on a symplectic geometry approach where we show, after a thorough analysis of a Hamiltonian group action and the introduction of an important bilinear form on its Lie algebra, what are the necessary conditions required to put the rhs of \eqref{contactt} in the quadratic moment form displayed on the lhs of \eqref{contactt}. Here, we also take the opportunity to refine and improve the discussion of \cite{Yo} and modify several of the results presented there that, at the end, are the ones responsible for the resolution of the issues mentioned above in I) and II). We complete the discussion about the conditions for the gauge symmetry invariance of all actions functionals introduced and also include the symplectic properties of (co)-adjoint orbits defects in this section. In $\S \eqref{6}$, we formally define the path integral measures along the integration domains $\overline{\mathcal{A}}$ and $L\mathcal{O}_{\lambda}$, in terms of Liouville volume forms, via exponentials of their associated symplectic forms. In $\S \eqref{7}$, we present the result \eqref{MR}, which is actually generalized by the presence of (co)-adjoint orbit insertions, as shown by the main result \eqref{main result}. In the limit $\zeta \rightarrow 0$, we propose that this formula relates the path integral of some 2d non-ultralocal IFT's to a canonical symplectic integral expression that, in principle, can be evaluated exactly via non-Abelian localization. In $\S \eqref{8}$, we finish with some remarks and comments on open problems to be considered in the future. 

\section{Contact 4d CS theory as a dual action}\label{2}

We introduce, in the usual 4d CS theory, certain modifications that are necessary for applying the method of non-Abelian localization. We start with a review of the conventional 4d CS theory, then construct a regularization thereof that allows the implementation of the ideas of \cite{NA loc CS} and finish, via duality manipulations, with the introduction of the so-called contact 4d CS theory. Furthermore, following \cite{Wilson NA loc}, we include (co)-adjoint orbits defects of the 1d CS theory type that will modify, in a consistent way, the path integral formulation of the regularized 4d CS theory.

Before starting, a comment is in order. In formulating the original 4d CS theory, known  boundary eom are imposed from the outset upon the space of gauge connections where the theory is defined. As our main interest is to introduce a consistent modification of the latter that localizes, we will keep throughout the whole analysis, all boundary contributions until we can safely discard them and by this we mean proving that the same boundary eom valid in the original 4d CS theory can be enforced as well upon the field content in all action functionals used in our approach.

\subsection{Bulk data}

Here we cover the 4d CS theory, the regularized 4d CS theory, the extended 4d CS theory and finally, the dual or contact 4d CS theory. The analysis here is similar to the one of \cite{Yo}, with the fundamental difference that the regularized 4d CS theory is defined now on a new deformed quotient space, not considered in \cite{Yo}, containing the usual 4d CS theory in a particular corner. 

\subsubsection{4d CS theory}\label{2.1.1}

The 4d CS theory action functional is defined \cite{CY} by the expression 
\begin{equation}
S(\mathbb{A})_{4d\text{-CS}}:=ic\dint\nolimits_{\mathbb{M}}\omega_{C} \wedge CS(\mathbb{A}), \label{4d CS}
\end{equation}
where $\mathbb{M}=\Sigma \times C$, $\Sigma=\mathbb{R}\times S^{1}$ is a cylinder (also referred to as the world-sheet), $C$ is a genus $g$ Riemann surface with local holomorphic coordinate $z$ (to be identified with the IFT spectral parameter), $\omega_{C}\in \Omega_{C}^{1,0}$ is a meromorphic $(1,0)$-form on $C$ (the twist form) with a set of zeroes $\mathfrak{z}$ and poles $\mathfrak{p}$ obeying the constraint
\begin{equation}
n_{\mathfrak{p}}-n_{\mathfrak{z}}=2-2g,\label{RRR}
\end{equation} 
where $n_{\mathfrak{p}}$ is the number of poles and $n_{\mathfrak{z}}$ is the number of zeroes, both counted with multiplicity. Also, $CS(\mathbb{A})\in \Omega_{\mathbb{M}}^{3}$ is the Chern-Simons 3-form
\begin{equation}
CS(\mathbb{A})=\text{Tr}\left( \mathbb{A}\wedge d_{\mathbb{M}}\mathbb{A}+\frac{2}{3} \mathbb{A}\wedge \mathbb{A}\wedge \mathbb{A} \right), \label{CS 3-form}
\end{equation}
$\mathbb{A}\in \Omega_{\mathbb{M}}^{1}\otimes \mathfrak{g}$ is a gauge field on $\mathbb{M}$ and $c\in \mathbb{R}$ is a constant\footnote{Usually, $1/ 2\pi$ or $1/ 4\pi$.}. Moreover, $\mathfrak{g}$ is the Lie algebra of a complex, connected, simply-connected and semi-simple Lie group $G$, $\text{Tr}(\cdot,\cdot):\mathfrak{g} \times \mathfrak{g}\rightarrow \mathbb{C}$ denotes an ad-invariant, symmetric and non-degenerate bilinear form on $\mathfrak{g}$ and $d_{\mathbb{M}}$ is the exterior derivative on $\mathbb{M}$. Reality conditions for the action \eqref{4d CS}, are studied in \cite{unifying} but, for the sake of completeness, these will be reviewed below in \S \eqref{3}.

From the general variation of the action \eqref{4d CS}, i.e.
\begin{equation}
\delta S(\mathbb{A})_{4d\text{-CS}}=2ic \dint_{\mathbb{M}}\omega_{C}\wedge \text{Tr}\left(\delta\mathbb{A}\wedge F_{\mathbb{A}}  \right)+ic \dint_{\mathbb{M}}d_{C}\omega_{C}\,\text{Tr} \left(\delta \mathbb{A} \wedge \mathbb{A}   \right),
\end{equation}
where $F_{\mathbb{A}}=d_{\mathbb{M}}\mathbb{A}+\mathbb{A} \wedge \mathbb{A}\in \Omega_{\mathbb{M}}^{2}\otimes \mathfrak{g}$ is the field strength of $\mathbb{A}$ and $d_{C}$ is the exterior derivative on $C$, we read off the bulk eom of the theory and the boundary eom that are to be imposed upon the gauge field $\mathbb{A}$ as well. They are given, respectively, by
\begin{equation}
\omega_{C}\wedge F_{\mathbb{A}}=0
\end{equation}
and
\begin{equation}
\dint_{\mathbb{M}}d_{C}\omega_{C}\,\text{Tr} \left(\delta \mathbb{A} \wedge \mathbb{A}   \right)=0.\label{BC C}
\end{equation}
In \eqref{BC C} we have to use the fact that the 2-form $d_{C}\omega_{C}$ is actually a Dirac delta distribution with support at the set of poles $\mathfrak{p}$ and, in practice, it reduces the 4d integral in \eqref{BC C}, over $\mathbb{M}$, to a 2d integral along the so-called pole-defect surface $\Sigma \times \mathfrak{p}$, see \cite{CY}. By a boundary contribution we mean any expression including the 2-form $d_{C}\omega_{C}$, hence the name boundary eom to \eqref{BC C}. This lower dimensional integral is the one responsible for the boundary eom to be imposed on field content of the 4d theory that ultimately will provide the 2d field degrees of freedom (dof) of an associated IFT living on the pole-defect surface. The very form of the resulting 2d IFT action functional is quite sensitive to the solutions to equation \eqref{BC C} and several examples can be found in \cite{CY} and also in \cite{unifying}. A sample of these IFT's will be re-derived later on in \S \eqref{4}, within our approach.

The action \eqref{4d CS} is manifestly invariant under the $(1,0)$-shifts given by
\begin{equation}
^{\chi_{C}}\mathbb{A}:=\mathbb{A}+s\chi_{C} , \label{chi}
\end{equation}
where $s \in \Omega_{\mathbb{M}}^{0}\otimes \mathfrak{g}$ is a Lie algebra valued function and $\chi_{C} \in \Omega_{C}^{1,0}$ a $(1,0)$-form. As a consequence, the gauge field component $A_{z}$ completely decouples from the theory. This can be seen from the local expressions $\mathbb{A}=A_{\tau}d\tau + A_{\sigma}d\sigma + A_{z}dz + A_{\overline{z}}d\overline{z}$ and $\omega_{C}=\varphi(z)dz$, where $-\infty < \tau< \infty$ and $0 \leq \sigma \leq 2\pi$ are coordinates on $\Sigma$. Then, the theory \eqref{4d CS} is naturally defined on the quotient space formed by the space of gauge fields on $\mathbb{M}$ modulo the space of gauge fields of the form $s\chi_{C}$.

\subsubsection{Regularized 4d CS theory}\label{2.1.2}

The idea now is to decouple a second component of $\mathbb{A}$ in order to obtain an equivalent formulation of the theory \eqref{4d CS} that is amenable to the use of the method of non-Abelian localization. See \cite{NA loc CS} for the original idea applied to conventional 3d CS theories on Seifert manifolds. However, in order to implement this idea in our context and in the correct way, the action \eqref{4d CS} must be \textit{regularized} first and now we proceed to show how this is done. 

Following \cite{Yo}, we start by replacing $\mathbb{M}=\Sigma \times C$ by $\mathbb{M}=\mathbb{R}\times \text{M}$, where M is now seen as the total space of a degree $n\neq 0$ non-trivial circle bundle over $C$, i.e.
\begin{equation}
S^{1}\overset{n}{\longhookrightarrow} \text{M}\overset{\underline{\pi}}{\longrightarrow} C. \label{n bundle}
\end{equation}
In practice, this simply means that $\Sigma \times C$ and $\mathbb{R}\times \text{M}$ only agree locally when restricted to any chart $\mathcal{U}\subset C$ and this is the only new working hypothesis we need in relation to the original theory, which is recovered when we consider the trivial bundle, i.e. $n=0$.
Next, we introduce two 1-forms $\kappa, \omega_{\zeta} \in \Omega_{\mathbb{M}}^{1}$, with $\zeta \in \mathbb{R}$ and a vector field $X\in \mathfrak{X}_{\mathbb{M}}$, satisfying the following defining relations
\begin{equation}
\iota_{X}\kappa=1,\text{ \qquad \qquad }\iota_{X}\omega_{\zeta}=0,\text{ \qquad \qquad  }\iota_{X}( d_{\mathbb{M}} \omega_{\zeta} )=0, \label{Omega X conditions}
\end{equation}
where $\iota_{X}$ is the interior product or contraction with $X \in \mathfrak{X}_{\mathbb{M}}$. Notice that $\pounds_{X}\omega_{\zeta}=0$, where
$\pounds_{X}=d_{\mathbb{M}}\circ \iota_{X}+\iota_{X}\circ d_{\mathbb{M}}$ is the Lie derivative along $X$. Now, the last step in regularizing the action \eqref{4d CS} consists in substituting in \eqref{4d CS}, the twist form $\omega_{C}$ by $ \omega_{\zeta}$ and in extending the $(1,0)$-shift symmetry \eqref{chi} to a new $\omega_{\zeta}$-shift symmetry defined by 
\begin{equation}
^{\omega_{\zeta}}\mathbb{A}:=\mathbb{A}+s \omega_{\zeta}, \label{Omega shift symmetry}
\end{equation}
which, of course, is still responsible for decoupling one of the components of the gauge field $\mathbb{A}$, i.e. the theory is now defined on the quotient space formed by the space of gauge fields on $\mathbb{M}$ modulo the space of gauge fields of the form $s\omega_{\zeta}$. We also require that in the $\zeta \rightarrow 0$ limit, we get
\begin{equation}
\omega_{\zeta=0}= \omega:= \underline{\pi}^{\ast}(\omega_{C}), \label{zeta=0 limit}
\end{equation}
where $\omega$ is the pull-back, under the projection map $\underline{\pi}$, of the twist form $\omega_{C}$ from the base manifold $C$ to the total space M. Later on, we will show that
\begin{equation}
\omega_{\zeta}=\omega + 2\zeta \alpha_{\tau}  d\tau, \label{spoiler sol iii}
\end{equation} 
where $\alpha_{\tau}\neq 0 \in \mathbb{R}$. It is not difficult to see \cite{Yo}, that the natural $\omega_{\zeta}$-shift invariant extension of \eqref{4d CS} is given by the regularized action given by
\begin{equation}
S(\mathbb{A})_{\text{reg}}:=ic\dint\nolimits_{\mathbb{M}}\omega_{\zeta} \wedge CS\left( 
\mathbb{A}\right) +ic\dint\nolimits_{\mathbb{M}}d_{\mathbb{M}}\omega_{\zeta}\wedge \kappa\wedge \text{Tr}\left( \mathbb{A}\iota_{X}\mathbb{A}  \right). \label{Omega shift inv}
\end{equation}
By regularized, we mean the two steps considered so far:
\begin{itemize}
\item The change of the 4d manifold $\Sigma \times C \rightarrow \mathbb{R} \times \text{M}$.
\item The replacement $\omega_{C} \rightarrow \omega_{\zeta}$, plus the $\omega_{\zeta}$-shift symmetry extension.
\end{itemize}
To show \eqref{Omega shift inv}, we start by assuming that $\omega_{\zeta}$ is an arbitrary 1-form on $\mathbb{M}$ and consider the expression
\begin{equation}
\omega_{\zeta}\wedge CS\big( ^{\omega_{\zeta}}\mathbb{A}\big)=\omega_{\zeta}\wedge CS\left( \mathbb{A}\right)+d_{\mathbb{M}}\omega_{\zeta}\wedge \omega_{\zeta} \wedge \text{Tr}\left(s \mathbb{A}   \right).
\end{equation}
Thus, in order to handle the second term on the rhs right above, we use the trivial 5-form contraction
\begin{equation}
0=\iota_{X}\Big(  d_{\mathbb{M}}\omega_{\zeta}\wedge \omega_{\zeta} \wedge \kappa \wedge \text{Tr}\left(s \mathbb{A}   \right) \Big),
\end{equation}
together with \eqref{Omega X conditions}, to write
\begin{equation}
d_{\mathbb{M}}\omega_{\zeta}\wedge \omega_{\zeta} \wedge \text{Tr}\left(s \mathbb{A}   \right)=-d_{\mathbb{M}}\omega_{\zeta}\wedge  \kappa \wedge \text{Tr}\big[(s\omega_{\zeta}) \iota_{X}\mathbb{A}   \big].
\end{equation}
The final step consists in writing $s\omega_{\zeta}=\,^{\omega_{\zeta}}\mathbb{A}-\mathbb{A}$ and using $\iota_{X}\big(^{\omega_{\zeta}}\mathbb{A}\big)=\iota_{X}\mathbb{A}$ to conclude that the action \eqref{Omega shift inv} is indeed invariant under the shift symmetry \eqref{Omega shift symmetry}, for generic $\omega_{\zeta}$. Of course, in this paper we will be mainly interested in the particular choice \eqref{spoiler sol iii}. 

The action \eqref{Omega shift inv} is also invariant under the $t$-rescalings
\begin{equation}
\kappa \rightarrow t \kappa,\text{\qquad \qquad  }X \rightarrow X/t,\text{ \qquad \qquad }\omega_{\zeta}\rightarrow \omega_{\zeta}, \label{rescalings}
\end{equation}
for any non-zero function $t\in \Omega_{\mathbb{M}}^{0}$. The reason for keeping $\omega_{\zeta}$ intact comes from the fact that the original action \eqref{4d CS} does not have an analogue $t$-rescaling symmetry of the form $\omega_{C} \rightarrow t \omega_{C}$. Furthermore, in the limit $\zeta \rightarrow 0$, the action \eqref{Omega shift inv} displays a $\chi$-shift invariance under the change
\begin{equation}
^{\chi}\mathbb{A}=\mathbb{A}+s\chi, \label{chi n}
\end{equation}
where $\chi=\underline{\pi}^{\ast}(\chi_{C})$. This corresponds to the $(1,0)$-shift invariance of the original 4d-CS theory on the trivial bundle $\Sigma \times C$, now lifted to the space $\mathbb{R}\times \text{M}$, with M as in \eqref{n bundle}. Thus, we conclude that the regularized theory is defined on a $\zeta$-dependent quotient\footnote{Later on, we will explore this quotient space a bit more in some interesting limits.} space formed by the space of gauge fields on $\mathbb{M}$ modulo the space of gauge fields of the form $s\omega_{\zeta}$, up to equivalence $\kappa \sim t\kappa$. In addition, from the contraction
\begin{equation}
\iota_{Y}\big( ^{\omega_{\zeta}}\mathbb{A}  \big)=\iota_{Y}\mathbb{A}+\zeta s, 
\end{equation} 
against the vector field $Y:=\frac{1}{2 \alpha_{\tau}}\partial_{\tau}$, we notice that, for $\zeta \neq 0$, an admissible $\omega_{\zeta}$-shift symmetry gauge fixing condition is
\begin{equation}
\iota_{\partial_{\tau}}\mathbb{A}=0. \label{tau gauge fixing}
\end{equation}
For $\zeta\rightarrow 0$, we can use instead the $\chi$-shift symmetry to fix the gauge in a way similar to the 4d CS theory (1,0)-shift symmetry \eqref{chi} gauge fixing, e.g. locally in the form
\begin{equation}
A_{z}=0. \label{Azeta=0gauge}
\end{equation}

From the general variation of the regularized 4d CS theory action \eqref{Omega shift inv}, i.e.
\begin{equation}
\delta S(\mathbb{A})_{\text{reg}}=2ic \dint_{\mathbb{M}}\omega_{\zeta}\wedge \text{Tr}\left(\delta\mathbb{A}\wedge F_{\mathbb{A}}  \right)+2ic \dint\nolimits_{%
\mathbb{M}}d_{\mathbb{M}}\omega_{\zeta} \wedge \text{Tr}%
\Big[ \delta \mathbb{A\wedge }\Big( \mathbb{A }-\frac{1}{2}\kappa\iota_{X}%
\mathbb{A}\Big) \Big] ,
\end{equation}
we read off the bulk eom and the boundary eom of the theory. They are given, respectively, by
\begin{equation}
\omega_{\zeta}\wedge F_{\mathbb{A}}=0
\end{equation}
and the vanishing of 
\begin{equation}
\mathcal{B}(\mathbb{A})=\dint\nolimits_{%
\mathbb{M}}d_{\mathbb{M}}\omega_{\zeta} \wedge \text{Tr}%
\Big[ \delta \mathbb{A\wedge }\Big( \mathbb{A }-\frac{1}{2}\kappa\iota_{X}%
\mathbb{A}\Big) \Big] . \label{BC M}
\end{equation}

Below in \S \eqref{4}, we will explain why the usual solutions to the boundary eom \eqref{BC C}, are also solutions to the regularized 4d CS theory boundary eom \eqref{BC M} and furthermore, we will show as well that the second contribution to the rhs of \eqref{Omega shift inv} can be put to zero, precisely by virtue of the vanishing of the boundary eom \eqref{BC M}, turning the regularized and the conventional 4d CS theories essentially the same theory, in form. The only difference being their underlying 4d manifolds which are given, respectively, by $\Sigma\times C$ and $\mathbb{R}\times \text{M}$. For the time being, let us notice from the $\zeta$-independent expression\footnote{To see this, use $d_{\mathbb{M}}=d\tau \wedge \partial_{\tau}+d_{\text{M}}$, the identity $d_{\text{M}}\circ \underline{\pi}^{\ast}=\underline{\pi}^{\ast} \circ d_{C}$ and the fact that $\alpha_{\tau}, \zeta$ are constants.} $d_{\mathbb{M}}\omega_{\zeta}=\underline{\pi}^{\ast}(d_{C}\omega_{C})$, that the integral \eqref{BC M} reduces, as well, to a 2d integral along the pole-defect surface $\Sigma \times \mathfrak{p}$ where, locally, $\Sigma$ is constructed out of the time direction $\mathbb{R}$ and the circle fibers $S^{1}$ on top of the poles $\mathfrak{p}\in \mathcal{U}$ contained in some chart $\mathcal{U}\subset C$. This is the reason why \eqref{BC M} is a boundary eom as well, not receiving any contribution from the bulk eom whatsoever. More on this below.

\subsubsection{Extended and contact 4d CS theory} \label{2.1.3}

After having regularized the 4d-CS theory, we proceed now to decouple a second component of the gauge field $\mathbb{A}$. Following \cite{NA loc CS}, we introduce the extended 4d CS theory action 
\begin{equation}
S(\mathbb{A},\Phi)_{\text{ext}}:=S(\mathbb{A}-\kappa \Phi  )_{\text{reg}}. \label{ext action}
\end{equation}
This action is manifestly invariant under the $\kappa$-shift symmetry defined by
\begin{equation}
^{\kappa}\mathbb{A}:=\mathbb{A}+s \kappa ,\text{ \qquad \qquad  }^{\kappa}\Phi:=\Phi + s, \label{kappashift}
\end{equation}
where $\Phi \in \Omega_{\mathbb{M}}^{0}\otimes \mathfrak{g}$ is an adjoint scalar field and it is also invariant under the $t$-rescalings \eqref{rescalings} and $\omega_{\zeta}$-shifts, provided we postulate the following transformation rules
\begin{equation}
\Phi \rightarrow \Phi /t,\text{\qquad \qquad  }^{\omega_{\zeta}}\Phi=\Phi. \label{postulate}
\end{equation}
The action \eqref{ext action} is then defined on the quotient space formed by the space of gauge fields on $\mathbb{M}$ modulo the space of gauge fields of the form $s\kappa + s' \omega_{\zeta}$, up to equivalence $\kappa \sim t \kappa$. Explicitly, we have that
\begin{equation}
S(\mathbb{A},\Phi)_{\text{ext}}=S(\mathbb{A})_{\text{reg}}+ic\dint\nolimits_{\mathbb{M}}\gamma_{\text{top}} \text{Tr}(\Phi^{2})-2ic\dint\nolimits_{\mathbb{M}}\text{Tr} \Big[\Phi \big(\omega_{\zeta}\wedge \kappa \wedge F_{\mathbb{A}} +d_{\mathbb{M}}\omega_{\zeta} \wedge \kappa \wedge \mathbb{A} \big)   \Big], \label{double shift inv}
\end{equation}
where we have introduced a top form $\gamma_{\text{top}}\in \Omega_{\mathbb{M}}^{4}$, defined by
\begin{equation}
\gamma_{\text{top}}:=\omega_{\zeta} \wedge \kappa \wedge d_{\mathbb{M}}\kappa. \label{gamma top'}
\end{equation}
This extended action is gauge equivalent, under $\kappa$-shifts, to the regularized action \eqref{Omega shift inv} by taking the gauge
\begin{equation}
\Phi=0. \label{Phi=0}
\end{equation}
Notice in passing that, from \eqref{kappashift} and \eqref{Omega X conditions}, we get
\begin{equation}
\iota_{X}\big(^{\kappa}\mathbb{A} \big)=\iota_{X} \mathbb{A}  +s,
\end{equation}
meaning that an admissible $\kappa$-shift symmetry gauge fixing condition is
\begin{equation}
\iota_{X} \mathbb{A}=0. \label{sigma gauge fixing}
\end{equation} 
We will keep the gauge fixing conditions \eqref{tau gauge fixing} and \eqref{sigma gauge fixing} aside, to be used later on.

Continuing, the general variation of the extended 4d CS theory action \eqref{double shift inv}, is given by
\begin{equation}\begin{aligned}
\delta S(\mathbb{A},\Phi)_{\text{ext}}=&-2ic \dint_{\mathbb{M}} \text{Tr}\Big[\delta\mathbb{A}\wedge\big(\omega_{\zeta}\wedge F_{\mathbb{A}-\kappa \Phi}  \big) \Big]+2ic \dint_{\mathbb{M}}\text{Tr}\Big[\delta \Phi \big(\gamma_{\text{top}}   \Phi -\omega_{\zeta}\wedge \kappa \wedge F_{\mathbb{A}}\big)    \Big]\\
&+2ic \dint\nolimits_{%
\mathbb{M}}d_{\mathbb{M}}\omega_{\zeta} \wedge \text{Tr}%
\Big[ \delta \mathbb{A\wedge }\Big( \mathbb{A }-\frac{1}{2}\kappa\iota_{X}%
\mathbb{A}\Big)-\delta \Phi \big(\kappa \wedge \mathbb{A}  \big) \Big]. \label{ext act var}
\end{aligned}
\end{equation}
The bulk eom for the gauge field $\mathbb{A}$ reads
\begin{equation}
\omega_{\zeta}\wedge F_{\mathbb{A}-\kappa \Phi} =0, \label{X}
\end{equation}
where we have used the identity
\begin{equation}
F_{\mathbb{A}-\kappa \Phi}=F_{\mathbb{A}}+\kappa \wedge d_{\mathbb{A}}\Phi -d_{\mathbb{M}}\kappa \Phi
\end{equation}
and $d_{\mathbb{A}}:=d_{\mathbb{M}}+[\mathbb{A}, \cdot\,]$. In order to find the bulk eom for the field $\Phi$, we make now the important assumption that the top form $\gamma_{\text{top}}$ is globally defined and nowhere vanishing on $\mathbb{M}$, i.e. we assume it is a volume form. Thus, if this is the case, the field $\Phi$ is quadratic all over $\mathbb{M}$ and the bulk eom are simply given by\footnote{Any 4-form $\gamma \in \Omega_{\mathbb{M}}^{4}$ is proportional to $\gamma_{\text{top}}$ and can be written as $\gamma= \phi \gamma_{\text{top}}$, for some $\phi \in \Omega_{\mathbb{M}}^{0}$, hence `dividing' by $\gamma_{\text{top}}$ simply means picking $\phi$. This notation, originally introduced in \cite{NA loc CS}, is actually quite useful to perform explicit calculations.}
\begin{equation}
\Phi_{\text{on}} =
\frac{\omega_{\zeta} \wedge \kappa \wedge F_{\mathbb{A}}}{\gamma_{\text{top}} }. \label{Phi eom}
\end{equation}
Notice that the expression \eqref{X} when wedged against the 1-form $\kappa$, imply \eqref{Phi eom}. The boundary eom are now given by the vanishing of
\begin{equation}
\mathcal{B}(\mathbb{A}, \Phi)=\mathcal{B}(\mathbb{A})-\dint\nolimits_{%
\mathbb{M}}d_{\mathbb{M}}\omega_{\zeta} \wedge \kappa \wedge \text{Tr}%
\big( \mathbb{A} \delta \Phi    \big). \label{full BC M}
\end{equation}

By inserting \eqref{Phi eom} back into the action \eqref{double shift inv}, we obtain a dual theory that is classically equivalent to the regularized theory. We quickly find that
\begin{equation}
S(\mathbb{A})_{\text{dual}} =S(\mathbb{A})_{\text{reg}} -ic\dint\nolimits_{\mathbb{M}}\gamma_{\text{top}} \text{Tr}\left( \Phi_{\text{on}} ^{2}%
\right),\label{dual}
\end{equation}%
where $\Phi_{\text{on}}$ right above is given by the on-shell expression \eqref{Phi eom}. The dual action is invariant under $\omega_{\zeta}$-shifts and $\kappa$-shifts as well as under the $t$-rescalings \eqref{rescalings}. The latter statement follows from the fact that
\begin{equation}
\omega_{\zeta} \wedge \kappa \wedge d_{\mathbb{M}}\kappa \longrightarrow t^{2}\omega_{\zeta} \wedge \kappa \wedge d_{\mathbb{M}}\kappa .
\end{equation}
Clearly, this action is also defined on the quotient space formed by the space of gauge fields on $\mathbb{M}$ modulo the space of gauge fields of the form $s\kappa + s' \omega_{\zeta}$, up to equivalence $\kappa \sim t \kappa$. In what follows, we will also refer to the dual theory \eqref{dual} as the contact 4d-CS theory\footnote{This name, taken from \cite{Mickler}, was initially coined by the author of that reference to refer to the dual action of a conventional 3d-CS theory on a Seifert manifold.}. 

The general variation of the dual 4d-CS theory action is
\begin{equation}
\delta S(\mathbb{A})_{\text{dual}} =-2ic \dint_{\mathbb{M}} \text{Tr}\Big[\delta\mathbb{A}\wedge\big(\omega_{\zeta}\wedge F_{\mathbb{A}-\kappa \Phi_{\text{on}}}  \big) \Big]+2ic \dint\nolimits_{%
\mathbb{M}}d_{\mathbb{M}}\omega_{\zeta} \wedge \text{Tr}%
\Big[ \delta \mathbb{A\wedge }\Big( \mathbb{A }-\frac{1}{2}\kappa\iota_{X}%
\mathbb{A}-\kappa \Phi_{\text{on}}\Big)\Big].
\end{equation}
From this result, we extract the bulk eom
\begin{equation}
\omega_{\zeta}\wedge F_{\mathbb{A}-\kappa \Phi_{\text{on}}}=0
\end{equation}
and, as before, the boundary eom which are now given by the vanishing of 
\begin{equation}
\mathcal{B}(\mathbb{A})_{\text{dual}}=\mathcal{B}(\mathbb{A})+\dint\nolimits_{%
\mathbb{M}}d_{\mathbb{M}}\omega_{\zeta} \wedge \kappa\wedge \text{Tr}%
\big( \delta \mathbb{A }\Phi_{\text{on}}\big). \label{BC dual 1}
\end{equation}

It is important to emphasize that the main and only reason for regularizing the original action \eqref{4d CS} in the form \eqref{Omega shift inv} is to guarantee, as we shall review below, that the top form $\gamma_{\text{top}}$ defined in \eqref{gamma top'} is, up to a multiplicative constant, a genuine volume form on $\mathbb{M}$. Notice that in the action \eqref{dual}, we still can fix the $\kappa$-shift symmetry via the gauge fixing condition
\begin{equation}
\Phi_{\text{on}}=0, \label{Phion=0}
\end{equation}
hence recovering the regularized action \eqref{Omega shift inv} and this is because of the on-shell field \eqref{Phi eom} satisfies
\begin{equation}
^{\kappa}\Phi_{\text{on}} =\Phi_{\text{on}} +s,\text{ \qquad \qquad  }^{\omega_{\zeta}}\Phi_{\text{on}} =\Phi_{\text{on}}, \label{key rel}
\end{equation}
showing that this gauge is also admissible. After getting back the action \eqref{Omega shift inv}, we implement the limit $\zeta \rightarrow  0$, to undo the initial substitution $\omega_{C} \rightarrow \omega_{\zeta}$, in order to recover the original action \eqref{4d CS} up to a boundary term, which will vanish at the end by impositions of appropriate solutions to the boundary eom. This is a partial de-regularization though, because of the change $\Sigma \times C \rightarrow \mathbb{R} \times \text{M}$, can not be undone in a continuous way due to obvious topological reasons.

We end this section by summarizing our findings related to the field theory duality properties of the addressed 4d theories  in the following diagram:
\begin{equation*}
\begin{array}{ccc}
S(\mathbb{A})_{4d\text{-CS}}=\small{\text{action\,}} \eqref{4d CS} & \qquad \underset{\omega_{\zeta} \text{-shift
extension}}{\overset{\Sigma \times C \rightarrow \mathbb{R}\times \text{M} }{\xrightarrow{\hspace*{2.5cm}}  } }& \qquad  \boxed{S(\mathbb{A})_{\text{reg}}=\small\text{action\,} \eqref{Omega shift inv}}
\\ 
&  &  \\ 
\; \;  \textcolor{red}{\Bigg\uparrow}
\begin{array}{c}
\footnotesize\text{step I: }\, \scriptstyle{\Phi_{\text{on}}=0}  \\ 
\! \! \! \! \! \! \! \! \!  \quad  \footnotesize\text{step II: }\, \scriptstyle{\zeta \rightarrow 0}%
\end{array}
&  & \qquad \Bigg\downarrow \, \kappa \footnotesize\text{-shift extension} \\ 
&  &  \\ 
\boxed{S(\mathbb{A})_{\text{dual}}=\small\text{action\,}\eqref{dual}}
& \qquad  \overset{\Phi \text{-integration}}{\xleftarrow{\hspace*{2.5cm}} } & \qquad  S(\mathbb{A},\Phi )_{\text{ext}}=%
\small\text{action\,} \eqref{double shift inv}%
\end{array}
\end{equation*}

The boxed action functionals in the diagram, are dual to each other and are classically equivalent. After implementing steps I and II, we do not recover the original theory \eqref{4d CS}, but obtain instead a variant thereof defined on $\mathbb{R}\times \text{M}$ plus a boundary term, hence the red arrow on the left. We keep all boundary terms until we conclude, that the $(\mathbb{A}, \Phi)$ boundary eom solutions and any eventual new supplementary condition to be imposed upon the 1-form $\kappa$, are simultaneously satisfied and consistent with the original CY 4d CS theory boundary eom solutions. 

\subsubsection{Gauge invariance I}\label{2.1.4}

Here, we briefly consider the behavior of the action functionals introduced so far, namely \eqref{Omega shift inv}, \eqref{double shift inv} and \eqref{dual}, under the action of the group $\mathcal{G}$ of formal gauge transformations. Thus, under the usual changes
\begin{equation}
^{g}\mathbb{A}=g^{-1}\mathbb{A}g+g^{-1}d_{\mathbb{M}}g,\text{ \qquad \qquad  }^{g}\Phi=g^{-1}\Phi g 
\end{equation}
and
\begin{equation}
^{g}\Phi_{\text{on}}=g^{-1}\Phi_{\text{on}}g,
\end{equation}
with $g:\mathbb{M}\rightarrow G $, i.e. an element of $ \mathcal{G}$, the regularized 4d CS theory action transforms as follows
\begin{equation}
S\big(\! \,^{g}\mathbb{A}  \big)_{\text{reg}}=S\left(\mathbb{A}  \right)_{\text{reg}}+ic\dint\nolimits_{\mathbb{M}}d_{\mathbb{M}}\omega_{\zeta}\wedge \kappa\wedge \text{Tr}\big(\mathbb{J}W_{\text{reg}}\big)+ ic\dint\nolimits_{\mathbb{M}}\omega_{\zeta}\wedge \chi(g), \label{reg gauge transf}
\end{equation}
where $W_{\text{reg}}\in \Omega_{\mathbb{M}}^{0}\otimes \mathfrak{g}$ is given by
\begin{equation}
W_{\text{reg}}=2\iota_{X}\mathbb{A}+\iota_{X}\mathbb{J},
\end{equation} 
$\mathbb{J}=d_{\mathbb{M}}gg^{-1}\in \Omega_{\mathbb{M}}^{1}\otimes \mathfrak{g}$ and $\chi(g)\in \Omega_{\mathbb{M}}^{3}$ is the Wess-Zumino (WZ) 3-form given by
\begin{equation}
\chi(g)=-\frac{1}{3}\text{Tr}(\mathbb{J}\wedge \mathbb{J}\wedge \mathbb{J}).
\end{equation}
In showing \eqref{reg gauge transf}, one is to use the identity
\begin{equation}
CS(^{g}\mathbb{A})=CS(\mathbb{A})+d_{\mathbb{M}}\text{Tr}(\mathbb{A}\wedge \mathbb{J})+\chi(g) \label{CS gauge}
\end{equation} 
and the trivial 5-form contraction
\begin{equation}
0=\iota_{X}\Big( d_{\mathbb{M}}\omega_{\zeta}\wedge \kappa \wedge \text{Tr}\big(\mathbb{A}\wedge \mathbb{J}\big)  \Big),
\end{equation}
in order to write
\begin{equation}
d_{\mathbb{M}}\omega_{\zeta}\wedge \text{Tr}\big(\mathbb{A}\wedge \mathbb{J}\big)=d_{\mathbb{M}}\omega_{\zeta}\wedge \kappa \wedge \text{Tr}\big(\mathbb{J}\iota_{X} \mathbb{A}- \mathbb{A}\iota_{X} \mathbb{J}\big).
\end{equation}

For the extended and dual actions we find, respectively, that
\begin{equation}
\begin{aligned}
S\big(\! \,^{g}\mathbb{A},\! \, ^{g}\Phi  \big)_{\text{ext}}&=S\big(\mathbb{A}, \Phi  \big)_{\text{ext}}+ic\dint\nolimits_{\mathbb{M}}d_{\mathbb{M}}\omega_{\zeta}\wedge \kappa\wedge \text{Tr}\big(\mathbb{J}W_{\text{ext}}\big)+ ic\dint\nolimits_{\mathbb{M}}\omega_{\zeta}\wedge \chi(g),\\
S\big(\! \,^{g}\mathbb{A} \big)_{\text{dual}}&=S\big(\mathbb{A}\big)_{\text{dual}}+ic\dint\nolimits_{\mathbb{M}}d_{\mathbb{M}}\omega_{\zeta}\wedge \kappa\wedge \text{Tr}\big(\mathbb{J}W_{\text{reg}}\big)+ ic\dint\nolimits_{\mathbb{M}}\omega_{\zeta}\wedge \chi(g),
\end{aligned}
\end{equation}
where
\begin{equation}
W_{\text{ext}}=W_{\text{reg}}-2\Phi.
\end{equation}

In all three cases we have the following general structure
\begin{equation}
^{g}S=S+ic\dint\nolimits_{\mathbb{M}}d_{\mathbb{M}}\omega_{\zeta}\wedge \kappa\wedge \text{Tr}(\mathbb{J}W)+ ic\dint\nolimits_{\mathbb{M}}\omega_{\zeta}\wedge \chi(g), \label{general gauge transf}
\end{equation}
for $W\in \Omega_{\mathbb{M}}^{0}\otimes \mathfrak{g}$. Below we will find, in a systematic way, what are the appropriate conditions to be imposed upon the gauge group elements $g\in \mathcal{G}$ and the gauge field $\mathbb{A}$ that turn any of the three action functionals introduced above genuinely gauge invariant. However, in order to show this, a detailed study of the boundary eom solution space is required first. Thus, we will postpone the analysis of gauge invariance until this task is completed. For the time being, let us denote by $\mathcal{G}_{0}\subset \mathcal{G}$, the subgroup of the group of formal gauge transformations ensuring that $^{g}S=S$ and refer to it simply as the gauge group, which is the true physical gauge symmetry group.

\subsection{Defect data}\label{2.2}

Here we introduce, in the regularized 4d CS theory, classical (co)-adjont orbit insertions of the 1d CS theory type. They modify, as we shall see below in $\S \eqref{7}$, the quantum field theory formulation of the regularized theory via path integrals.

\subsubsection{1d CS theory}

Let us pick $N$ points $z_{j}\in C$, $j=1,...,N$ on the base manifold $C$ and consider the circle fibers $S^{1}$ on top of them in the total space $\text{M}\subset \mathbb{M}$. We assume that these points are rather generic and do not belong to $\mathfrak{z}$, $\mathfrak{p}$ or are orbifolds points in $C$ in the general case of a generic Seifert manifold. Attach a defect action functional of the 1d CS theory type to each one of these circles, defined by   
\begin{equation}
S(U_{j}, \mathbb{A})_{1d\text{-CS}}:=2cl \oint_{S^{1}}\text{Tr}(\lambda_{j}f^{-1}d_{\mathbb{A}}f),\text{ \qquad \qquad  }j=1,...,N \label{1d CS}
\end{equation}
where $d_{\mathbb{A}}:=d_{\mathbb{M}}+\mathbb{A}|_{S^{1}}\cdot$, $f: S^{1}\rightarrow G$ belongs to the loop group $LG$, any $\lambda_{j} \in \mathfrak{t}\subset \mathfrak{g}$ is a constant element in the Cartan subalgebra $\mathfrak{t}$ of $\mathfrak{g}$ and $l\in \mathbb{R}$ is a constant. Notice that the lhs in \eqref{1d CS} exhibits a dependence on $(\tau,z_{j})$, because of the gauge field $\mathbb{A}\in \Omega_{\mathbb{M}}^{1}\otimes \mathfrak{g}$ is only partially integrated against $S^{1}$. The $j$-th defect data is then encoded in the pair $(z_{j},\lambda_{j})$. 

This defect theory is invariant under the action of the formal gauge group $\mathcal{G}$
\begin{equation}
^{g}\mathbb{A}=g^{-1}\mathbb{A}g+g^{-1}d_{\mathbb{M}}g,\text{ \qquad \qquad  }^{g}f=g^{-1}|_{S^{1}}\cdot f,  \label{gauge defects}
\end{equation}
where $g\in \mathcal{G}$ and it is also invariant under the local action of the stabilizer group $G_{\lambda}\subset G$ of $\lambda$, that fixes it under the adjoint action, i.e.
\begin{equation}
^{h}f=fh,
\end{equation}
where $h:S^{1}\rightarrow G_{\lambda}$, provided the quantity
\begin{equation}
2cl\oint_{S^{1}}d_{\mathbb{M}}\text{Tr}(\lambda_{j}\eta)
\end{equation} 
vanishes or it is defined modulo $2\pi$. In computing the latter, we have used $h=e^{\eta}$, with $\eta: S^{1}\rightarrow \mathfrak{g}_{\lambda}$ being a map from $S^{1}$ to the Lie algebra of $G_{\lambda}$. Furthermore, the action is invariant under the $\omega_{\zeta}$-shifts \eqref{Omega shift symmetry} as well, with $\omega_{\zeta}$ as in \eqref{spoiler sol iii} and this is because none of the differential 1-forms, $d\tau$ or $\omega=\underline{\pi}^{\ast}(\omega_{C})$ entering $\omega_{\zeta}$, have support along the $S^{1}$ fiber that provides the integration domain. In this way, the defect action is valued on the same $\zeta$-dependent quotient space of the regularized 4d CS theory. 

For example, when the Lie group $G$ is real and compact, the 1d CS theory action functional \eqref{1d CS} for $\mathbb{A}=0$, describes \cite{Jones} the low-energy limit of a 1d sigma model describing the propagation of a particle, with periodic `time', on a background space given by the adjoint orbit $\mathcal{O}_{\lambda}$, defined by the constant element $\lambda \in \mathfrak{t}$ and the adjoint action of $G$. We can identify the adjoint and the coadjoint orbits by virtue of the isomorphism $\mathfrak{g}^{\ast}\cong \mathfrak{g}$ induced by the bilinear form Tr and in what follows we will use both names interchangeably, unless this lead to any confusion. This 1d sigma model is defined in terms of the embedding map
\begin{equation}
U_{\lambda}:S^{1}\rightarrow \mathcal{O}_{\lambda},\text{ \qquad \qquad }U_{\lambda}=f\lambda f^{-1},
\end{equation} 
where the last expression shows the explicit relation between the maps $U$ and $f$ used on the lhs of \eqref{1d CS}. Notice that $U_{\lambda}\in L\mathcal{O}_{\lambda}$, i.e. it is actually an element of the loopspace of $\mathcal{O}_{\lambda}$. The left action of the Lie group $G$ on $f$ is a global symmetry of this system and by minimally coupling the field $f$ to the gauge field $\mathbb{A}$ restricted to $S^{1}$, we promote the global symmetry group $G$ to the formal gauge group $\mathcal{G}$, as seen in the expression \eqref{gauge defects}. Notice that here we are using the formal gauge group $\mathcal{G}$ instead of $\mathcal{G}_{0}$. 

The action functional \eqref{1d CS} is constructed in terms of quantities all belonging to a complex Lie group and as a consequence, the orbit $\mathcal{O}_{\lambda}$ is actually a complex Lie group (co)-adjoint orbit. These kind or orbits are of a slightly different nature when compared to their counterparts based on compact real Lie groups, which are quite well-understood \cite{Besse}. Complex (co)-adjoint orbits possesses more than one complex structure leading to the existence of holomorphic pseudo-K\"ahler structures \cite{Wagner} which, roughly speaking, define K\"ahler metrics that are not necessarily positive-definite. For the time being we only need to pay attention to the reality properties of the action \eqref{1d CS}, to be studied in $\S \eqref{3}$, postponing the review of complex coadjoint orbits, at least in some particular cases, to $\S \eqref{6.1.3}$, where the understanding of their structural aspects is required for constructing their associated loop-space path integral Liouville volume forms. 

For further reference, it is useful to introduce the following integration formulas
\begin{equation}
\oint_{S^{1}} \gamma=\dint\nolimits_{\mathbb{M}}\delta_{S^{1}}\wedge \gamma,\text{ \qquad \qquad  } \oint_{S^{1}}\kappa\, \iota_{X}\gamma=\oint_{S^{1}}\gamma,\label{integral formula}
\end{equation}
where $\gamma \in \Omega_{\mathbb{M}}^{1}$ is any 1-form on $\mathbb{M}$ and $\delta_{S^{1}}\in \Omega_{\mathbb{M}}^{3}$ is a 3-form with Dirac delta support along $S^{1}$ that represents the Poincare dual of $S^{1}$ in $\mathbb{M}$. The first integral formula is useful when we want to extend quantities defined on $S^{1}$ to the whole 4d manifold $\mathbb{M}$. For example, with the help of \eqref{integral formula}, a general variation of the 1d CS theory action \eqref{1d CS}, can be written in the form
\begin{equation}
\delta S(U_{j}, \mathbb{A})_{1d\text{-CS}}=2cl\dint\nolimits_{\mathbb{M}}\delta_{S^{1}}\wedge \text{Tr}(\delta \mathbb{A}U_{j})-2cl \oint\nolimits_{S^{1}}\text{Tr}\left(\delta ff^{-1}d_{\mathbb{A}}U_{j}   \right),\label{1d CS var}
\end{equation} 
where $d_{\mathbb{A}}:=d_{\mathbb{M}}+[\mathbb{A}|_{S^{1}},\, \cdot\,]$ and\footnote{We find more convenient to use, in $U$, the label $j$ rather than $\lambda_{j}$.} $U_{j}=f \lambda_{j}f^{-1}$. From the first integration formula in \eqref{integral formula}, it is understood that $\delta_{S^{1}}$ is attached to the index $j$, although we denote it generically by $\delta_{S^{1}}$, this should be clear from the context. We will use this variation to compute the bulk eom and the boundary eom of the contact 4d CS theory with $N$ (co)-adjoint orbit defect insertions. 

\subsection{Dual action with defects}

Introduce now the regularized action \eqref{Omega shift inv} with the $N$ defects 
\eqref{1d CS}, defined by
\begin{equation}
S(\mathbb{A},U)_{\text{reg}}:=S(\mathbb{A})_{\text{reg}}+\sum_{j=1}^{N} S(U_{j},\mathbb{A})_{1d\text{-CS}}. \label{reg action N}
\end{equation}

A general variation of \eqref{reg action N}, is given by
\begin{equation}
\delta S(\mathbb{A},U)_{\text{reg}}= 2ic\dint_{\mathbb{M}}  \text{Tr} \Big[\delta \mathbb{A}\wedge \big(-\omega_{\zeta}\wedge F_{\mathbb{A}}+\delta_{S^{1}}\cdot \tilde{U}  \big)   \Big]-2cl \sum_{j=1}^{N}\oint\nolimits_{S^{1}}\text{Tr}\left(\delta ff^{-1}d_{\mathbb{A}}U_{j}   \right)+2ic\, \mathcal{B}(\mathbb{A}), \label{var reg N}
\end{equation}
where we have used \eqref{BC M}, \eqref{1d CS var} and defined the quantity
\begin{equation}
\tilde{U}= il\sum_{j=1}^{N} U_{j}.
\end{equation}
The boundary eom and the bulk eom are given, respectively, by the vanishing of \eqref{BC M} and 
\begin{equation}
\omega_{\zeta}\wedge F_{\mathbb{A}}=\delta_{S^{1}}\cdot \tilde{U} ,\text{ \qquad \qquad  }d_{\mathbb{A}}U_{j}=0, \text{ \qquad \qquad  }j=1,...,N\label{ext reg eom}
\end{equation}
Notice that the curvature $F_{\mathbb{A}}$ is zero only away the $N$ circles located at the insertion points $z_{j}$ and that the (co)-adjoint orbits $U_{j}$ are covariantly constant, for any $j$.

A natural extension to the action \eqref{reg action N}, that is invariant under $\kappa$-shifts as well, is given by the extended action with $N$ defect insertions, defined by
\begin{equation}
\begin{aligned}
S(\mathbb{A},\Phi,U )_{\text{ext}}&:=S(\mathbb{A}-\kappa \Phi,U)_{\text{reg}} \label{ext action N}\\
&=S(\mathbb{A},\Phi)_{\text{ext}}+\sum_{j=1}^{N} S(U_{j},\mathbb{A})_{1d\text{-CS}}-2cl\sum_{j=1}^{N}\oint_{S^{1}}\kappa\, \text{Tr}\left(\Phi U_{j}  \right),
\end{aligned}
\end{equation}
where we have assumed that the $\kappa$-shifts act trivially on $U$. As done before, we can recover \eqref{reg action N} in the gauge $\Phi=0$ or, alternatively, obtain a classically equivalent dual action after using the quadratic field $\Phi$ eom, as we shall see right below.

An arbitrary variation of the action \eqref{ext action N}, is easily computed with the help of the results \eqref{ext act var} and \eqref{1d CS var} found previously. We find that
\begin{equation}
\begin{aligned}
\delta S(\mathbb{A},\Phi,U )_{\text{ext}}=\, & 2ic \dint_{\mathbb{M}} \text{Tr}\Big[\delta\mathbb{A}\wedge\Big(-\omega_{\zeta}\wedge F_{\mathbb{A}-\kappa \Phi} +\delta_{S^{1}}\cdot \tilde{U} \Big) \Big]+2ic\, \mathcal{B}(\mathbb{A},\Phi) \\
&+2ic \dint_{\mathbb{M}}\text{Tr}\Big[\delta \Phi \Big(\gamma_{\text{top}}   \Phi -\omega_{\zeta}\wedge \kappa \wedge F_{\mathbb{A}}+\delta_{S^{1}}\wedge \kappa \cdot \tilde{U}\Big)    \Big] \\
&-2cl \sum_{j=1}^{N}\oint_{S^{1}}\text{Tr}\big( \delta f f^{-1}d_{\mathbb{A}-\kappa \Phi}U_{j}  \big),
\end{aligned}
\end{equation} 
where we have used the variation
\begin{equation}
\delta \oint_{S^{1}}\kappa\, \text{Tr}\left(\Phi U_{j}\right)=\dint_{\mathbb{M}} \delta_{S^{1}}\wedge \kappa\, \text{Tr}\left(\delta\Phi U_{j}  \right)-\oint_{S^{1}}\kappa \, \text{Tr} \Big(\delta f f^{-1}[\Phi,U_{j}]   \Big)
\end{equation}
in order to deal with the last contribution in \eqref{ext action N} and the extended 4d CS theory boundary eom expression \eqref{full BC M}. From this, we read off all the bulk eom. For instance, the field $\Phi$ eom is 
\begin{equation}
\tilde{\Phi}_{\text{on}} =
\frac{\omega_{\zeta} \wedge \kappa \wedge F_{\mathbb{A}}-\delta_{S^{1}}\wedge \kappa \cdot \tilde{U}}{\gamma_{\text{top}} }.\label{Phi defect  eom}
\end{equation}
The other two bulk eom being given by
\begin{equation}
\omega_{\zeta}\wedge F_{\mathbb{A}-\kappa \Phi} =\delta_{S^{1}}\cdot \tilde{U},\text{ \qquad \qquad  }d_{\mathbb{A}-\kappa \Phi}U_{j}=0,\text{ \qquad \qquad  }j=1,...,N.
\end{equation}
As before, we can obtain \eqref{Phi defect eom} by wedging the first equation right above against $\kappa$. The boundary eom of the theory thus correspond to the vanishing of \eqref{full BC M}.

After putting the on-shell expression $\tilde{\Phi}_{\text{on}}$ back into the action \eqref{ext action N}, we get the contact 4d-CS theory action with $N$ defect insertions, namely
\begin{equation}
S(\mathbb{A},U)_{\text{con}} =S(\mathbb{A},U)_{\text{reg}} -ic\dint\nolimits_{\mathbb{M}}\gamma_{\text{top}} \text{Tr}\left( \tilde{\Phi}_{\text{on}} ^{2}%
\right).\label{defect dual}
\end{equation} 
Finally and for the sake of completeness, we write the general variation of the action \eqref{defect dual}. Using the variation \eqref{var reg N}, we get 
\begin{equation}
\begin{aligned}
\delta S(\mathbb{A},U)_{\text{con}}=2ic&\dint\nolimits_{\mathbb{M}} \text{Tr}\Big[ \delta 
\mathbb{A} \wedge \Big(-\omega_{\zeta}\wedge F_{\mathbb{A}-\kappa \tilde{\Phi}_{\text{on}}}+\delta_{S^{1 }}\cdot \tilde{U} \Big)\Big] +2ic \,\tilde{\mathcal{B}}(\mathbb{A})_{\text{dual}} \\ 
&-2cl\sum_{j=1}^{N}\oint_{S^{1}}\text{Tr}\Big(\delta ff^{-1}d_{\mathbb{A}-\kappa \tilde{\Phi}_{\text{on}}}U_{j}  \Big).
\end{aligned} 
\end{equation}
The bulk eom are then given by
\begin{equation}
\omega_{\zeta}\wedge F_{\mathbb{A}-\kappa \tilde{\Phi}_{\text{on}}} =\delta_{S^{1}}\cdot \tilde{U},\text{ \qquad \qquad  }d_{\mathbb{A}-\kappa \tilde{\Phi}_{\text{on}}}U_{j}=0,\text{ \qquad \qquad  }j=1,...,N
\end{equation}
while the boundary eom correspond to the vanishing of
\begin{equation}
\tilde{\mathcal{B}}(\mathbb{A})_{\text{dual}}=\mathcal{B}(\mathbb{A})+\dint\nolimits_{%
\mathbb{M}}d_{\mathbb{M}}\omega_{\zeta} \wedge \kappa \wedge \text{Tr}%
\big( \delta \mathbb{A} \tilde{\Phi}_{\text{on}}\big). \label{BC dual defect}
\end{equation}
Basically, the only difference between \eqref{BC dual 1} and \eqref{BC dual defect} is the change $\Phi_{\text{on}}\rightarrow \tilde{\Phi}_{\text{on}}$.


\section{Reality conditions}\label{3}

In this section we briefly review the reality conditions \cite{unifying} that are necessary to ensure that all relevant objects considered in our discussion are indeed real-valued. The only new ingredient to be added here and that was not considered previously in \cite{Yo}, is the association of an equivariant differential form to a real-valued differential form. An example of this will be given in \S \eqref{5}, for the case of a real contact form. 

Recall that the Lie algebra $\mathfrak{g}$ is assumed to be complex. Let $\hat{\tau}:\mathfrak{g}\rightarrow \mathfrak{g} $ be an anti-linear involutive automorphism. It provides $\mathfrak{g}$ with an action of the cyclic group $\mathbb{Z}_{2}$. Its fixed point
subset is a real Lie subalgebra $\mathfrak{g}^{\mathbb{R}}$ of $\mathfrak{g}$, regarded itself as a real Lie algebra. The
anti-linear involution $\hat{\tau}$ is compatible with the bilinear form on $\mathfrak{g}$, in the sense that
\begin{equation}
\overline{\text{Tr}(ab)}=\text{Tr}\big(\hat{\tau} (a)\hat{\tau} (b)\big),
\end{equation}
for any elements $a,b$ in the Lie algebra $\mathfrak{g}$. 
Denote by $x$ a set of local coordinates on the manifold $\mathbb{M}$ and endow it with a complex structure. The complex conjugation $x\rightarrow \overline{x}$ defines an involution $\nu : \mathbb{M}\rightarrow \mathbb{M}$, which also provides $\mathbb{M}$ with a $\mathbb{Z}_{2}$ action. We then require that any differential form $\gamma \in \Omega_{\mathbb{M}}^{\bullet}$ and any Lie algebra valued differential form $\rho\in \Omega_{\mathbb{M}}^{\bullet}\otimes \mathfrak{g}$, is equivariant under the action of $\mathbb{Z}_{2}$, in the sense of the following commutative diagrams
\begin{equation*}
\begin{array}{ccc}
\mathbb{M} & \overset{\nu}{{\xrightarrow{\hspace*{1.5cm}}  }} & \mathbb{M} \\ 
\! \! \! \!    \gamma\, \Bigg\downarrow &  &\! \! \! \! \gamma\, \Bigg\downarrow \\ 
\Omega_{\mathbb{M}}^{\bullet}  & \overset{\nu}{{\xrightarrow{\hspace*{1.5cm}}  }} & \Omega_{\mathbb{M}}^{\bullet} 
\end{array}%
\text{ \qquad \qquad \ \ \ \ \ \ }%
\begin{array}{ccc}
\text{\ }\mathbb{M} & \overset{\nu}{{\xrightarrow{\hspace*{1.5cm}}  }} & \mathbb{M} \\ 
\! \! \rho\, \Bigg\downarrow &  &\! \! \! \rho\, \Bigg\downarrow \\ 
\Omega_{\mathbb{M}}^{\bullet}\otimes \mathfrak{g}  & \overset{\hat{\tau}}{{\xrightarrow{\hspace*{1.5cm}}  }} & \Omega_{\mathbb{M}}^{\bullet}\otimes \mathfrak{g} 
\end{array}%
\end{equation*}
\break
Above, we have interpreted $\gamma$ and $\rho$ as maps (sections) from $\mathbb{M}$ to $\Omega_{\mathbb{M}}^{\bullet}$ and $\Omega_{\mathbb{M}}^{\bullet}\otimes \mathfrak{g}$, respectively. In practice, we require that 
\begin{equation}
\overline{\gamma(x)}=\gamma(\overline{x})=(\nu^{\ast}\gamma)(x), \text{ \qquad \qquad } \hat{\tau}\big(\rho(x)\big)=\rho(\overline{x})=(\nu^{*}\rho)(x). \label{equiv cond}
\end{equation}
It follows from \eqref{equiv cond}, that the exterior derivatives of $\gamma$ and $\rho$ are equivariant as well, i.e.
\begin{equation}
\overline{d_{\mathbb{M}}\gamma(x)}=(d_{\mathbb{M}}\gamma)(\overline{x})=(\nu^{\ast}d_{\mathbb{M}}\gamma)(x), \text{ \qquad \qquad } \hat{\tau}\big(d_{\mathbb{M}}\rho(x)\big)=(d_{\mathbb{M}}\rho)(\overline{x})=(\nu^{*}d_{\mathbb{M}}\rho)(x). \label{equiv cond'}
\end{equation}
Furthermore, any map $u:\mathbb{M}\rightarrow G$ like, for example, an element of the formal gauge group $\mathcal{G}$, is also required to be equivariant, i.e.
\begin{equation}
\hat{\tau}\big( u(x)\big)=u(\overline{x}),\label{tau on G}
\end{equation}
with $\hat{\tau}$ denoting the lift of the antilinear involutive automorphism to the group $G$. This is to guarantee, of course, that gauge transformations also preserve the reality conditions if, for instance, we consider that $\rho=\mathbb{A}$ is a gauge field.

For example, if we consider 1-forms and Lie algebra valued 1-forms, in the local coordinates $x^{\tilde{\mu}}$, with $\tilde{\mu}=0,1,2,3$, the expressions \eqref{equiv cond} and \eqref{equiv cond'} become
\begin{equation}
\overline{\gamma_{\tilde{\mu}}(x)}=\gamma_{\tilde{\mu}}(\overline{x}),\text{\qquad \qquad }\hat{\tau}\big(\rho_{\tilde{\mu}}(x)  \big)=\rho_{\tilde{\mu}}(\overline{x}), \label{compo}
\end{equation}
and
\begin{equation}
\overline{(\partial_{\tilde{\mu}}\gamma_{\tilde{\nu}})(x)}=\big(\nu^{\ast}\partial_{\tilde{\mu}}\gamma_{\tilde{\nu}}\big)(x) ,\text{\qquad \qquad } \hat{\tau}\big[(\partial_{\tilde{\mu}}\rho_{\tilde{\nu}})(x)\big]=  \big(\nu^{\ast}\partial_{\tilde{\mu}}\rho_{\tilde{\nu}}\big)(x),            \label{compo'}
\end{equation}
respectively.

One particular case of interest, refers to those elements $\gamma$ and $\rho$ having coordinate-independent components. In this case, we have that
\begin{equation}
\overline{\gamma_{\tilde{\mu}}}=\gamma_{\tilde{\mu}},\text{\qquad \qquad}\hat{\tau}\big(\rho_{\tilde{\mu}} \big)=\rho_{\tilde{\mu}}. \label{result equi}
\end{equation}
The first expression imply that $\gamma_{\tilde{\mu}}$ must be real, while the second one imply that $\rho_{\tilde{\mu}}$ must be $\hat{\tau}$-invariant, hence belonging to the real form $\mathfrak{g}^{\mathbb{R}}$. Another situation of interest, refers to the case when a real differential form is given but one desires instead to construct an equivariant extension thereof. In other words, given $\overline{\gamma(x)}=\gamma(x)$, we are interested in associating an equivariant extension $\gamma_{\text{eq}}$, such that $\overline{\gamma_{\text{eq}}(x)}=\gamma_{\text{eq}}(\overline{x})$. Fortunately, we will only need to do this when $\gamma=\alpha_{r}$ and $\alpha_{r}$ is a real contact form structure defined on $\text{M}\subset \mathbb{M}$. We will show this explicitly later on in $\S \eqref{5}$.

We are now ready to apply the reality conditions to the action functionals introduced above. On the bulk side, at the end of the day, we only need to show that the extended action \eqref{double shift inv} is real-valued, as it reproduces, via duality manipulations, the regularized and the contact 4d CS theory actions. Thus, as in the conventional 4d CS theory \cite{unifying}, the 1-form $\omega$ is required to be equivariant and from \eqref{spoiler sol iii}, we notice that $\omega_{\zeta}$ is the sum of an equivariant 1-form and a 1-form with a constant real coefficient, which by virtue of the first relation in \eqref{result equi} is equivariant as well. In addition, the Lie algebra valued fields $\Phi$ and $\mathbb{A}$ are also required to be equivariant. All these requirements imply that the first term on the rhs of \eqref{double shift inv}, i.e. the action $S(\mathbb{A})_{\text{reg}}$, is a real number. For the second contribution, one faces the necessity of requiring that top form $\gamma_{\text{top}}$ is equivariant as well, while for the third term, one then requires the same for the 1-form $\kappa$. Because of $\gamma_{\text{top}}=\omega_{\zeta}\wedge \kappa \wedge d_{\mathbb{M}}\kappa$, we only need to demand that $\kappa$ is an equivariant 1-form. However, as we shall see below, Cf. \eqref{solutions omega, kappa}, we have that
\begin{equation}
\kappa=\zeta'\left(\alpha_{\tau}d\tau+\alpha   \right), \label{capa}
\end{equation}
where $\alpha_{\tau}$ and $\zeta'$ are real constants and $\alpha \in \Omega_{\text{M}}^{1}$ is an equivariant extension of a real contact form $\alpha_{r}$ defined on M. In conclusion, if we write the extended action \eqref{double shift inv} in the form
\begin{equation}
S(\Phi,\mathbb{A})_{\text{ext}}=ic\dint_{\mathbb{M}}\mathcal{L}(x),
\end{equation}
where $\mathcal{L} \in \Omega_{\mathbb{M}}^{4}$ is an equivariant 4-form, we get that
\begin{equation}
\overline{S(\Phi,\mathbb{A})_{\text{ext}}}=-ic\dint_{\mathbb{M}}\overline{\mathcal{L}(x)}=-ic\dint_{\mathbb{M}}(\nu^{\ast}\mathcal{L})(x)=-ic\dint_{\nu(\mathbb{M})}\mathcal{L}(x)=S(\Phi,\mathbb{A})_{\text{ext}},
\end{equation}
where we have used $\nu({\mathbb{M}})=-\mathbb{M}$, i.e. $\mathbb{M}$ with its orientation being reversed. Furthermore, the requirement that the 1-forms $\omega_{\zeta},\kappa$ are equivariant is a necessary condition for preserving the nature of the gauge connection under $\omega_{\zeta}$-shifts and $\kappa$-shifts. Indeed, we have that
\begin{equation}
\hat{\tau}\left( ^{\omega_{\zeta}}\mathbb{A}  \right)=\hat{\tau}\left( \mathbb{A} +s\omega_{\zeta} \right)=\nu^{\ast} \left( ^{\omega_{\zeta}}\mathbb{A} \right),\text{ \qquad \qquad}\hat{\tau}\left( ^{\kappa}\mathbb{A}  \right)=\hat{\tau}\left( \mathbb{A} +s\kappa \right)=\nu^{\ast} \left( ^{\kappa}\mathbb{A} \right),
\end{equation} 
where one requires that $s\in \Omega_{\mathbb{M}}^{0}\otimes \mathfrak{g}$ is, of course, equivariant too. 

On the defect side, consider the (co)-adjoint orbit defect action \eqref{1d CS} 
\begin{equation}
S(U_{j}, \mathbb{A})(\tau, z_{j})=2cl \oint_{S^{1}}\text{Tr}(\lambda_{j}f^{-1}d_{\mathbb{A}}f), 
\end{equation} 
where, on the lhs, we have dropped the label 1d-CS and exhibited explicitly its $(\tau,z_{j})$ dependence that comes from the restriction of the connection\footnote{We have used local bundle coordinates $(\sigma,z)$ on M plus time, i.e. $x=(\tau,\sigma,z)$.} $\mathbb{A}(\tau,\sigma,z_{j})$ to $S^{1}$. Its complex conjugate is given by
\begin{equation}
\overline{S(U_{j}, \mathbb{A})(\tau, z_{j})}=2cl\oint_{S^{1}}\text{Tr}\Big[\hat{\tau}\big(\lambda_{j}\big)\hat{\tau}\big(f^{-1}\big)d_{\hat{\tau}(\mathbb{A})}\hat{\tau}\big(f\big) \Big].
\end{equation}
Now, because of
\begin{equation}
\hat{\tau}\Big(\mathbb{A}(\tau,\sigma,z) \Big|_{S^{1}}\Big)=\mathbb{A}(\tau,\sigma,\overline{z})\Big|_{S^{1}},
\end{equation}
we find that
\begin{equation}
\overline{S(U_{j}, \mathbb{A})(\tau, z_{j})}=S\big(\hat{\tau}(U_{j}), \mathbb{A}\big)(\tau, \overline{z}_{j})
\end{equation}
Thus, for the defect action contribution to be real, the number of (co)-adjoint orbit insertions must, naively, be an even number $N\in 2\mathbb{Z}_{+}$ and the defect data has now the following structure
\begin{equation}
\big(z_{k},z_{k+1}=\overline{z}_{k}\big),\text{ \qquad \qquad}\big(U_{k},U_{k+1}=\hat{\tau}(U_{k})\big),\text{ \qquad \qquad}k=1,...N/2. \label{defect data'}
\end{equation} 

In summary, we have already presented all the 4d action functionals of relevance to our approach, briefly analyzed how they behave under gauge symmetries and even specified their reality conditions. As commented before in the introduction, Cf. II), the generalization of the 4d CS theory introduced in \cite{Yo} does not reproduce the same boundary eom of the original CY 4d CS theory. Thus, an important question that raises now is the following: to what extend the regularized 4d CS theory reproduces the 2d IFT's described by the original 4d CS theory?. Answering this, is the topic of the next section.

\section{Regularized 4d CS theory and 2d IFT's} \label{4}

We explore now the relation between the regularized action \eqref{Omega shift inv} and the 2d IFT's, mostly integrable $\sigma$-models, that are known to live on the pole-defect surface $\Sigma\times \mathfrak{p}$. The latter being extracted from the original action \eqref{4d CS}, after gauge fixing and integration over the `spectral' space\footnote{Thus, $n_{\mathfrak{p}}=n_{\mathfrak{z}}+2$, Cf. \eqref{RRR} with $g=0$.} $C=\mathbb{CP}^{1}$ in all cases.  We start from a total space M defined by $S^{1}\overset{n}{\longhookrightarrow} \text{M}\overset{\underline{\pi}}{\longrightarrow} \mathbb{CP}^{1}$ and explore how a sample of these well-known IFT's emerge in our approach in the $\zeta\rightarrow 0$ limit. We follow \cite{unifying} closely, in regard to the treatment of the boundary eom solutions and the use of some useful general formulas computed there. We will mainly dwell on the key differences between our approach and the original treatment of IFT's based on the 4d CS theory point of view. We finish this section with a brief study of the boundary eom for the extended and contact 4d CS theories as well in the generic case when $\zeta \neq 0$.

\subsection{2d integrable field theories}

As mentioned in $\S (\ref{2})$, the solutions to the bulk eom of the action \eqref{Omega shift inv}, i.e. 
\begin{equation}
S(\mathbb{A})_{\text{reg}}=ic\dint\nolimits_{\mathbb{M}}\omega_{\zeta} \wedge CS\left( 
\mathbb{A}\right) +ic\dint\nolimits_{\mathbb{M}}d_{\mathbb{M}}\omega_{\zeta}\wedge \kappa\wedge \text{Tr}\left( \mathbb{A}\iota_{X}\mathbb{A}  \right), \label{reg N action}
\end{equation}
which are given by 
\begin{equation}
\omega_{\zeta}\wedge F_{\mathbb{A}}=0, \label{reg N eom}
\end{equation}
must also satisfy the boundary eom \eqref{BC M}, namely
\begin{equation}
\mathcal{B}(\mathbb{A})= \dint\nolimits_{%
\mathbb{M}}d_{\mathbb{M}}\omega_{\zeta} \wedge \text{Tr}%
\big( \delta \mathbb{A}\wedge \hat{\mathbb{A}} \big) =0, \text{ \qquad \qquad }\hat{\mathbb{A}}:=\mathbb{A }-\frac{1}{2}\kappa\iota_{X}%
\mathbb{A}. \label{BC}
\end{equation}
When $n=0$, $\kappa =0$ and $\zeta \rightarrow 0$, we get that $\mathbb{M}$, $\omega_{\zeta}$ and \eqref{BC} reduce to the usual 4d CS theory expressions $\Sigma \times C$, $\omega_{C}$ and \eqref{BC C}, respectively.

The goal now is to find solutions to the boundary eom \eqref{BC} for $n\ne 0$, $\kappa \neq 0$ and $\zeta \rightarrow 0$, while we fix all the gauge symmetries of the theory and integrate over the base space $\mathbb{CP}^{1}$, as done in the conventional 4d CS theory \cite{CY}. To do this, we choose coordinates $(\tau,\sigma, z)$ on $\mathbb{M}$, with the pair $(\sigma, z)$ being local bundle coordinates on M. $z$ is a holomorphic coordinate on a chart $\mathcal{U}\subset \mathbb{CP}^{1}$ centered at the origin, while $\sigma \in [0,2\pi]$ is an angular coordinate along the circle fibers $S^{1}$ on top of $\mathcal{U}$. In this way, we have a local description of $\mathbb{R} \times \text{M}$ of the form $\Sigma \times \mathcal{U}$. To describe the point $z=\infty$, we use instead the coordinates $(\sigma',w)$, with $w=1/z$ on a chart $\mathcal{U}' \subset \mathbb{CP}^{1}$ centered at infinity and $\sigma' \in [0,2\pi]$, hence describing $\mathbb{R} \times \text{M}$ locally in the form $\Sigma' \times \mathcal{U}'$. The principal $G$-bundle over $\mathbb{M}$, where the regularized 4d CS theory is defined, is trivial\footnote{We will comment more about this in $\S \eqref{5}$.} and the restrictions of the connection $\mathbb{A}$ to $\Sigma \times \mathcal{U}$ and $\Sigma' \times \mathcal{U}'$ are trivially glued at points $p\in \mathcal{U}\cap \mathcal{U}'$ where $\sigma'-\sigma \sim n$. As in the case $n=0$, for the examples to be considered here, the twist form $\omega \big|_{\mathcal{U}'}$ has a double pole at $w=0$, i.e. $z=\infty $ and $\mathbb{A}\big|_{\Sigma' \times \mathcal{U}'}$ is required to vanish there. Hence, as $p\in \mathcal{U} \rightarrow \infty$ the latter condition requires that $\mathbb{A}\big|_{\Sigma \times \mathcal{U}}$ goes to zero in the limit $z \rightarrow \infty$ and in practice, when solving the boundary eom \eqref{BC}, we only need to consider $\mathbb{A}\big|_{\Sigma \times \mathcal{U}}$. Because of there are no physical degrees of freedom associated to the double pole located at $z=\infty$, the associated IFT's Lax connections are, in principle\footnote{It would be interesting to construct, perhaps, new IFT's on non-trivial principal $G$-bundles on top of $\mathbb{R}\times \text{M}$ for generic M. The author thanks M. Ashwinkumar for raising this point.}, independent of $n$. 

We will refer to the pair of coordinates $x^{\mu}=(\tau,\sigma)$ of $\Sigma$ as the fiber coordinates and this is because the time direction $\mathbb{R}$, in the decomposition $\mathbb{M}=\mathbb{R}\times \text{M}$, can be seen as the global fiber of a trivial $U(1)$ bundle over $C$. It is also useful to introduce light-cone variables, given by $x^{\pm}=\tau \pm \sigma$ and $A_{\pm}=\frac{1}{2}(A_{\tau}\pm A_{\sigma})$. Notice that \eqref{BC} is independent of the deformation parameter $\zeta$, as $d_{\mathbb{M}}\omega_{\zeta}=d_{\text{M}}\omega=\underline{\pi}^{\ast}(d_{C}\omega_{C})$. This was already noticed below \eqref{BC M}.

The action \eqref{reg N action} is invariant under the gauge transformations generated by $\mathcal{G}_{0}$ and the $\omega_{\zeta}$-shifts \eqref{Omega shift symmetry} and in the limit $\zeta \rightarrow 0$, the latter symmetry reduces to the $\chi$-shifts \eqref{chi n}, which is the bundle counterpart of the $(1,0)$-shift invariance \eqref{chi} present in the original theory \eqref{4d CS}. In the $\zeta \rightarrow 0$ limit, these two gauge symmetries can be gauge fixed in a similar way as they are treated in the conventional 4d CS theory. To show this, we first denote by $(^{g}\mathbb{A},\mathbb{A})$ the pair of gauge connections related by a formal gauge transformation, i.e.
\begin{equation}
^{g}\mathbb{A}=g^{-1}\mathbb{A}g+g^{-1}d_{\mathbb{M}}g,\text{ \qquad \qquad }g\in \mathcal{G}.
\end{equation}
In order to match our notation with the notation of \cite{unifying}, where the fundamental 2d field $g:\Sigma \rightarrow G^{\mathbb{R}}$ of the IFT's defined on the pole-defect surface $\Sigma \times \mathfrak{p}$ is denoted by the same symbol, we consider instead the pair $(^{\hat{g}^{-1}}\mathbb{A},\mathbb{A})$, but now renamed as $(\mathbb{A},\mathscr{L})$. Thus, we write
\begin{equation}
\mathbb{A}=\hat{g}\mathscr{L}\hat{g}^{-1}-d_{\mathbb{M}}\hat{g}\hat{g}^{-1},\text{ \qquad \qquad }\hat{g}\in \mathcal{G}, \label{rotation}
\end{equation}
where $\mathscr{L}$ will, at the end, play the r\^ole of the Lax connection of the IFT living on $\Sigma \times \mathfrak{p}$ and depending only on the restriction $\hat{g}\big|_{\Sigma \times \mathfrak{p}}$. Furthermore, the field $\hat{g}$ is assumed to be of the archipelago type, in the sense of \cite{unifying}. The expression \eqref{rotation} is to be understood as a parameterization of the connection $\mathbb{A}$ in terms of the pair $(\hat{g}, \mathscr{L})$ and the goal now is to gauge fix all local symmetries, while solving the boundary eom \eqref{BC} in the process, by imposing conditions upon $\mathbb{A}$ and $\kappa$. Notice that the action \eqref{reg N action} is not invariant under the change \eqref{rotation}, but its bulk eom \eqref{reg N eom} are and in the limit $\zeta \rightarrow 0$, they take the more familiar form
\begin{equation}
\omega\wedge F_{\mathscr{L}}=0. \label{L eom}
\end{equation}

On the one hand, in the local coordinates $(\tau, \sigma, z)$ on $\mathbb{R}\times \mathcal{U}$, the $\chi$-shift invariance is a $(1,0)$-shift invariance and the $A_{z}$ component of the gauge connection
\begin{equation}
\mathbb{A}=A_{\tau}d\tau+ A_{\sigma}d\sigma+ A_{z}dz+A_{\overline{z}}d\overline{z},
\end{equation}
completely decouples from the theory. Thus, a natural gauge fixing condition for this symmetry is to take $A_{z}=0$, Cf. \eqref{Azeta=0gauge}. On the rhs of \eqref{rotation}, this requires $\mathscr{L}_{z}=\hat{g}^{-1}\partial_{z}\hat{g}$. On the other hand, the gauge symmetry $\mathcal{G}_{0}$ is gauge fixed, as usual, by taking the condition $A_{\overline{z}}=-\partial_{\overline{z}}\hat{g}\hat{g}^{-1}$, which on the rhs of \eqref{rotation}, translates into having $\mathscr{L}_{\overline{z}}=0$. Summarizing, we have
\begin{equation}
\begin{aligned}
\chi \text{-shift symmetry:}&\qquad \qquad  A_{z}=0,\qquad \mathscr{L}_{z}=\hat{g}^{-1}\partial_{z}\hat{g}, \\ 
\mathcal{G}_{0}\text{ gauge symmetry:}&\qquad \qquad  A_{\overline{z}}=-\partial_{\overline{z}}\hat{g}\hat{g}^{-1} ,\qquad \mathscr{L}_{\overline{z}}=0.%
\end{aligned} \label{gauge fixing cond}
\end{equation}
All this means that $\mathscr{L}$ has components only along the 2d fiber directions, i.e. $x^{\mu}=(\tau,\sigma)$, that when combined look like the cylinder $\Sigma$ and this is because when $\zeta\rightarrow 0$, the theory is defined on the quotient space $\mathcal{A}/s\chi$, where we can disregard its $z$-component. The fact that $\mathscr{L}$ is two-dimensional is precisely one of the main properties of a Lax connection. Another property being its dependence on a spectral parameter $z$, which is determined by \eqref{L eom}, decomposed now in the form
\begin{equation}
\varphi F_{\tau \sigma}=0,\text{ \qquad \qquad }\varphi \partial_{\overline{z}}\mathscr{L}_{\mu}=0,\text{ \qquad \qquad }\mu=\tau,\sigma, \label{IFT eom}
\end{equation}
where 
\begin{equation}
F_{\tau \sigma}=\partial_{\tau}\mathscr{L}_{\sigma}-\partial_{\sigma}\mathscr{L}_{\tau}+[\mathscr{L}_{\tau},\mathscr{L}_{\sigma}]
\end{equation}
and $\omega_{C}=\varphi(z) dz$. The function $\varphi(z)$ is the so-called twist function and it is deeply related to the associated 2d IFT on $\Sigma\times \mathfrak{p}$. The second equation in \eqref{IFT eom} is fulfilled by requiring that the Lax connection $\mathscr{L}_{\mu}$ is meromorphic with a set of poles located at the zeroes $\mathfrak{z}$ of the twist form $\omega_{C}$. In relation to \eqref{gauge fixing cond}, we consider now the following transformation on the pair $(\hat{g},\mathscr{L})$, given by
\begin{equation}
\hat{g}'= u\hat{g}h,\text{ \qquad \qquad}\mathscr{L}'=h^{-1}\mathscr{L}h+h^{-1}d_{\Sigma}h, \label{true gauge sym}
\end{equation} 
where $d_{\Sigma}=d\tau \wedge \partial_{\tau}+d\sigma \wedge \partial_{\sigma}$ is the exterior differential on $\Sigma$, $u\in \mathcal{G}_{0}$ and $h:\Sigma \rightarrow G^{\mathbb{R}}$. The elements $h$ belong to the subgroup $\mathcal{G}_{\Sigma}\in \mathcal{G}$ of formal gauge transformations $\mathcal{G}$, that are independent of $(z,\overline{z})$ and real, i.e. $\hat{\tau}(h)=h$, in order to preserve the equivariance properties of $\hat{g}$. The action of $\mathcal{G}_{\Sigma}$ is a redundancy of the second line of the gauge fixing conditions \eqref{gauge fixing cond} and the action of $u$ implements, as seen before, a physical gauge transformations on $\mathbb{A}$. Its action on the first line is irrelevant and it is not a source of concern as the theory is defined on the quotient space $\mathcal{A}/s\chi$. The action of $\mathcal{G}_{\Sigma}$ turns out to be a symmetry of the resulting 2d IFT defined on the pole-defect surface and it is implemented via restriction of the first term in \eqref{true gauge sym} to  $\Sigma \times \mathfrak{p}$. Further details can be found in \cite{unifying}.

To find the 2d IFT action functional on $\Sigma \times \mathfrak{p}$, we use the parameterization \eqref{rotation} in the regularized action \eqref{reg N action} and subsequently solve for the components of the 2d Lax connection $\mathscr{L}=\mathscr{L}_{\mu}dx^{\mu}$ in terms of a collection of fields $\hat{g}\big|_{\Sigma \times \mathfrak{p}}$ determined by gauge fixing conditions, and then inserting the boundary eom \eqref{BC} solution $\mathbb{A}$ into the lhs of the parameterization \eqref{rotation} restricted to $\Sigma \times \mathfrak{p}$. This mechanism produces a plethora of 2d IFT's action functionals and their corresponding on-shell flat Lax connections. Their flatness being equivalent to the usual Euler-Lagrange eom of the associated 2d IFT action functional just derived.

The first step is then to use the parameterization \eqref{rotation} into the action \eqref{reg N action}. After using the result \eqref{CS gauge}, we find that
\begin{equation}
\begin{aligned}
S(\hat{g},\mathscr{L})_{\text{reg}}=\, & ic\dint_{\mathbb{M}}d_{\mathbb{M}}\omega_{\zeta}\wedge \text{Tr}\left(\hat{g}^{-1}d_{\mathbb{M}}\hat{g}  \wedge \mathscr{L} \right)-ic\dint_{\mathbb{M}}\omega_{\zeta}\wedge\chi(\hat{g})\\
 &+ ic\dint\nolimits_{\mathbb{M}}\omega_{\zeta} \wedge  CS\left( 
\mathscr{L}\right)+ic\dint\nolimits_{\mathbb{M}}d_{\mathbb{M}}\omega_{\zeta}\wedge \kappa\wedge \text{Tr}\left( \mathbb{A}\iota_{X}\mathbb{A}  \right), 
\end{aligned} \label{rotated action}
\end{equation}
where $\mathbb{A}$ in the last term right above is given by \eqref{rotation}. We will see below, that this whole contribution vanishes by virtue of the conditions imposed on $\mathbb{A}$ that solve the boundary eom \eqref{BC}. The first, second\footnote{See the discussion below eq. \eqref{WZ term}.} and fourth terms on the rhs in \eqref{rotated action} are independent of the deformation parameter $\zeta$ and in the limit $\zeta \rightarrow 0$, the third term vanishes by virtue of the second eom in \eqref{IFT eom}, i.e. $\varphi\partial_{\overline{z}}\mathscr{L}_{\mu}=0$. Below, we provide some explicit formulas, imported from \cite{unifying}, for the first two contributions in \eqref{rotated action} that ultimately will provide the 2d IFT action functional defined on $\Sigma \times \mathfrak{p}$.

In order to continue, we need to solve the boundary eom \eqref{BC} by specifying $\mathbb{A}\big|_{\mathfrak{p}}$. To do this, we start with a twist form $\omega_{C}$ on $C$ that has a number of zeroes $\mathfrak{z}$ and poles $\mathfrak{p}$ that are constrained by the Riemann-Roch theorem. Let us focus now on the pole structure of $\omega_{C}$, i.e. its set of poles $\mathfrak{p}$ and suppose there are $M$ of them denoted collectively by $z_{r}\in \mathfrak{p}$, $r=1,...,M$, each one of order $m_{r}$ and all contained in the chart $\mathcal{U}$. Thus, around these poles, the pole part of the twist form can be written as \cite{unifying}
\begin{equation}
\omega_{C}=\sum_{r=1}^{M}\sum_{p=0}^{m_{r}-1}\frac{k_{p}^{(r)}}{(z-z_{r})^{p+1}}dz,\text{ \qquad \qquad }k_{p}^{(r)}:=\text{res}_{z=z_{r}}\big[(z-z_{r})^{p}\omega_{C}   \big]. \label{twist on U}
\end{equation}
If there is an order $m_{\infty}$ pole at $z=\infty$, we use, on the chart $\mathcal{U}'$, the corresponding expression
\begin{equation}
\omega_{C}=\sum_{p=0}^{m_{\infty}-1}\frac{k_{\infty}^{(r)}}{w^{p+1}}dw,\text{ \qquad \qquad }w=1/z.
\end{equation}
From \eqref{twist on U}, it follows that
\begin{equation}
d_{C}\omega_{C}=2\pi i\left( \sum_{r=1}^{M}\sum_{p=0}^{m_{r}-1}\frac{(-1)^{p}}{p!}k_{p}^{(r)}\partial^{p}_{z}\delta_{zz_{r}}\right)dz \wedge d\overline{z},
\end{equation}
where $d_{C}=dz \wedge \partial_{z}+d\overline{z}\wedge \partial_{\overline{z}}$ and
\begin{equation}
\delta_{zz'}=\delta(z-z')=-\frac{1}{2\pi i}\partial_{\overline{z}}\left(\frac{1}{z-z'}   \right)
\end{equation}
is the usual Dirac delta distribution on $\mathcal{U}$. From all this, we obtain the general integration formula
\begin{equation}
\dint_{\mathcal{U}}d_{C}\omega_{C}f=2\pi i\sum_{r=1}^{M}\sum_{p=0}^{m_{r}-1}\frac{k_{p}^{(r)}}{p!}\partial^{p}_{z}f \big|_{z=z_{r}},\label{formula BC}
\end{equation}
for an arbitrary smooth function $f:\mathcal{U}\rightarrow \mathbb{C}$. Some solutions to the boundary eom, using this formula, will be reviewed below in well-known standard integrable $\sigma$-model examples.

Explicit expressions for the first two terms on the rhs of \eqref{rotated action}, after integration over the base space $C=\mathbb{CP}^{1}$, were already computed in \cite{unifying} and are given by
\begin{equation}
\dint_{\mathbb{M}}\omega\wedge \text{Tr}\left[\left(\hat{g}^{-1}d_{\mathbb{M}}\hat{g}\right)^{3} \right]=2\pi i\sum_{r=1}^{M}\big(\text{res}_{z=z_{r}}\omega_{C}\big)\dint_{\Sigma \times [0,R_{z_{r}}]}\text{Tr}\left[ \left( \hat{g}_{z_{r}}^{-1}d_{\Sigma'}\hat{g}_{z_{r}}  \right)^{3}  \right]\label{I1}
\end{equation}
and
\begin{equation}
\dint_{\mathbb{M}}d_{\text{M}}\omega\wedge \text{Tr}\big( \hat{g}^{-1}d_{\mathbb{M}}\hat{g}\wedge \mathscr{L}  \big)=-2\pi i \sum_{r=1}^{M} \dint_{\Sigma}\text{Tr}\Big[g^{-1}_{z_{r}}d_{\Sigma}g_{z_{r}}\wedge \big(\text{res}_{z=z_{r}}\omega_{C}\wedge\mathscr{L}   \big)  \Big], \label{I2}
\end{equation}
where $\hat{g}_{z_{r}}=\hat{g}\big|_{U_{z_{r}}}$ is the restriction of $\hat{g}$ to a disc $U_{z_{r}}$ of radius $R_{z_{r}}$ centered at the pole $z_{r}$ and $g_{z_{r}}=\hat{g}\big|_{z=z_{r}}$ is the restriction of $\hat{g}$ to the pole $z_{r}$. We have used the shorthand notation $\mathbb{I}^{3}=\mathbb{I} \wedge \mathbb{I} \wedge \mathbb{I}$ in the WZ term and extended the $d_{\Sigma}$ differential from $\Sigma$ to $d_{\Sigma'}$ on $\Sigma \times [0,R_{z_{r}}]$. Furthermore,
\begin{equation}
\text{res}_{z=z_{r}}\omega_{C}=k_{0}^{(r)},\text{\qquad \qquad }\text{res}_{z=z_{r}}\omega_{C}\wedge\mathscr{L}=\sum_{p=0}^{m_{r}-1}
\frac{k_{p}^{(r)}}{p!}\big(\partial_{z}^{p}\mathscr{L}   \big)\big|_{z=z_{r}}.
\end{equation}

We still need to provide an explicit expression to the quadratic boundary contribution given by the fourth term on the rhs of \eqref{rotated action}, i.e. the integral that depends of the contact 1-form $\alpha$ via $\kappa$. Define 
\begin{equation}
\mathcal{Q}(\mathbb{A}):=\dint\nolimits_{\mathbb{M}}d_{\mathbb{M}}\omega_{\zeta}\wedge \kappa\wedge \text{Tr}\left( \mathbb{A}\iota_{X}\mathbb{A}  \right). \label{Q term}
\end{equation}
We will come back to this expression soon.

\subsection{Boundary eom solutions and some 2d IFT's} \label{4.2}


There are two standard pole configurations of interest that we will consider in what follows and a sample of IFT's examples associated to them. They appear as a consequence of the first equivariance condition \eqref{equiv cond} applied to $\omega_{C}$, that requires $\overline{\varphi(z)}=\varphi(\overline{z})$. The first one is when we have a real double pole $z_{r}$, i.e. $m_{r}=2$ and the second one is when we have two simple poles $(z_{1},z_{2})$, i.e. $m_{1}=m_{2}=1$ fitting into two categories: i) both poles are real $(\overline{z}_{1},\overline{z}_{2})=(z_{1},z_{2})$ and ii) both poles are complex conjugate to each other $(\overline{z}_{1},\overline{z}_{2})=(z_{2},z_{1})$. Furthermore, we will consider examples with the following properties: a) at a pair of simple real poles we have that $k_{0}^{(2)}=-k_{0}^{(1)}$, with $k_{0}^{(1)}\neq 0 \in \mathbb{R}$. b) at a pair of complex conjugate simple poles we have that $k_{0}^{(2)}=-k_{0}^{(1)}$, $k_{0}^{(2)}=\overline{k_{0}^{(1)}}$, with $k_{0}^{(1)}\neq 0 \in i\mathbb{R}$ and c) at a real double pole we have that $k_{0}^{(r)},k_{1}^{(r)}\in \mathbb{R}$.   

For further reference, we write the modified classical Yang-Baxter (mCYB) equation restricted to $R$-matrices solutions that are skew-symmetric with respect to an inner product defined on $\mathfrak{g}^{\mathbb{R}}$, i.e.
\begin{equation}
[Rx,Ry]-R[x,y]_{R}=-\tilde{c}^{2}[x,y],\text{\qquad \qquad}\text{Tr}(x\, Ry)=-\text{Tr}(Rx\, y), \label{cmCYB eq}
\end{equation}
where $R\in \text{End}(\mathfrak{g}^{\mathbb{R}})$, $x,y \in \mathfrak{g}^{\mathbb{R}}$, $[x,y]_{R}=[Rx,y]+[x,Ry]$ is the $R$-bracket and $\tilde{c}=0,1,i$. When $\tilde{c}=0$, the mCYB equation reduces to the classical Yang-Baxter (CYB) equation. Also, $\text{Re}$ and $\text{Im}$ denotes the real and the imaginary parts of a complex number $z=a+ib\in \mathbb{C}$. For example, if $x=x_{1}+ix_{2}$ and $x'=x'_{1}+ix'_{2}$ with $x_{1},x_{2},x_{1}',x_{2}'\in \mathfrak{g}^{\mathbb{R}}$ belong to $\mathfrak{g}$. Then, 
\begin{equation}
\text{Tr}(xx')=\text{Re}\big[\text{Tr}(xx')  \big]+i\text{Im}\big[\text{Tr}(xx')  \big],
\end{equation}
where
\begin{equation}
\text{Re}\big[\text{Tr}(xx')\big]=\text{Tr}_{\mathbb{R}}(x_{1}x_{1}'-x_{2}x_{2}'), \text{\qquad \qquad} \text{Im}\big[\text{Tr}(xx') \big]=\text{Tr}_{\mathbb{R}}(x_{1}x_{2}'+x_{2}x_{1}'), \label{ReIm parts}
\end{equation}
where we have denoted the restriction of $\text{Tr}$ to $\mathfrak{g}^{\mathbb{R}}$ by $\text{Tr}_{\mathbb{R}}$. We will use \eqref{ReIm parts} later on when we consider norms of gauge fields belonging to the space of connections and norms of vector fields defined on adjoint orbits of complex Lie groups. 

Now, the gauge field $\mathbb{A}$ boundary eom solutions \cite{unifying} are specified in terms of a Manin triple denoted here, generically by $(\underline{\mathfrak{g}},\underline{\mathfrak{t}},\underline{\mathfrak{p}})$, where $\underline{\mathfrak{t}}$ and $\underline{\mathfrak{p}}$ are Lagrangian subalgebras of $\underline{\mathfrak{g}}$ with respect to a well-defined bilinear form $\langle\! \langle \cdot,\cdot \rangle \!\rangle_{\underline{\mathfrak{g}}}:\underline{\mathfrak{g}} \times \underline{\mathfrak{g}} \rightarrow \mathbb{R}$ on $\underline{\mathfrak{g}}$ defined by $\omega$ (through the coefficients $k_{p}^{(r)}$) and that are also complementary to each other, in the sense that $\underline{\mathfrak{g}}=\underline{\mathfrak{t}}\oplus \underline{\mathfrak{p}}$, with $\oplus$ understood as a direct sum of vector spaces. See also $\S 2$ of \cite{Doubles}. The algebraic data for each pole configuration to be considered below is then the following:

$\bullet$ At a pair of real simple poles, we have 
\begin{equation}
(\underline{\mathfrak{g}},\underline{\mathfrak{t}},\underline{\mathfrak{p}})=\big(\mathfrak{d},\mathfrak{g}^{\delta},\mathfrak{g}_{R}\big),
\end{equation}
where $\mathfrak{d}=\mathfrak{g}^{\mathbb{R}}\oplus \mathfrak{g}^{\mathbb{R}}$ is the real double of $\mathfrak{g}^{\mathbb{R}}$,
\begin{equation}
\mathfrak{g}^{\delta}:=\bigl\{(x,y)\in \mathfrak{d}\, \big| y=x   \bigr\},\text{\qquad \qquad}\mathfrak{g}_{R}:=\bigl\{(x,y)\in \mathfrak{d}\, \big| (R-1)y=(R+1)x  \bigr\}\label{Lag for real poles}
\end{equation}
and
\begin{equation}
\big\langle\! \! \big\langle (x,y),(x',y') \big\rangle \! \! \big\rangle _{\mathfrak{d}}:=k_{0}^{(1)}\text{Tr}\big(x x'\big)+k_{0}^{(2)}\text{Tr}\big(y y'\big), \label{Inner R}
\end{equation}
for any $x,y,x',y' \in \mathfrak{g}^{\mathbb{R}}$. To verify that $\mathfrak{g}^{\delta}$ and $\mathfrak{g}_{R}$ are Lagrangian, set $k_{0}^{(2)}=-k_{0}^{(1)}$ and use
\begin{equation}
\text{Tr}\big(x x'-y y'\big)=\frac{1}{2}\text{Tr}\big[(x+y)(x'-y')+(x-y)(x'+y')\big].
\end{equation} 
Moreover, the $R$-matrix in \eqref{Lag for real poles}, solving the mCYB equation \eqref{cmCYB eq} with $\tilde{c}=1$, ensures that $\mathfrak{g}_{R}$ is a Lie subalgebra of $\mathfrak{d}$, i.e. we have that
\begin{equation}
\big[(R-1)y,(R-1)y'   \big]=\big[(R+1)x,(R+1)x'   \big],
\end{equation}
for $(x,y), (x',y')\in \mathfrak{g}_{R}$. Equivalently, 
\begin{equation}
(R-1)[y,y']_{R}=(R+1)[x,x']_{R}.\label{R1}
\end{equation}

$\bullet$ At a pair of complex conjugate simple poles, we have
\begin{equation}
(\underline{\mathfrak{g}},\underline{\mathfrak{t}},\underline{\mathfrak{p}})=\big(\mathfrak{g},\mathfrak{g}^{\mathbb{R}},\mathfrak{g}_{R}\big),
\end{equation}
where
\begin{equation}
\mathfrak{g}_{R}:=\bigl\{x\in \mathfrak{g} \, \big|(R+i)x\in \mathfrak{g}^{\mathbb{R}}  \bigr\},\label{Lag for com poles}
\end{equation}
with $\mathfrak{g}^{\mathbb{R}}\subset \mathfrak{g}$ denoting the real subalgebra of $\mathfrak{g}$ regarded itself as a real Lie algebra. Furthermore, the inner product is given by
\begin{equation}
\big\langle\! \! \big\langle x,x' \big\rangle \! \! \big\rangle _{\mathfrak{g}}:=k_{0}^{(1)}\text{Tr}\big(x x'\big)+k_{0}^{(2)}\overline{\text{Tr}\big(x x'\big)},\label{Inner C1}
\end{equation}
for any $x,x' \in \mathfrak{g}$. Alternatively, the definition \eqref{Lag for com poles} is equivalent to
\begin{equation}
\mathfrak{g}_{R}:=\bigl\{x, y=\hat{\tau}(x) \in \mathfrak{g} \, \big|(R-i)y=(R+i)x  \bigr\}, \label{Lag for com poles'}
\end{equation} 
where $\hat{\tau}$ is the anti-linear involutive automorphism $\hat{\tau}$ acting on $\mathfrak{g}$ introduced in $\S \eqref{3}$, with the pair $(\hat{\tau},R)$ satisfying the commutative relation $\hat{\tau}\circ R=R\circ \hat{\tau}$. To verify that $\mathfrak{g}^{\mathbb{R}}$ and $\mathfrak{g}_{R}$ are Lagrangian, take $k_{0}^{(2)}=-k_{0}^{(1)}$ and write
\begin{equation}
\text{Im}\big[\text{Tr}(xx')\big]  =\frac{1}{4i}\text{Tr}\big[(x+y)(x'-y')+(x-y)(x'+y')\big].
\end{equation}
Finally, the $R$-matrix entering \eqref{Lag for com poles}, solving the mCYB \eqref{cmCYB eq} with $\tilde{c}=i$, ensures that $\mathfrak{g}_{R}$ is a Lie subalgebra of $\mathfrak{g}$, i.e. we verify that
\begin{equation}
\big[(R-i)\hat{\tau}(x),(R-i)\hat{\tau}(y)   \big]=\big[(R+i)x,(R+i)y  \big],
\end{equation}
for $x,\hat{\tau}(x), y,\hat{\tau}(y) \in \mathfrak{g}$. Equivalently,
\begin{equation}
(R-i)\hat{\tau}\big([x,y]_{R}\big)=(R+i)[x,y]_{R}.\label{R2}
\end{equation} 

$\bullet$ At a real double pole when $k_{0}^{(r)}=0$, we have
\begin{equation}
(\underline{\mathfrak{g}},\underline{\mathfrak{t}},\underline{\mathfrak{p}})=(\mathfrak{d}_{\text{ab}},\mathfrak{g}_{R},\mathfrak{g}^{\mathbb{R}}\ltimes \{0\}),
\end{equation}
where $\mathfrak{d}_{\text{ab}}=\mathfrak{g}^{\mathbb{R}}\ltimes \mathfrak{g}^{\mathbb{R}}_{\text{ab}}$, with $\mathfrak{g}^{\mathbb{R}}_{\text{ab}}$ being an abelian copy of $\mathfrak{g}^{\mathbb{R}}$ on which $\mathfrak{g}^{\mathbb{R}}$ acts by the adjoint action. That is, $\mathfrak{d}_{\text{ab}}$ is isomorphic to the direct sum $\mathfrak{g}^{\mathbb{R}}\oplus \mathfrak{g}^{\mathbb{R}}$ as a vector space, but with a Lie bracket defined by
\begin{equation}
\big[(x,y),(x',y')\big]:=\big([x,x'], [x,y']-[x',y] \big), \label{Bracket str}
\end{equation}
for any $x,y,x',y' \in \mathfrak{g}^{\mathbb{R}}$. In addition, we have 
\begin{equation}
\mathfrak{g}_{R}:=\bigl\{(x',y') \in \mathfrak{d}_{\text{ab}} \, \big| (x',y')=(-Rx,x) \bigr\}, \label{Lag for double pole}
\end{equation}
with $x\in \mathfrak{g}^{\mathbb{R}}$. Furthermore, the inner product is in general, defined by
\begin{equation}
\big\langle\! \! \big\langle (x,y),(x',y') \big\rangle \! \! \big\rangle _{\mathfrak{d}_{\text{ab}}}^{(r)}:=k_{0}^{(r)}\text{Tr}\big( xx'  \big)+k_{1}^{(r)}\text{Tr}\big( xy'+x'y  \big).\label{Inner double}
\end{equation}
Moreover, the $R$-matrix entering \eqref{Lag for double pole}, solving the CYB equation \eqref{cmCYB eq} with $\tilde{c}=0$, ensures that $\mathfrak{g}_{R}$ is a Lie subalgebra of $\mathfrak{d}_{\text{ab}}$ under the bracket \eqref{Bracket str}. Indeed, we get that
\begin{equation}
\big[(-Rx,x),(-Ry,y) \big]=\big( -Rz,z  \big),
\end{equation}
where $z:=-[x,y]_{R}$.

When $k_{0}^{(r)}\neq 0$, $\underline{\mathfrak{t}}:=\{0\}\ltimes \mathfrak{g}^{\mathbb{R}}_{\text{ab}} \subset \mathfrak{d}_{\text{ab}}$ is a Lagrangian subalgebra formed by the elements $(0,y)\in \mathfrak{d}_{\text{ab}}$, for any $y\in \mathfrak{g}^{\mathbb{R}}_{\text{ab}}$. This subalgebra plays a central r\^ole when boundary eom are considered at infinity in $C=\mathbb{CP}^{1}$, i.e. for $z=\infty$ or $w=0$. 

Let us now briefly consider in our setup, these three pole configurations and their associated IFT's on $\Sigma \times \mathfrak{p}$. The key goal now is to further identify what are the conditions to be imposed upon the 1-form $\kappa$, allowing the regularized 4d CS theory to reproduce the same IFT's extracted from the original 4d CS theory or, at least, a sample of them that are of our particular interest.  

\subsubsection{Solutions for a real double pole} 

In this case, applying the formula \eqref{formula BC} to \eqref{BC}, gives
\begin{equation}
\mathcal{B}(\mathbb{A})=2\pi i \dint_{\Sigma}d\tau \wedge d\sigma\Big[ k_{0}^{(r)} \epsilon^{\mu \nu}\text{Tr}\big(\delta A_{\mu}\hat{A}_{\nu}\big)\big|_{z=z_{r}}+k_{1}^{(r)}\epsilon^{\mu \nu}\partial_{z}\text{Tr} \big( \delta A_{\mu}\hat{A}_{\nu} \big) \big|_{z=z_{r}} \Big]=0 , \label{BC again}
\end{equation}
where $\mu=\tau, \sigma$ and $\epsilon ^{\tau \sigma}=1$. By the equivariance properties \eqref{equiv cond}, it follows that $A_{\mu}$ and $\partial_{z}A_{\mu}$ (see \eqref{compo'}), are both valued in $\mathfrak{g}^{\mathbb{R}}$. Then, the vanishing of \eqref{BC again} boils down to the condition
\begin{equation}
\epsilon^{\mu \nu}\Big\langle\! \! \Big\langle \Big(\delta A_{\mu}\big|_{z=z_{r}},\delta \big( \partial_{z}A_{\mu}\big)\big|_{z=z_{r}}   \Big),\Big( \hat{A}_{\nu}\big|_{z=z_{r}},\partial_{z}\hat{A}_{\nu}\big|_{z=z_{r}}   \Big) \Big\rangle\! \! \Big\rangle _{\mathfrak{d}_{\text{ab}}}^{(r)}=0,\label{double real pole BC}
\end{equation}
where we have used the inner product defined in \eqref{Inner double}. In reaching the form \eqref{double real pole BC}, we have imposed a condition on top of the 1-form $\kappa$, given by
\begin{equation}
\kappa_{\mu} \big|_{z=z_{r}}\in \mathbb{R}. \label{cond kappa 0}
\end{equation}

Two known examples of integrable $\sigma$-models with this pole structure are the PCM with WZ term and the Homogeneous Yang-Baxter $\sigma$-model. In the first case we have $k_{0}^{(r)}\neq 0$, $k_{1}^{(r)}\neq 0$ and \eqref{double real pole BC} is satisfied if we impose a trivial boundary eom solution, i.e.
\begin{equation}
A_{\mu}\big|_{z=z_{r}}=0, \label{double poles BC 1}
\end{equation} 
with the same condition being valid for its variation, i.e. $\Big(A_{\mu}\big|_{z=z_{r}},\partial_{z} A_{\mu}\big|_{z=z_{r}}   \Big)$ is required to be an element of $\underline{\mathfrak{t}}$, as defined below eq. \eqref{Inner double}. In the second case, we have $k_{0}^{(r)}= 0$, $k_{1}^{(r)}\neq 0$ and \eqref{double real pole BC} is satisfied as well, if this time we require that $\Big(A_{\mu}\big|_{z=z_{r}},\partial_{z} A_{\mu}\big|_{z=z_{r}}   \Big)$ and its variation both belong to the Lagrangian subalgebra $\mathfrak{g}_{R}$, defined in \eqref{Lag for double pole}. Thus, we demand that
\begin{equation}
A_{\mu}\big|_{z=z_{r}}=-R\big(\partial_{z}A_{\mu}  \big)\big|_{z=z_{r}}, \label{double poles BC 2}
\end{equation}
supplemented with an additional condition imposed upon $\kappa$, given by
\begin{equation}
\partial_{z} \kappa_{\mu}\big|_{z=z_{r}}=0. \label{cond kappa 1}
\end{equation}
If the double pole is located at infinity, we always use the trivial condition \eqref{double poles BC 1}. In these proofs we have assumed that the contraction $\iota_{X}\mathbb{A}$ is proportional, with a real constant coefficient, to the fiber component $A_{\sigma}$ when, in the coordinate system $(\tau, \sigma, z)$, we write
\begin{equation}
\kappa=\kappa_{z}dz+\kappa_{\overline{z}}d\overline{z}+\kappa_{\tau}d\tau+ \kappa_{\sigma}d\sigma, \text{ \qquad \qquad }X=(1/\kappa_{\sigma})\partial_{\sigma}, \label{Reeb components}
\end{equation}
with $\kappa_{\sigma} \neq 0$. Of course, this assumption will be justified below. Both boundary eom solutions $\mathbb{A}$ must be preserved by the action of the gauge group $\mathcal{G}_{0}$. 

The quadratic contribution \eqref{Q term} to the action \eqref{rotated action}, then becomes
\begin{equation}
\mathcal{Q}(\mathbb{A})=2\pi i\dint_{\Sigma}d\tau \wedge d\sigma \Big[k_{0}^{(r)} \epsilon^{\mu \nu}\kappa_{\mu}\text{Tr}\big(A_{\nu}\iota_{X}\mathbb{A}\big)\big|_{z=z_{r}}+k_{1}^{(r)}\epsilon^{\mu \nu}\partial_{z}\text{Tr} \big( \kappa_{\mu}A_{\nu}\iota_{X}\mathbb{A} \big) \big|_{z=z_{r}}   \Big]
\end{equation}
and vanishes in both situations, i.e. when we consider the solutions \eqref{double poles BC 1} or \eqref{double poles BC 2}. 

A typical example with the pole structure considered here is provided by the PCM with WZ term twist form \cite{unifying}, see also \cite{CY}. Here we simply invoke their results and the reader is referred to those references for further details. It is given by
\begin{equation}
\omega_{C}=a\frac{1-z^{2}}{(z-k)^{2}}dz,
\end{equation}
where\footnote{We have set $K\rightarrow a$, where $K$ is the constant used in \cite{unifying}.} $a,k \in \mathbb{R}$. It has two simple zeroes at $\mathfrak{z}=\{+1,-1\}$ and two double poles at $\mathfrak{p}=\{k,\infty\}$. Using the equation \eqref{double poles BC 1}, for $z_{1}=k,z_{2}=\infty$, in the restriction of \eqref{rotation} to $\Sigma\times \mathfrak{p}$ and
\begin{equation}
\hat{g}\big|_{\Sigma \times \mathfrak{p}}=\hat{g} \big|_{\Sigma \times \{k, \infty \}}=\left(\hat{g}\big|_{z=k},\hat{g}\big|_{z=\infty}\right):=(g,1),
\end{equation}
for some $g: \Sigma \rightarrow G^{\mathbb{R}}$, where the freedom of acting with $\mathcal{G}_{\Sigma}$ is used to fix the value of $\hat{g}$ at $\infty$ to the identity, we find the following Lax connection
\begin{equation}
\mathscr{L}=\left(\frac{k-1}{z-1}\right)J_{+}d\sigma^{+}+\left(\frac{k+1}{z+1}\right)J_{-}d\sigma^{-},
\end{equation}
where $J=g^{-1}d_{\Sigma}g \in \Omega_{\Sigma}^{1}\otimes \mathfrak{g}^{\mathbb{R}} $ are the 2d $\sigma$-model currents. Notice that $J$ is indeed a 1-form valued on $\mathfrak{g}^{\mathbb{R}}$ and this follows from the second equivariance condition introduced in \eqref{equiv cond}, requiring $\hat{\tau}(J_{\pm})=J_{\pm}$. Now, after using
\begin{equation}
k_{0}^{(k)}=-2ak,\text{ \qquad \qquad }k_{1}^{(k)}=a(1-k^{2}) \label{residues 1}
\end{equation}
and the Lax connection right above in the integral formulas \eqref{I1}, \eqref{I2}, we find that the action functional \eqref{rotated action}, becomes
\begin{equation}
S(g)_{\text{PCM},k}=-4 \pi ca \left\{ \dint_{\Sigma} d\sigma^{+} \wedge d\sigma^{-}\, \text{Tr}(J_{+}J_{-}) +kI_{\text{WZ}}(g)  \right\},
\end{equation} 
where
\begin{equation}
I_{\text{WZ}}(g)=-\frac{1}{3}\dint_{\Sigma \times [0,R_{k}]}\text{Tr}\left[ \left( \hat{g}_{k}^{-1}d_{\Sigma'}\hat{g}_{k}  \right)^{3}  \right]. \label{g WZ}
\end{equation}

The homogeneous Yang-Baxter sigma model corresponds to taking $k=0$, Cf. \eqref{residues 1}, but this time using the boundary eom solutions \eqref{double poles BC 2} and $\partial_{z} \hat{g}\big|_{z=0}=0$, consequence of the archipelago conditions. The rest of the analysis follows, up to some minor details, the same lines we have just considered and further details can be found in \cite{unifying}. The Lax connection and action functional are given, respectively, by
\begin{equation}
\mathscr{L}=-\left( \frac{1}{z-1}\right)\frac{1}{1+R_{g}}J_{+}d\sigma^{+}+\left( \frac{1}{z+1}  \right)\frac{1}{1-R_{g}}J_{-}d\sigma^{-},
\end{equation}
where $R_{g}:=Ad_{g^{-1}}\circ R\circ Ad_{g}$, $J$ as above and
\begin{equation}
S(g)_{\text{hYB-PCM}}=-4 \pi c \dint_{\Sigma} d\sigma^{+} \wedge d\sigma^{-}\, \text{Tr}\left( J_{+}\frac{1}{1-R_{g}}J_{-}\right).
\end{equation}

\subsubsection{Solutions for two real simple poles}\label{4.2.2}

In this case, the integral formula \eqref{formula BC} applied to \eqref{BC}, gives
\begin{equation}
\mathcal{B}(\mathbb{A})=2\pi i \dint_{\Sigma}d\tau \wedge d\sigma\Big[ k_{0}^{(1)} \epsilon^{\mu \nu}\text{Tr}\big(\delta A_{\mu}\hat{A}_{\nu}\big)\big|_{z=z_{1}}+k_{0}^{(2)}\epsilon^{\mu \nu}\text{Tr} \big( \delta A_{\mu}\hat{A}_{\nu} \big) \big|_{z=z_{2}} \Big] . \label{BC real poles}
\end{equation}
By the equivariance properties \eqref{equiv cond} applied to $A_{\mu}$, it follows that the components $A_{\mu} \big|_{z=z_{r}}$, for $r=1,2$, are valued in $\mathfrak{g}^{\mathbb{R}}$. Thus, the vanishing of \eqref{BC real poles} boils down to the condition
\begin{equation}
\epsilon^{\mu \nu}\Big\langle\! \! \Big\langle \Big(\delta A_{\mu}\big|_{z=z_{1}},\delta A_{\mu}\big|_{z=z_{2}}   \Big),\Big( \hat{A}_{\nu}\big|_{z=z_{1}},\hat{A}_{\nu}\big|_{z=z_{2}}   \Big) \Big\rangle\! \! \Big\rangle _{\mathfrak{d}}=0,\label{simple real poles BC}
\end{equation}
where we have used the inner product defined in \eqref{Inner R}. 

Let us solve the boundary eom \eqref{simple real poles BC} by requiring that $\Big(A_{\mu}\big|_{z=z_{1}},A_{\mu}\big|_{z=z_{2}}   \Big)$ and its variation both belong, respectively, to the Lagrangian subalgebras $\mathfrak{g}^{\delta}$ and $\mathfrak{g}_{R}$, defined in \eqref{Lag for real poles}. In the first case, the solution takes the form
\begin{equation}
A_{\mu}\big|_{z=z_{1}}=A_{\mu}\big|_{z=z_{2}}, \label{real cond gauge}
\end{equation}
while in the second case they are given by
\begin{equation}
\big(R-1  \big)A_{\mu}\big|_{z=z_{2}}=\big( R+1  \big)A_{\mu}\big|_{z=z_{1}}. \label{real cond gauge'}
\end{equation}
Thus, in order for the inner product \eqref{simple real poles BC} to vanish, we must impose an additional condition over the 1-form $\kappa$, given by
\begin{equation}
\kappa_{\mu}\big|_{z=z_{1}}=\kappa_{\mu}\big|_{z=z_{2}}. \label{cond kappa 2}
\end{equation}
These boundary eom solutions must be preserved by the action of the gauge group $\mathcal{G}_{0}$. Furthermore, the quadratic contribution \eqref{Q term} to the action \eqref{rotated action}, given by
\begin{equation}
\mathcal{Q}(\mathbb{A})=2\pi i\dint_{\Sigma}d\tau \wedge d\sigma \Big[k_{0}^{(1)} \epsilon^{\mu \nu}\kappa_{\mu}\text{Tr}\big(A_{\nu}\iota_{X}\mathbb{A}\big)\big|_{z=z_{1}}+k_{0}^{(2)} \epsilon^{\mu \nu}\kappa_{\mu}\text{Tr}\big(A_{\nu}\iota_{X}\mathbb{A}\big)\big|_{z=z_{2}}  \Big], \label{QQ}
\end{equation}
vanish trivially for both solutions.

A first typical example with this pole structure and associated to the Lagrangian subalgebra $\mathfrak{g}^{\delta}$, is provided by the $\lambda$-deformation of the PCM \cite{Sfetsos}. As before, the reader is referred to \cite{unifying} for further details. In this case the twist form, is given by
\begin{equation}
\omega_{C}=\frac{b}{1-a^{2}}\frac{1-z^{2}}{z^{2}-a^{2}}dz,
\end{equation}
where\footnote{We have set $\alpha \rightarrow a$ and  $K \rightarrow b$, where $\alpha, K$ are the constants used in \cite{unifying}.} $a,b \in \mathbb{R} $. Also, we introduce the deformation parameter $\lambda=(1+a)/(1-a)$. The twist form has two simple zeroes at $\mathfrak{z}=\{+1,-1\}$, two simple real poles at $\pm a$ and a double pole at infinity\footnote{This pole is treated as in the last example, i.e. $A_{\mu}|_{\infty}=0$.}, i.e. $\mathfrak{p}=\{+a,-a,\infty\}$. This time, using the relation \eqref{real cond gauge}, for $z_{1}=a, z_{2}=-a$, in the restriction of \eqref{rotation} to $\Sigma \times \mathfrak{p}$ and
\begin{equation}
\hat{g}\big|_{\Sigma \times \mathfrak{p}}=\hat{g} \big|_{\Sigma \times \{a,-a, \infty \}}=\left(\hat{g}\big|_{z=a},\hat{g}\big|_{z=-a},\hat{g}\big|_{z=\infty}\right):=(g,1,1),
\end{equation}
where the freedom of acting with $\mathcal{G}_{\Sigma}$ is used to fix the value of $\hat{g}$ at $\infty$ to the identity, we get the following Lax connection
\begin{equation}
\mathscr{L}=\left(\frac{a-1}{z-1}\right)\frac{\lambda}{\lambda-Ad_{g^{-1}}}J_{+}d\sigma^{+}+\left( \frac{a+1}{z+1}  \right)\frac{\lambda^{-1}}{\lambda^{-1}-Ad_{g^{-1}}}J_{-}d\sigma^{-}
\end{equation} 
where $J=g^{-1}d_{\Sigma}g \in \Omega_{\Sigma}^{1}\otimes \mathfrak{g}^{\mathbb{R}} $. Now, after using
\begin{equation}
k_{0}^{(1)}=b/2a,\text{ \qquad \qquad }k_{0}^{(2)}=-b/2a
\end{equation}
and the Lax connection right above in the integral formulas \eqref{I1}, \eqref{I2}, we find that the action functional \eqref{rotated action} takes the explicit form
\begin{equation}
S(g)_{\lambda\text{-PCM}}=-\frac{\pi bc}{a}\left\{ \dint_{\Sigma}d\sigma^{+}\wedge d\sigma^{-} \text{Tr}\Big[J_{+}\big(\hat{G}+\hat{B}\big)J_{-}   \Big ]   \right\},
\end{equation}
where
\begin{equation}
\hat{G}=(\lambda^{2}-1)\frac{1}{\lambda -Ad_{g^{-1}}}\circ \frac{1}{\lambda -Ad_{g}}
\end{equation}
implements the deformation of the background space metric and
\begin{equation}
\hat{B}=B_{0}+\frac{\lambda}{\lambda-Ad_{g}}-\frac{\lambda}{\lambda-Ad_{g^{-1}}}
\end{equation}
of the $B$-field. The $B_{0}$ indicates the contribution from the WZ term \eqref{g WZ} with $k \rightarrow a$. This is a classical approach and, as a direct consequence, there is not emerging dilaton field.

A second typical example with this pole structure and associated, this time, to the Lagrangian subalgebra $\mathfrak{g}_{R}$, is provided by the Yang-Baxter sigma model. However, this example will be reviewed in the next section, for a reason to be explained there. 

\subsubsection{Solutions for complex conjugate simple poles.}

In this case, we also have the expression \eqref{BC real poles} but by the equivariance properties \eqref{equiv cond} applied to $A_{\mu}$, requiring that $\hat{\tau}\big(A_{\mu}\big|_{z=z_{1}}\big)=A_{\mu}\big|_{z=z_{2}}$, it follows that the vanishing of \eqref{BC real poles} boils down, this time, to the condition
\begin{equation}
\epsilon^{\mu \nu}\Big\langle\! \! \Big\langle \delta A_{\mu}\big|_{z=z_{1}}, \hat{A}_{\nu}\big|_{z=z_{1}} \Big\rangle\! \! \Big\rangle _{\mathfrak{g}}=0,\label{simple complex poles BC}
\end{equation}
where, in order to reach the inner product form \eqref{simple complex poles BC}, besides using \eqref{cond kappa 2}, we have further imposed the additional condition on the 1-form $\kappa$, given by
\begin{equation}
\kappa_{\mu}\big|_{z=z_{1}}\in \mathbb{R}. \label{cond kappa 3}
\end{equation}

The equation \eqref{simple complex poles BC} can be solved if we require that $A_{\mu}\big|_{z=z_{1}}$ and its variation both belong to the Lagrangian subalgebra $\mathfrak{g}_{R}$, defined in \eqref{Lag for com poles}. Alternatively, Cf. \eqref{Lag for com poles'}, this is equivalent to 
\begin{equation}
\big(R-i  \big)A_{\mu}\big|_{z=z_{2}}=\big( R+i  \big)A_{\mu}\big|_{z=z_{1}}. \label{complex cond gauge}
\end{equation} 
This solution must be preserved by the action of the gauge group $\mathcal{G}_{0}$. Furthermore, notice that the quadratic contribution \eqref{Q term}, taking the form \eqref{QQ}, also vanishes by virtue of \eqref{complex cond gauge}.

A typical example with this pole structure is provided by the Yang-Baxter $\sigma$-model \cite{Klimcik,eta-def bos}, also known as the $\eta$-deformed PCM. As before, the reader is referred to \cite{unifying} for further details. In this case the twist form, is given by
\begin{equation}
\omega_{C}=\frac{a}{1-\tilde{c}^{2}\eta^{2}}\frac{1-z^{2}}{z^{2}-\tilde{c}^{2}\eta^{2}}dz,
\end{equation}
where\footnote{We have set $K \rightarrow a$ and  $c \rightarrow \tilde{c}$, where $K, c$ are the constants used in \cite{unifying}.} $a,\eta \in \mathbb{R} $. The twist form has two simple zeroes at $\mathfrak{z}=\{+1,-1\}$, two simple poles at $\pm \tilde{c}\eta$ and a double pole at infinity\footnote{This pole is treated as in the first example, i.e. $A_{\mu}|_{\infty}=0$.}, i.e. $\mathfrak{p}=\{\tilde{c}\eta,-\tilde{c}\eta,\infty\}$. For $\tilde{c}=1$, the pair of simple poles are real hence belonging to the category $\S \eqref{4.2.2}$. We have included this pole configuration here simply because its analysis can be performed in parallel to the configuration of complex conjugate pair of simple poles corresponding to the choice $\tilde{c}=i$. 

With the identification $z_{1}=\tilde{c}\eta, z_{2}=-\tilde{c}\eta$, both solutions \eqref{real cond gauge'} and \eqref{complex cond gauge} can be written in the form
\begin{equation}
\big(R+\tilde{c}\big)A_{\mu}\big|_{z=z_{1}}=\big(R-\tilde{c}\big)A_{\mu}\big|_{z=z_{2}}. \label{TTT}
\end{equation}
After applying \eqref{TTT} to the restriction of \eqref{rotation} to $\Sigma \times \mathfrak{p}$ and using
\begin{equation}
\hat{g}\big|_{\Sigma \times \mathfrak{p}}=\hat{g} \big|_{\Sigma \times \{\tilde{c}\eta,-\tilde{c}\eta, \infty \}}=\left(\hat{g}\big|_{z=\tilde{c}\eta},\hat{g}\big|_{z=-\tilde{c}\eta},\hat{g}\big|_{z=\infty}\right):=(g,g,1),
\end{equation}
for some $g: \rightarrow G^{\mathbb{R}}$, where the freedom of acting with $\mathcal{G}_{\Sigma}$ is used to fix the value of $\hat{g}$ at $\infty$ to the identity, we get the following Lax connection
\begin{equation}
\mathscr{L}=\left(\frac{1}{z-1}\right)\frac{\tilde{c}^{2}\eta^{2}-1}{1+\eta R_{g}}J_{+}d\sigma^{+}-\left(\frac{1}{z+1}  \right)\frac{\tilde{c}^{2}\eta^{2}-1}{1-\eta R_{g}} J_{-}d\sigma^{-}.
\end{equation}
As before, $J=g^{-1}d_{\Sigma}g \in \Omega_{\Sigma}^{1}\otimes \mathfrak{g}^{\mathbb{R}} $. Now, using
\begin{equation}
k_{0}^{(1)}=a/2\tilde{c}\eta,\text{ \qquad \qquad }k_{0}^{(2)}=-a/2\tilde{c}\eta
\end{equation}
and the Lax connection above in the integral formulas \eqref{I1}, \eqref{I2}, we find that the action functional \eqref{rotated action}, takes the final form
\begin{equation}
S(g)_{\eta\text{-PCM}}=-4\pi ca \dint_{\Sigma} d\sigma^{+} \wedge d\sigma^{-}\, \text{Tr}\left( J_{+}\frac{1}{1-\eta R_{g}}J_{-}\right),
\end{equation}
where $R_{g}=Ad_{g^{-1}}\circ R\circ Ad_{g}$. 

Now we briefly comment on the conditions \eqref{cond kappa 0}, \eqref{cond kappa 1}, \eqref{cond kappa 2} and \eqref{cond kappa 3} imposed on top of the 1-form $\kappa$. Clearly, they are all easily satisfied if we require that the fiber components of $\kappa$, i.e. $\kappa_{\mu}$, $\mu=\tau,\sigma$ are real constants. Remarkably, this is precisely what happens for $\mu=\sigma$, when we consider a contact structure on a Seifert manifold M, as will be reviewed in $\S \eqref{5}$. For $\mu=\tau$, this is trivially fulfilled by choice. Furthermore, having $\kappa_{\mu}\in \mathbb{R}$ also ensures that the restriction $\kappa\big|_{\Sigma}$ is an equivariant 1-form, Cf. the first equation in \eqref{result equi}. The equivariance properties of $\kappa \big|_{C}$, will be addressed later on in $\S \eqref{5}$, as well.

Finally, notice that in the presence of defect insertions at the points $z_{j}$, $j=1,...,N$ and in the limit $\zeta \rightarrow 0$, the bulk eom \eqref{ext reg eom} change in a nontrivial way, namely
\begin{equation}
\omega\wedge F_{\mathbb{A}}=\delta_{S^{1}}\cdot \tilde{U} ,\text{ \qquad \qquad }d_{\mathbb{A}}U_{j}=0, \text{ \qquad \qquad }j=1,...,N
\end{equation}
It would be interested to explore how the Lax connections and their associated 2d IFT action functionals are modified in this case. We leave this as an open problem to be considered elsewhere in the future. 

\subsection{Extended and contact 4d CS theories boundary eom}

We investigate now other boundary eom that appear when we consider the variations of the extended and the contact 4d CS theories actions. We do this when $\zeta \neq 0$ and, for simplicity, in the absence of adjoint orbit insertions. For the extended 4d CS theory, we assume that the gauge field $\mathbb{A}$ boundary eom solutions now apply. In the case of the contact 4d CS theory, boundary eom are to be solved differently by choosing convenient gauge fixing conditions for the $\omega_{\zeta}$-shift and $\kappa$-shift symmetries. In both cases we assume that the 1-form $\kappa$ conditions found before, also hold. 

Consider first the extended 4d CS theory. The boundary eom for this theory are given by the solutions to the equation  \eqref{full BC M}, i.e. 
\begin{equation}
\mathcal{B}(\mathbb{A},\Phi)=\mathcal{B}(\mathbb{A})-\dint\nolimits_{%
\mathbb{M}}d_{\mathbb{M}}\omega_{\zeta} \wedge \kappa \wedge \text{Tr}%
\big( \mathbb{A} \delta \Phi    \big)=0, \label{Again}
\end{equation}
where we put to zero the first term on the rhs, as it corresponds to the regularized 4d CS theory boundary eom term \eqref{BC M}. The strategy now is to seek for boundary eom solutions for the adjoint scalar field $\Phi$, induced by the ones already imposed upon the gauge field $\mathbb{A}$, thus ensuring the vanishing of the remaining contribution. This strategy, by no means, exhausts all possibilities as boundary eom are to be solved under gauge fixing conditions for the whole set of local symmetries. As a guide of what conditions to impose upon $\Phi$, we use the $\omega_{\zeta}$-shift and $\kappa$-shift invariant field $\tilde{\mathbb{A}}:=\mathbb{A}-\kappa \Phi$ entering the extended action \eqref{ext action}. The idea is to require that the fiber components $A_{\mu}$ and $\tilde{A}_{\mu}$, both belong to the same space of boundary eom solutions. In this way, we stay as close as possible to the regularized 4d CS theory boundary eom solutions considered before. 

\textbf{Solutions for a real double pole}. We have that 
\begin{equation}
\mathcal{B}(\mathbb{A}, \Phi)=-2\pi i \dint_{\Sigma}d\tau \wedge d\sigma \, k_{1}^{(r)}\epsilon^{\mu \nu}\kappa_{\mu}\text{Tr} \big(\partial_{z}A_{\nu}\delta \Phi \big) \big|_{z=z_{r}}, \label{BC APhi1}
\end{equation}
for $k_{0}^{(r)}\neq 0$, $k_{1}^{(r)}\neq 0$, corresponding to the PCM with WZ term and
\begin{equation}
\mathcal{B}(\mathbb{A}, \Phi)=-2\pi i\dint_{\Sigma}d\tau \wedge d\sigma \, k_{1}^{(r)}\epsilon^{\mu \nu}\kappa_{\mu}\text{Tr} \Big[\partial_{z}A_{\nu}\delta\Big( \Phi+R(\partial_{z}\Phi)   \Big) \Big] \big|_{z=z_{r}}, \label{BC APhi2}
\end{equation}
for $k_{0}^{(r)}= 0$, $k_{1}^{(r)}\neq 0$, corresponding to the homogeneous Yang-Baxter sigma model. In the first case, the expression \eqref{BC APhi1} can be put to zero if we impose
\begin{equation}
\partial_{z}A_{\mu}\big|_{z=z_{r}}=0\text{\qquad \qquad  or \qquad \qquad }\Phi \big|_{z=z_{r}}=0, \label{x}
\end{equation}
while in the second case, the expression \eqref{BC APhi2} vanishes if we demand that\footnote{If the real double pole is at infinity, we always use $\Phi \big|_{\infty}=0$ in any case.}
\begin{equation}
\Phi\big|_{z=z_{r}}=-R(\partial_{z}\Phi)\big|_{z=z_{r}}. \label{y}
\end{equation}
In the latter case, one is to take $\partial_{z}A_{\nu}\big|_{z=z_{r}}\neq 0$. Thus, in order for $\tilde{A}_{\mu}$ to belong to the same space of boundary eom solutions as $A_{\mu}$, we impose the second relation in \eqref{x} and \eqref{y}. The same holding for their variations. These solutions must be preserved by the action of the gauge group $\mathcal{G}_{0}$ as well.

\textbf{Solutions for two real simple poles}. We get
\begin{equation}
\mathcal{B}(\mathbb{A}, \Phi)=-2\pi i\dint_{\Sigma}d\tau \wedge d\sigma \,k_{0}^{(1)}\epsilon^{\mu \nu}\kappa_{\mu} \text{Tr} \Big[A_{\nu}\big|_{z=z_{1}}\delta\Big( \Phi\big|_{z=z_{1}}-\Phi \big|_{z=z_{2}}  \Big) \Big] ,
\end{equation}
corresponding to the $\lambda$-deformed PCM. This boundary eom expression vanish when
\begin{equation}
\Phi\big|_{z=z_{1}}=\Phi \big|_{z=z_{2}} \text{ \qquad \qquad }\text{or}\text{ \qquad \qquad }\Phi \big|_{z=z_{r}}=0, \text{ \qquad }r=1,2 \label{z}
\end{equation}
where we have assumed that $A_{\nu}\big|_{z=z_{1}} \neq 0$. For the Yang-Baxter $\sigma$-model with $\tilde{c}=1$, we obtain instead
\begin{equation}
\mathcal{B}(\mathbb{A}, \Phi)=-\pi i\dint_{\Sigma}d\tau \wedge d\sigma \,k_{0}^{(1)}\epsilon^{\mu \nu}\kappa_{\mu} \text{Tr} \Big[\Big( A_{\nu}^{1}-A_{\nu}^{2} \Big)\delta\Big( \Phi^{1}+\Phi^{2}+R\big( \Phi^{1}-\Phi^{2}  \big)  \Big) \Big] , 
\end{equation}
with $A_{\nu}^{1}\neq A_{\nu}^{2}$ and where we have used the shorthand notation $F^{r}=F \big|_{z=z_{r}}$, for $r=1,2$ and $F=A_{\mu}, \Phi$. This boundary eom is solved if the impose that
\begin{equation}
\big(R+1\big)\Phi \big|_{z=z_{1}}=\big(R-1\big)\Phi\big|_{z=z_{2}}.\label{z'}
\end{equation} 
Clearly, we select the first option in \eqref{z} and \eqref{z'} to ensure that the pair $(A_{\mu},\tilde{A}_{\mu})$ belong to the same boundary eom solutions space as $A_{\mu}$. Both must be preserved by the action of $\mathcal{G}_{0}$, of course.

\textbf{Solutions for two complex conjugate simple poles}. We obtain
\begin{equation}
\mathcal{B}(\mathbb{A}, \Phi)=-2\pi i\dint_{\Sigma}d\tau \wedge d\sigma \,k_{0}^{(1)}\epsilon^{\mu \nu}\kappa_{\mu} \text{Tr} \Big[A_{\nu}\delta \Phi \big|_{z=z_{1}}-A_{\nu}\delta \Phi \big|_{z=z_{2}}\Big] ,
\end{equation}
corresponding to the Yang-Baxter $\sigma$-model. Following the same steps, this boundary eom vanish when we impose
\begin{equation}
\big(R+i\big)\Phi\big|_{z=z_{1}}=\big(R-i \big)\Phi\big|_{z=z_{2}},\label{w}
\end{equation}
where we have used the fact that $A_{\nu}^{1}\neq A_{\nu}^{2}$.

The structure for the adjoint scalar field $\Phi$ boundary eom solutions is clear and should not come as a surprise. Indeed, if $\mathbb{A} \big|_{\Sigma \times \mathfrak{p}}\in \Omega_{\Sigma }^{1}\otimes \underline{\mathfrak{l}}$, with $\underline{\mathfrak{l}}=\underline{\mathfrak{t}},\underline{\mathfrak{p}}$ being Lagrangian subalgebras of $\underline{\mathfrak{g}}$, is a solution to the boundary eom, then the condition $\tilde{\mathbb{A}}\big|_{\Sigma \times \mathfrak{p}}\in \Omega_{\Sigma}^{1}\otimes \underline{\mathfrak{l}}$, requires that
\begin{equation}
\Phi \big|_{\Sigma \times \mathfrak{p}} \in \Omega_{\Sigma }^{0}\otimes \underline{\mathfrak{l}}, \text{\qquad \qquad}\text{or}\text{\qquad \qquad}\Phi \big|_{\Sigma \times \mathfrak{p}} =0. \label{BC Phi off}
\end{equation}
We can understand the latter condition as originating from the $\kappa$-shift symmetry gauge fixing condition $\Phi=0$, Cf. \eqref{Phi=0}, restricted to the pole-defect surface. However, in this gauge the extended 4d CS theory action reduces to the action of the regularized 4d CS theory and no new structure is revealed. 

Now, we move to find the contact 4d CS theory boundary eom, which are found when \eqref{BC dual 1}, i.e.
\begin{equation}
\mathcal{B}(\mathbb{A})_{\text{dual}}=\mathcal{B}(\mathbb{A})+\dint\nolimits_{%
\mathbb{M}}d_{\mathbb{M}}\omega_{\zeta} \wedge \kappa \wedge \text{Tr}%
\big( \delta \mathbb{A} \Phi_{\text{on}}\big), \label{dual BC complete'}
\end{equation}
vanishes. We shall try two ways to solve them. The first one assumes that the first term on the rhs of \eqref{dual BC complete'} is already zero, while the second 
exploits the existing similarities to the 3d CS theory case \cite{NA loc CS}, in the sense of exploiting a different gauge fixing conditions for the shift symmetries. We will see why the second way is more natural from the contact 4d CS theory point of view.

In the first case, from the similarity between \eqref{dual BC complete'} and \eqref{Again}, and the relation between $\Phi$ and $\Phi_{\text{on}}$, we propose that $\Phi_{\text{on}}$ must possess, in principle, the following properties:

\textbf{At a real double pole:}

For $k_{0}^{(r)}\neq 0$, $k_{1}^{(r)}\neq 0$ and  $k_{0}^{(r)}= 0$, $k_{1}^{(r)}\neq 0$ we require, respectively, that
\begin{equation}
\Phi_{\text{on}}\big|_{z=z_{r}}=0\text{ \qquad \qquad  }\text{and}\text{\qquad \qquad}\Phi_{\text{on}}\big|_{z=z_{r}}=-R(\partial_{z}\Phi_{\text{on}})\big|_{z=z_{r}}.\label{q}
\end{equation}

\textbf{At two real simple poles:}

We impose that
\begin{equation}
\Phi_{\text{on}}\big|_{z=z_{1}}=\Phi_{\text{on}}\big|_{z=z_{2}}\text{ \qquad \qquad }\text{and}\text{ \qquad \qquad } \big(R+1\big)\Phi_{\text{on}} \big|_{z=z_{1}}=\big(R-1\big)\Phi_{\text{on}}\big|_{z=z_{2}},\label{qq}
\end{equation}
corresponding to the $\lambda$-deformed PCM and the Yang-Baxter $\sigma$-model with $\tilde{c}=1$, respectively.

\textbf{At two complex conjugate simple poles:}

For the Yang-Baxter $\sigma$-model with $\tilde{c}=i$, we demand that
\begin{equation}
\big(R+i\big)\Phi_{\text{on}} \big|_{z=z_{1}}=\big(R-i\big)\Phi_{\text{on}}\big|_{z=z_{2}}.\label{qqq}
\end{equation}

It is possible to eliminate the second contribution to the rhs in \eqref{dual BC complete'} from the very beginning, if we invoke the $\kappa$-shift symmetry gauge fixing condition $\Phi_{\text{on}}=0$, Cf. \eqref{Phion=0}, restricted to the pole-defect surface, i.e.
\begin{equation}
\Phi_{\text{on}}\big|_{\Sigma \times \mathfrak{p}}=0. \label{kappa gauge BC}
\end{equation}
Cf. also step I in the diagram at the end of $\S \eqref{2.1.3}$. However, this choice of gauge fixing condition takes the contact 4d CS theory back to the regularized 4d CS theory.

Nevertheless, it is interesting to explore if $\Phi_{\text{on}}$ inherits the demanded properties from the gauge field $\mathbb{A}$ boundary eom solutions, namely, \eqref{q}, \eqref{qq} and \eqref{qqq}, as $\Phi_{\text{on}}$ depends on $\mathbb{A}$ via the field strength $F_{\mathbb{A}}=d_{\mathbb{M}}\mathbb{A}+\mathbb{A}\wedge \mathbb{A}$ and the equivariant differential forms $\omega_{\zeta}, \kappa$ and $\gamma_{\text{top}}$, Cf. eq. \eqref{Phi eom}. 

Let us begin with the equivariance properties. The second relation in \eqref{equiv cond'} with $\rho=\mathbb{A}$ means that $F_{\mathbb{A}}$ is a Lie algebra valued equivariant 2-form and from this follows that $\Phi_{\text{on}}$ is a Lie algebra valued equivariant $0$-form, a property already required for $\Phi$. Indeed, we have that
\begin{equation}
\hat{\tau}\big( \Phi_{\text{on}}  \big)\overline{\gamma}_{\text{top}}=\overline{\omega_{\zeta}}\wedge \overline{\kappa}\wedge \hat{\tau}\big( F_{\mathbb{A}} \big)=\nu^{\ast}\big(\Phi_{\text{on}}   \big)\nu^{\ast}\big( \gamma_{\text{top}}   \big)=\nu^{\ast}\big(\Phi_{\text{on}}\big)  \overline{\gamma}_{\text{top}},
\end{equation} 
implying
\begin{equation}
\hat{\tau}\big( \Phi_{\text{on}}(x)\big)=\Phi_{\text{on}}(\overline{x})=\big(\nu^{\ast}\Phi_{\text{on}}  \big)(x).
\end{equation}
See \eqref{equiv cond} with $\rho=\Phi_{\text{on}}$. Thus, at the level of equivariance properties, the consistency is clear.  

Consider now the behavior induced on $F_{\tilde{\mu}\tilde{\nu}}$, which depends on the whole set of components of the gauge field $A_{\tilde{\mu}}$, with $\tilde{\mu}=\tau, \sigma, z, \overline{z}$. It is then natural to extend the boundary eom solutions from the fiber components $A_{\mu}$, with $\mu=\tau, \sigma$ to the whole gauge field $A_{\tilde{\mu}}$. 
However, the difficulties start to appear even when considering simple solutions like \eqref{double poles BC 2}, corresponding to a real double pole. Indeed, they imply that
\begin{equation}
F_{\tilde{\mu}\tilde{\nu}}\big|_{z=z_{r}}=-R\big(\partial_{z}F_{\tilde{\mu}\tilde{\nu}}  \big)\big|_{z=z_{r}},
\end{equation}
by virtue of the YB equation, $\tilde{c}=0$ in \eqref{cmCYB eq}, but for $\Phi_{\text{on}}$ we get instead
\begin{equation}
\Phi_{\text{on}}\big|_{z=z_{r}}=-R\big(\partial_{z}\Phi_{\text{on}}\big)\big|_{z=z_{r}}+\partial_{z}b_{\tilde{\mu}\tilde{\nu}}R\big(F_{\tilde{\mu}\tilde{\nu}}\big)\big|_{z=z_{r}},
\end{equation}
where we have written $\Phi_{\text{on}}=b_{\tilde{\mu}\tilde{\nu}}F_{\tilde{\mu}\tilde{\nu}}$, with $b_{\tilde{\mu}\tilde{\nu}}$ constructed out of the $\omega_{\zeta}, \kappa$ and $\gamma_{\text{top}}$ components. Thus, the second way for solving the boundary eom \eqref{Again} now enters the picture.  

In the second case, we exploit the fact that the contact 4d CS theory is invariant under $\omega_{\zeta}$ and $\kappa$ shifts. Thus, instead of using the $\kappa$-shift gauge fixing condition $\Phi_{\text{on}}=0$, Cf. \eqref{Phion=0}, we use $\iota_{X} \mathbb{A}=0$, Cf. \eqref{sigma gauge fixing}. Now, from \eqref{Reeb components} we get the component expression 
\begin{equation}
\iota_{\partial_{\sigma}} \mathbb{A}=0. \label{t}
\end{equation}
In the 3d CS theory case on M, it becomes $\iota_{R}A=0$, where $R$ is the Reeb field \cite{NA loc CS} associated to the real contact form $\alpha_{r}$ on M. More on this below. For the time being, for the $\omega_{\zeta}$-shift symmetry (recall that $\zeta \neq 0$), we use the gauge fixing condition \eqref{tau gauge fixing}, i.e.
\begin{equation}
\iota_{\partial_{\tau}}\mathbb{A}=0. \label{s}
\end{equation}
This means that in this gauge the fiber components of $\mathbb{A}$ are completely eliminated. Thus, the contact 4d CS theory is then more naturally defined in terms of the remaining gauge field components $(A_{z}, A_{\overline{z}})$ and \eqref{dual BC complete'} trivially vanishes even though $\Phi_{\text{on}}\neq 0$. The fact that the theory is more naturally defined on $C$ is not accidental. We will comment a bit more about this in the concluding remarks.

Finally, the proof that any of the boundary eom solutions $\mathbb{A}$ and $\Phi$ are preserved by the action of the gauge group $\mathcal{G}_{0}$, will be provided below in $\S \eqref{5.1.3}$, after a detailed study of the symmetry properties of the quotient space $\overline{\mathcal{A}}$ in \eqref{MR}, where we formally identify $\mathcal{G}_{0}$, is performed. We will also address \eqref{t} and \eqref{s} and its relation to $\mathcal{G}_{0}$, in more generality in $\S \eqref{6}$.

 \section{Contact 4d CS theory as a quadratic action}\label{5}

The focus of this section is to refine and generalize the results originally presented in \cite{Yo}. We review the geometric and algebraic structures that are necessary to conclude that the contact 4d CS theory has one of the main properties required by the method of non-Abelian localization of symplectic integrals, specifically, that its action functional is quadratic in the moment map associated to a symmetry group, acting in a Hamiltonian fashion, on a sub-space of the space of gauge connections defined on $\mathbb{M}$. We also include in the analysis, following \cite{Wilson NA loc}, the symplectic data associated to the (co)-adjoint orbit defect insertions introduced in \eqref{2.2}. 

\subsection{Bulk data}

We introduce the symplectic form, moment map and inner product required to write the contact 4d CS theory action in quadratic form. In particular, we identify formally the physical gauge symmetry group $\mathcal{G}_{0}\in \mathcal{G}$ and explore its relation to the action functionals and boundary terms considered in $\S \eqref{2}$ and $\S \eqref{4}$, respectively.

\subsubsection{Symplectic form and moment map}\label{5.1.1}

Consider a trivial principal $G$-bundle $P$, where $G$ is a Lie group with Lie algebra $\mathfrak{g}$ as defined in $\S \eqref{2.1.1}$, over the 4d orientable manifold $\mathbb{M}$, with $\partial \mathbb{M}=0$. Denote by $\mathcal{A}$ the space of connections $\mathbb{A}$ on $P$ and identify it with the space $\Omega_{\mathbb{M}}^{1}\otimes \mathfrak{g}$ of 1-forms on $\mathbb{M}$ taking values in the Lie algebra $\mathfrak{g}$. Recall that $\mathcal{G}$ is the group of formal gauge transformations acting on $\mathcal{A}$ and that its Lie algebra, denoted by $\mathcal{G}_{\text{Lie}}$, consists of elements in the space $\Omega_{\mathbb{M}}^{0}\otimes \mathfrak{g}$. The geometric data is then provided by the bundles $G\longhookrightarrow P \longrightarrow \mathbb{M}$, i.e. $P=\mathbb{M}\times G$ and $S^{1}\overset{n}{\longhookrightarrow} \text{M}\overset{\underline{\pi}}{\longrightarrow} C$, such that $\mathbb{M}=\mathbb{R}\times \text{M}$. As before, $\mathbb{R}$ corresponds to the global time direction. A word of caution, we are not specifying reality conditions on some quantities and as consequence, some objects that are usually expected to be real, will turn out to be purely imaginary. Thus, some `i' factors will be inserted at due time.

Introduce a pre-symplectic form $\hat{\Omega}\in\Omega_{\mathcal{A}}^{2}$ on the space of connections $\mathcal{A}$, defined by
\begin{equation}
\hat{\Omega}:=-\frac{1}{2}\dint\nolimits_{\mathbb{M}}\Omega \wedge
\kappa \wedge \text{Tr}\big( \hat{\delta }\mathbb{A}\wedge\hat{\delta }\mathbb{A}\big), \label{pre-symplectic}
\end{equation}%
where $\Omega, \kappa \in \Omega_{\mathbb{M}}^{1}$ are two globally defined and nowhere vanishing 1-forms on $\mathbb{M}$, satisfying the defining relations
\begin{equation}
i_{\mathcal{R}}\kappa =1,\text{ \qquad }i_{\mathcal{R}}\left( d_{\mathbb{M}
}\kappa \right) =0,\text{ \qquad }i_{\mathcal{R}}\Omega =0,\text{ \qquad }i_{\mathcal{R}}(d_{\mathbb{M}}\Omega) =0 \label{Conditions 1}
\end{equation}
and
\begin{equation}
i_{\mathcal{R}'}\kappa =0,\text{ \qquad }i_{\mathcal{R}'}\left( d_{\mathbb{M}%
}\kappa \right) =0,\text{ \qquad}i_{\mathcal{R}'}\Omega =1,\text{ \qquad }i_{\mathcal{R}'}(d_{\mathbb{M}}\Omega) =0. \label{Conditions 2}
\end{equation}
As a consequence, we have that $\Omega, \kappa$ are invariant under the actions of $\mathcal{R}$ and $\mathcal{R}'$, i.e.
\begin{equation}
\pounds _{\mathcal{R}}\kappa =0,\text{ \qquad }\pounds_{\mathcal{R}}\Omega  =0, \text{ \qquad }\pounds _{\mathcal{R}'}\kappa =0,\text{ \qquad}\pounds_{\mathcal{R}'}\Omega =0. \label{consequences}
\end{equation} 
Above, $\hat{\delta}$ represents the exterior derivative on $\mathcal{A}$, $\mathcal{R}, \mathcal{R}'\in \mathfrak{X}_{\mathbb{M}}$ are two commuting vector fields on $\mathbb{M}$, i.e. its Lie bracket $[\mathcal{R},\mathcal{R}']$ vanishes and Tr is the usual bilinear form on $\mathfrak{g}$.

The 2-form $\hat{\Omega}$ is closed and invariant under the action of $\mathcal{G}$ and, in particular, under the action of the group $\mathcal{S}=U(1)^{\times 2}$ generated by two independent shifts of the form  
\begin{equation}
^{\kappa}\mathbb{A}=\mathbb{A}+s\kappa ,\text{ \qquad \qquad }^{\Omega}\mathbb{A}=%
\mathbb{A}+s' \Omega , \label{shifts}
\end{equation}
where $s,s'\in \Omega_{\mathbb{M}}^{0}\otimes \mathfrak{g}$ are arbitrary and from this follows that the 2-form $\hat{\Omega}$ is degenerate along elements $\mathbb{A}$ of the form $\mathbb{A}=s \kappa$, $\mathbb{A}=s' \Omega$. Thus, we take the quotient of $\mathcal{A}$ by the action of the group $\mathcal{S}$ and define the symplectic space $\overline{\mathcal{A}}=\mathcal{A}/\mathcal{S}$. Under the quotient, the pre-symplectic form $\hat{\Omega}$ on $\mathcal{A}$ descends to a symplectic form on $\overline{\mathcal{A}}$. The action of the formal gauge group $\mathcal{G}$ on $\mathcal{A}$ also descends to an action on $\overline{\mathcal{A}}$ and the 2-form $\hat{\Omega}$ on $\overline{\mathcal{A}}$ is invariant under the action of $\mathcal{G}$. 

We will consider the following solutions \cite{Yo}, to the conditions \eqref{Conditions 1} and \eqref{Conditions 2}, given by\footnote{To match with the expressions used in \cite{Yo}, set $\zeta \rightarrow 2\zeta/\alpha_{\tau}$ and $\zeta'\rightarrow 1/2$.}
\begin{equation}
\Omega=\omega+\zeta\big(\alpha_{\tau}d\tau-\alpha\big),\text{ \qquad \qquad }\kappa=\zeta'\big(\alpha_{\tau}d\tau+\alpha\big) \label{solutions omega, kappa}
\end{equation} 
and
\begin{equation}
\mathcal{R}'=\frac{1}{2\zeta}\Big(T-R  \Big),\text{ \qquad }\mathcal{R}=\frac{1}{2\zeta'}\Big(T+R  \Big),\text { \qquad }T=\frac{1}{\alpha_{\tau}}\partial_{\tau}. \label{R and R'}
\end{equation}
Right above, the 1-form $\omega=\underline{\pi}^{\ast}(\omega_{C}) \in \Omega_{\text{M}}^{1}$ is the pull-back, under the projection map $\underline{\pi}$, of the twist form $\omega_{C}\in \Omega_{C}^{1,0}$ specifying the analytic structure of the underlying 2d IFT Lax connection, the 1-form $\alpha \in \Omega_{\text{M}}^{1}$ is an equivariant extension, in the sense introduced above in $\S \eqref{3}$ of a real contact form $\alpha_{r}$ defined on M, $R\in \mathfrak{X}_{\text{M}}$ is the Reeb vector field associated to $\alpha_{r}$, $\zeta, \zeta'$ are `deformation' parameters and $\alpha_{\tau}\neq 0 \in \mathbb{R}$ is a real constant. In general, the contact form $\alpha_{r}$ on M satisfy the fundamental defining relations\footnote{Consult also the references \cite{Martinet,Etnyre,Boothby-Wang}, for further details on contact manifolds.} \cite{Blair},\cite{Geiges}
\begin{equation}
\iota_{R}\alpha_{r}=1,\text{ \qquad }\iota_{R}(d_{\text{M}}\alpha_{r})=0,\text{ \qquad }d_{\text{M}}\alpha_{r}=n \underline{\pi}^{\ast}(\sigma_{C}), \text{ \qquad }\alpha_{r} \wedge d_{\text{M}}\alpha_{r} \neq 0, \label{contact defs}
\end{equation}
where $n$ is the degree of the bundle M, $\sigma_{C}\in \Omega_{\text{M}}^{1,1}$ is a symplectic form on the base space $C$ and $d_{\text{M}}$ represents the exterior differential on M in the decomposition $d_{\mathbb{M}}=d\tau\wedge \partial_{\tau}+d_{\text{M}}$. We also have that $\iota_{R}\omega=0$ and the following normalizations
\begin{equation}
\dint\nolimits_{S^{1}}\alpha_{r}=1,  \text{ \qquad }\dint\nolimits_{C}\sigma_{C}=1,\text{ \qquad }\dint\nolimits_{\text{M}}\alpha_{r} \wedge d_{\text{M}}\alpha_{r}=n. \label{normalizations}
\end{equation}
Actually, there is a generalization of the last formula right above when the base space is an orbifold of the genus $g$  Riemann surface $C$, but we will not cover those geometries in the present work. In fact, we will mostly operate as if $C=\mathbb{CP}^{1}$.

The 1-form $\alpha_{r}$ also defines an invariant $U(1)$-connection on M regarded now as the total space of a $U(1)$ principal bundle over $C$. The $U(1)$ action being generated by $R \in \mathfrak{X}_{\text{M}}$ and corresponding to rigid translations along the $S^{1}$ fibers. In particular, $\alpha_{r}$ associates a separation of $T_{p}\text{M}$, $p \in \text{M}$ into vertical and horizontal sub-spaces. This also applies to differential forms, for instance, if $\gamma=\underline{\pi}^{\ast}(\gamma_{C})$ for some $\gamma_{C} \in \Omega_{C}^{\ast}$, then $\iota_{R}\gamma=0$, i.e. $\gamma$ is horizontal. This is why $\iota_{R}\omega=0$ and $\iota_{R}(d_{\text{M}}\alpha_{r})=0$, above. Also, we have the important result that the top form
\begin{equation}
\gamma_{\text{top}}=\Omega\wedge \kappa \wedge d_{\mathbb{M}}\kappa=-(\zeta')^{2}\big( \alpha_{\tau} d\tau+\alpha \big)\wedge \omega \wedge d_{\text{M}}\alpha+2\alpha_{\tau}\zeta (\zeta')^{2}d\tau \wedge \alpha \wedge d_{\text{M}}\alpha \neq 0 \label{gamma Top}
\end{equation}
never vanishes unless\footnote{By the $t$-rescaling symmetry, the parameter $\zeta'$ is always assumed to be different from zero.} $\zeta\neq 0$, hence defining an equivariant volume form on $\mathbb{M}$ and validating what was assumed in the paragraph above equation \eqref{Phi eom}. The last result follows from the fact that
\begin{equation}
\omega\wedge d_{\text{M}}\alpha\sim n\underline{\pi}^{\ast}(\omega_{C}\wedge \sigma_{C})=0, \label{dim reason}
\end{equation}
by dimensional reasons, hence clarifying why in the limit $\zeta \rightarrow 0$, there is no notion of duality, as the field $\Phi$ entering the extended action \eqref{double shift inv} is linear instead of quadratic. In \eqref{gamma Top} we can appreciate the effect of the regularization procedure introduced in $\S \eqref{2.1.2}$, ensuring that $\gamma_{\text{top}}$ is a genuine volume form defined on $\mathbb{M}$ in every practical sense. Later on, we will show that $\Omega$ and $\omega_{\zeta}$ differ by a term proportional to $\kappa$ and this means that $\gamma_{\text{top}}$ can be defined using either $\Omega$ or $\omega_{\zeta}$. This also clarifies why the expression \eqref{spoiler sol iii} includes a $d\tau$ contribution of that particular form. Notice how important is to have $\zeta \neq 0$ in the definition of $\gamma_{\text{top}}$, so the limit $\zeta \rightarrow 0$ must be implemented carefully.

Let us briefly discuss now how to construct, in practice, an equivariant extension $\alpha$ of a given real contact form $\alpha_{r}$ defined on M. Suppose that in the local bundle coordinates $x=(z,\overline{z},\sigma)$ on $\mathcal{U}\subset \text{M}$, we have the following expression 
\begin{equation}
\alpha_{r}=\alpha_{rz}dz+\alpha_{r\overline{z}}d\overline{z}+\alpha_{\sigma}d\sigma, \label{alpha real 1}
\end{equation}
where $\alpha_{\sigma}\neq 0 \in \mathbb{R}$ is a real constant, 
\begin{equation}
\alpha_{rz}(z,\overline{z})=-\frac{i}{2}\partial_{z}\mathcal{K}(z,\overline{z}),\text{\qquad  \qquad}\alpha_{r\overline{z}}(z,\overline{z})=+\frac{i}{2}\partial_{\overline{z}}\mathcal{K}(z,\overline{z})\label{components}
\end{equation}
and the K\"ahler potential $\mathcal{K}:C\rightarrow \mathbb{R}$ is such that
\begin{equation}
\mathcal{K}(z,\overline{z})=\mathcal{K}(\overline{z},z),\text{ \qquad \qquad}\overline{\mathcal{K}(z,\overline{z})}=\mathcal{K}(z,\overline{z}).
\end{equation}
Notice that $d_{\text{M}}\alpha_{r}=i\partial \overline{\partial}\mathcal{K}$, where $\partial=dz\wedge\partial_{z}$ and $\overline{\partial}=d\overline{z}\wedge \partial_{\overline{z}}$, in consistency with the third relation in the defining expressions \eqref{contact defs}. It is not difficult to show that
\begin{equation}
\alpha=i\alpha_{rz}dz+i\alpha_{r\overline{z}}d\overline{z}+\alpha_{\sigma}d\sigma \label{equi contact}
\end{equation}
is equivariant, see the first equation in \eqref{equiv cond}, and that
\begin{equation}
\overline{\alpha(x)}=\alpha(\overline{x})=(\nu^{\ast}\alpha)(x).
\end{equation}
In showing this, we have used the first result in \eqref{result equi} in order to deal with the last contribution in \eqref{equi contact}. From this, it follows that the top form on M, given by
\begin{equation}
\alpha \wedge d_{\text{M}}\alpha =i\alpha_{r}\wedge d_{\text{M}}\alpha_{r}\neq 0,
\end{equation}
never vanishes, by virtue of the contact condition over $\alpha_{r}$, hence ensuring \eqref{gamma Top}. Thus, if we consider, for example, real and equivariant integrands $f_{r}$ and $f$, their corresponding integrals
\begin{equation}
\dint_{\text{M}}\alpha_{r}\wedge d_{\text{M}}\alpha_{r}f_{r},\text{ \qquad \qquad }\dint_{\text{M}}\alpha \wedge d_{\text{M}}\alpha f,
\end{equation}
belong to $\mathbb{R}$ and $i\mathbb{R}$, respectively. In addition, because of the fiber component remains unchanged, we have that
\begin{equation}
\iota_{R}\alpha_{r}=\iota_{R}\alpha=1. \label{contraction}
\end{equation}

As an example, take $\text{M}=S^{3}$, seen as $|z_{0}|^{2}+|z_{1}|^{2}=1$, for $(z_{0},z_{1})\in \mathbb{C}^{2}$. We have that
\begin{equation}
\begin{aligned}
\alpha_{r}&=\frac{i}{4\pi}\Big(z_{0}d\overline{z}_{0}-\overline{z}_{0}dz_{0}+ z_{1}d\overline{z}_{1}-\overline{z}_{1}dz_{1} \Big),\\
R&=2\pi i\Big( z_{0}\partial_{z_{0}}-\overline{z}_{0}\partial_{\overline{z}_{0}}+ z_{1}\partial_{z_{1}}-\overline{z}_{1}\partial_{\overline{z}_{1} }  \Big).
\end{aligned}
\end{equation}
In order to exhibit $\alpha_{r}$ in a clearer way, we introduce local Hopf coordinates defined by
\begin{equation}
\big(z,e^{i\sigma}\big)=\left( \frac{z_{0}}{z_{1}},\frac{z_{1}}{|z_{1}|}   \right),\text{ \qquad \qquad}\big(z_{0},z_{1}\big)=\frac{e^{i\sigma}}{\sqrt{1+|z|^{2}}}(z,1),
\end{equation}
where $z_{1}\neq 0$. Then, we get that $\alpha_{r}$, takes the form \eqref{alpha real 1} and \eqref{components}, with $\mathcal{K}=\frac{1}{2\pi}\text{ln}(1+|z|^{2})$, $\alpha_{\sigma}=\frac{1}{2\pi}$ and that the Reeb vector field is simply $R=2\pi \partial_{\sigma}$. Notice that because of $dz$ and $d\overline{z}$ form a basis of the space of horizontal differential 1-forms, we have that \eqref{contraction} holds. 

Let us consider now the symmetry structure of the quotient space $\overline{\mathcal{A}}$. The symmetry group $\mathcal{H}$ acting on the quotient space $\overline{\mathcal{A}}$ in a Hamiltonian way preserving the symplectic form $\hat{\Omega}$, depends on the combined action of the formal gauge group $\mathcal{G}$ and a vector field $X \in \mathfrak{X}_{\mathbb{M}}$ constructed out of $\mathcal{R}$ and $\mathcal{R}'$. We have that this group and its Lie algebra (as a vector space) are given, respectively, by $\mathcal{H}=U(1)^{\times2}\ltimes \tilde{\mathcal{G}}$ and $\mathfrak{h}=\mathbb{R}^{2}\oplus \tilde{\mathcal{G}}_{\text{Lie}}$, where $\tilde{\mathcal{G}}_{\text{Lie}}=\mathcal{G}_{\text{Lie}}\oplus \mathbb{C}$ is a central extension of the gauge algebra $\mathcal{G}_{\text{Lie}}$ defined by a Lie algebra co-cycle to be identified below. Above, in $\mathcal{H}$ we have that $U(1)^{\times 2}=U(1)_{\mathcal{R}}\times U(1)_{\mathcal{R}'}$. Elements in $\mathfrak{h}$ are triplets of the form $(X,\eta, a)$, where $X=p \mathcal{R}+p' \mathcal{R}' $, with $p,p' \in \mathbb{R}$, $\eta \in \mathcal{G}_{\text{Lie}}$ and $a \in \mathbb{C}$. In general, $\mathfrak{h}$ is equipped with a Lie algebra bracket defined by\footnote{The first entry on the rhs is zero for the particular form of the vector fields considered, where the Lie bracket $[\mathcal{R},\mathcal{R}']$ is always assumed to vanish, but in general it can be non-trivial for generic vector fields $X,Y$. This generalized Lie bracket was not considered initially in \cite{Yo} and is new. }
\begin{equation}
\Big[ \left( X,\eta ,a\right) ,\left( Y,\lambda ,b\right) \Big] :=\Big( [X,Y],\,%
\left[ \eta ,\lambda \right] +\pounds _{X}\lambda -\pounds _{%
Y}\eta ,\,c(\eta ,\lambda )\Big) , \label{total bracket}
\end{equation}%
where
\begin{equation}
c\left( \eta ,\lambda \right) =\dint\nolimits_{\mathbb{M}}d_{\mathbb{M}%
}\left( \Omega \wedge \kappa \right) \wedge \text{Tr}\left( \eta d_{\mathbb{%
M}}\lambda \right)  \label{co-cycle}
\end{equation}
is the Lie algebra co-cycle that defines the central extension of $\mathcal{G}_{\text{Lie}}$. The pairing between an element $\mu \in \mathfrak{h}^{\ast}$ and an element $(X,\eta,a)\in \mathfrak{h}$ is defined by
\begin{equation}
\left\langle \mu,(X,\eta,a) \right\rangle:=\left \langle \mu,(X,\eta,0) \right\rangle+a,
\end{equation}
where an explicit expression for $\left \langle \mu,(X,\eta,0) \right\rangle$, relevant to our discussion, will emerge below as we proceed.
 
The moment map $\mu: \overline{\mathcal{A}}\rightarrow \mathfrak{h}^{\ast}$ associated to the Hamiltonian action $\mathcal{H}\circlearrowright \overline{\mathcal{A}}$ of the symmetry group $\mathcal{H}$ on the quotient space $\overline{\mathcal{A}}$ is found by computing the contraction
\begin{equation}
-\iota_{V(X,\eta,a )}\hat{\Omega }=\hat{\delta }\big\langle \mu ,(X,\eta,a) \big\rangle, \label{def of moment}
\end{equation}%
where\footnote{The relative signs on the rhs of \eqref{vector on A}, have been chosen in order to make \eqref{total bracket} consistent with the calculations derived from the definitions of the moment map \eqref{def of moment} and the Poisson algebra \eqref{Poisson algebra} below. } 
\begin{equation}
\delta \mathbb{A}= V(X,\eta,a)=d_{\mathbb{A}}\eta-\pounds_{X}\mathbb{A} \label{vector on A}
\end{equation}
is the vector field on $\mathcal{A}$ induced\footnote{The $U(1)$ action on M generated by the vector field $R$, can be lifted to the space of connections $\mathcal{A}$ only when the $G$-bundle $P$ is trivial, see \cite{NA loc CS} and references therein. In addition, we also have to lift the action of a second $U(1)$ factor generated by the vector field $\partial_{\tau}$ along the global time direction $\mathbb{R}$, which is already topologically trivial and posses no problem. Furthermore, any $G$-bundle $P$ over a 3d manifold M is trivial if $G$ is simply-connected and, in our case, this is also true for the 4d manifold $\mathbb{M}=\mathbb{R}\times \text{M}$ as the global $\mathbb{R}$ factor and a point are of the same homotopy type. This is why we have restricted the 4d CS theory construction to complex Lie groups $G$ that are simply-connected. } by the infinitesimal action of $\mathcal{H}$. Notice that $a$ is assumed to act trivially. The first term on the rhs right above comes from a first order expansion of the action of the formal gauge group $\mathcal{G}$ on the elements of $\mathcal{A}$, given by 
\begin{equation}
^{g}\mathbb{A}=g^{-1}\mathbb{A}g+g^{-1}d_{\mathbb{M}}g, \label{gauge transf}
\end{equation}%
where $g=\exp \eta$ with $\eta\in \Omega_{\mathbb{M}}^{0}\otimes \mathfrak{g}$ and $d_{\mathbb{A}}=d_{\mathbb{M}}+\left[ \mathbb{A}, \cdot \; \right] $. Two immediate consequences of \eqref{def of moment} are that $\hat{\Omega}$ and $\mu$ are invariant under the action of $\mathfrak{h}$
\begin{equation}
\pounds_{V(\mathbf{\phi})} \hat{\Omega}=0, \text{ \qquad \qquad }\pounds_{V(\mathbf{\phi)}} \mu=0, \label{Inv}
\end{equation}
where $\pounds_{V(\mathbf{\phi})}=\hat{\delta}\circ \iota_{V(\mathbf{\phi})}+\iota_{V(\mathbf{\phi})}\circ \hat{\delta}$ is the Lie derivative along $V(\mathbf{\phi})$ and $\mathbf{\phi}=(X,\eta,a)$. After some algebra \cite{Yo}, we find that
\begin{equation}
\big\langle \mu ,\left(X,\eta ,a\right) \big\rangle =-\frac{1}{2}%
\dint\nolimits_{\mathbb{M}}\Omega \wedge \kappa \wedge \text{Tr}\left( 
\pounds _{X}\mathbb{A}\wedge \mathbb{A}\right) -\dint\nolimits_{%
\mathbb{M}}\Omega \wedge \kappa \wedge \text{Tr}\left( \eta F_{\mathbb{A}%
}\right) -\dint\nolimits_{\mathbb{M}}d_{\mathbb{M}}\left( \Omega \wedge
\kappa \right) \wedge \text{Tr}\left( \eta \mathbb{A}\right) +a, \label{total moment}
\end{equation}%
where $F_{\mathbb{A}}=d_{\mathbb{M}}\mathbb{A}+\mathbb{A}\wedge \mathbb{A}\in \Omega_{\mathbb{M}}^{2}\otimes \mathfrak{g}$ is the curvature of the connection $\mathbb{A}$. There is also a constant base-point connection with respect to $\hat{\delta}$ that appears in the third term on the rhs above, i.e. $\mathbb{A}\rightarrow \mathbb{A}-\mathbb{A}_{0}$, but it must be put to zero in order for the action $\mathcal{H}\circlearrowright \overline{\mathcal{A}}$ to be Hamiltonian \cite{NA loc CS}. The expression  \eqref{total moment} is invariant under the action of the shift group $\mathcal{S}$ and hence descends to $\overline{\mathcal{A}}$. 

The map $\mu: \overline{\mathcal{A}}\rightarrow \mathfrak{h}^{\ast}$ also obeys
\begin{equation}
\left \langle \mu(^{\mathfrak{h}} \mathbb{A})-\mu(\mathbb{A}),(Y,\lambda,b)\right \rangle=\Big \langle \mu(\mathbb{A}), ad_{(X,\eta,a)}(Y,\lambda,b)  \Big \rangle, 
\end{equation} 
where we have emphasized that $\mu$ depends on $\mathbb{A}$ and used the first order expression
\begin{equation}
^{\mathfrak{h}} \mathbb{A}=\mathbb{A}+d_{\mathbb{A}} \eta -\pounds_{X}\mathbb{A} 
\end{equation}
to perform the calculation. Recall that
\begin{equation}
ad_{(X,\eta,a)}(Y,\lambda,b)=\Big[ \left( X,\eta ,a\right) ,\left( Y,\lambda ,b\right) \Big]. 
\end{equation} 
Then, if the moment map $\mu$ is equivariant, in the sense that 
\begin{equation}
\mu(^{\mathfrak{h}} \mathbb{A})-\mu(\mathbb{A})=-ad^{*}_{(X,\eta,a)}\mu\left(\mathbb{A}\right),\label{equiv}
\end{equation}
where the rhs defines the co-adjoint action of $\mathfrak{h}$ on $\mathfrak{h}^{\ast}$, we obtain the expected result
\begin{equation}
\left \langle ad^{*}_{(X,\eta,a)}\mu,(Y,\lambda,b)\right \rangle=-\Big \langle \mu, ad_{(X,\eta,a)}(Y,\lambda,b)  \Big \rangle, \label{required}
\end{equation}
turning $\mathfrak{h}$-invariant the pairing between $\mathfrak{h}$ and $\mathfrak{h}^{*}$. We can, alternatively, start from the defining relation \eqref{required}, in order to conclude that \eqref{equiv} holds.

The Poisson bracket of two functionals on $\overline{\mathcal{A}}$ of the form \eqref{total moment} is computed from the basic relation
\begin{equation}
\Big\{\big\langle \mu ,(X,\eta,a) \big\rangle ,\big\langle \mu ,(Y,\lambda,b)
\big\rangle \Big\} =\hat{\delta }\big\langle \mu ,(Y,\lambda,b)
\big\rangle \big( V(X,\eta,a )\big). \label{Poisson algebra}
\end{equation}%
After some algebra \cite{Yo}, we find that
\begin{equation}
\Big\{ \big\langle \mu ,\left( X,\eta ,a\right) \big\rangle,\big\langle \mu ,\left(
Y,\lambda ,b\right) \big\rangle  \Big\} =\Big\langle \mu ,
\Big[ \left( X,\eta ,a\right) ,\left( Y,\lambda ,b\right) \Big]
\Big\rangle ,
\end{equation}%
where we have to use $[X,Y]=0$ at the end of the calculation, for the reason already explained before when introducing the bracket \eqref{total bracket}.

Because of the presence of two vector fields $\mathcal{R},\mathcal{R}'$ instead of a single one, namely $R$, when compared with \cite{NA loc CS}, introducing a hyperbolic, symmetric, ad-invariant and non-degenerate bilinear form $(\ast,\ast):\frak{h} \times \frak{h} \rightarrow \mathbb{C} $ on the Lie algebra $\frak{h}$, generalizing the one defined in $\S 4.9$ of \cite{Loop} is a delicate process and care must be taken at this crucial stage. Such an inner product is necessary in order to dualize the moment map, i.e. to find $\mu: \overline{\mathcal{A}}\rightarrow \mathfrak{h}$, and from it to construct the quadratic expression $(\mu,\mu)$, hence defining the would-be contact 4d CS theory. 

The following analysis concerning the definition of this important inner product was not covered by \cite{Yo} in full and is completely new. 

\subsubsection{Inner product}\label{5.1.2}

In order to find a well-defined inner product on the symmetry algebra acting on $\overline{\mathcal{A}}$ in a Hamiltonian way, we pursue the following strategy: first propose a symmetric bilinear form on $\mathfrak{h}$ and then look for ad-invariance conditions while preserving its non-degeneracy. The outcome is that this can be done only on a subalgebra $\overline{\mathfrak{h}} \subset \mathfrak{h}$ and as a consequence, $\mu$ is dualizable only when we consider the restricted map $\mu:\overline{\mathcal{A}}\rightarrow \overline{\mathfrak{h}}^{\ast}$. 

Define the inner product $(\cdot,\cdot):\frak{h} \times \frak{h} \rightarrow \mathbb{C} $ on $\mathfrak{h}$, by the formula 
\begin{equation}
\Big( \left( X,\eta ,a\right) ,\left( Y,\lambda ,b\right) \Big)
:=-\dint\nolimits_{\mathbb{M}}\gamma_{\text{top}}  \text{Tr}\left( \eta \lambda \right) -\beta(X)b-\beta(Y)a, \label{inner product}
\end{equation}%
where $\beta \in \Omega_{\mathbb{M}}^{1}$ is a 1-form on $\mathbb{M}$ constructed out of a linear combination of $\Omega$ and $\kappa$, i.e. $\beta=m\Omega + n \kappa$, $m,n\in \mathbb{R}$ with the important condition that $\beta(X)\neq 0$ and $\beta(Y)\neq 0$ for generic $X,Y$. From the reality conditions $\S \eqref{3}$, it is no difficult to show that if $\eta$ and $\lambda$ are equivariant Lie algebra valued functions and that if $a,b\in i\mathbb{R}$, then \eqref{inner product} is a purely imaginary number. In particular, the norm 
\begin{equation}
i\Big( \left( X,\eta ,a\right) ,\left( X,\eta ,a\right) \Big)
:=-i\dint\nolimits_{\mathbb{M}}\gamma_{\text{top}}  \text{Tr}\left( \eta \lambda \right)- 2i\beta(X)a   \label{facto 1}
\end{equation}
is real valued.

The first term on the rhs in \eqref{inner product} defines a metric-independent inner product on $\mathcal{G}_{\text{Lie}}$ and the other contributions are responsible for its hyperbolic nature. The bilinear form \eqref{inner product} is ad-invariant under the action of $\mathfrak{h}$, if and only if, we have that
\begin{equation}
\Big( \Big[ \left( X,\eta ,a\right) ,\left( Y,\lambda ,b\right) \Big]
,\left( Z,\phi ,c\right) \Big) =\Big( \left( X,\eta ,a\right) ,\Big[
\left( Y,\lambda ,b\right) ,\left( Z,\phi ,c\right) \Big] \Big) . \label{inv}
\end{equation}
The latter condition being equivalent to
\begin{equation}
d(\eta,\lambda,Z)=d(\lambda, \phi,X),\label{inv 2}
\end{equation}
where
\begin{equation}
d(\eta,\lambda, Z)=\beta(Z)c(\eta, \lambda)+\dint\nolimits_{\mathbb{M}}\gamma_{\text{top}}  \text{Tr}\left( \eta \pounds_{Z} \lambda  \right). \label{inv condition}
\end{equation}
We require \eqref{inv 2} to vanish independently on both sides and interpret the vanishing of \eqref{inv condition} as an equation imposing conditions upon the admissible vector fields and gauge parameters under which \eqref{inv} holds. In order to extract a more tangible expression, we write the co-cycle \eqref{co-cycle} in the form
\begin{equation}
c(\eta,\lambda)=-\dint\nolimits_{\mathbb{M}}\gamma_{\text{top}}\, \text{Tr}\left( \eta \pounds_{\mathcal{R}} \lambda  \right)-\dint\nolimits_{\mathbb{M}}d_{\mathbb{M}%
}\Omega \wedge \Omega \wedge \kappa \,  \text{Tr}\left( \eta \pounds_{\mathcal{R}'} \lambda  \right)
\end{equation}
and set $Z=r \mathcal{R}+ r' \mathcal{R}'$. Then, \eqref{inv condition} boils down to
\begin{equation}
\Big(\beta(Z)-r\Big)\dint\nolimits_{\mathbb{M}}\gamma_{\text{top}}  \text{Tr}\left( \eta \pounds_{\mathcal{R}} \lambda  \right)+\dint\nolimits_{\mathbb{M}}\Omega \wedge \kappa\wedge \Big(\beta(Z)d_{\mathbb{M}%
}\Omega  -r'd_{\mathbb{M}%
}\kappa    \Big)   \text{Tr}\left( \eta \pounds_{\mathcal{R}'} \lambda  \right)=0. \label{d condition}
\end{equation}

We do not intend to classify all possible solutions\footnote{This is equivalent to the classification of all subalgebras $\overline{\mathfrak{h}} \subset \mathfrak{h}$ solving \eqref{d condition}. They would produce inequivalent dualized moment maps, e.g. see \eqref{dual total moment}, as well as quadratic action functionals, e.g. see \eqref{the result}. It would be interesting to study what are the model building implications of this, at the level of their associated 2d IFT's. The author thanks the referee for raising this point.} to the equation \eqref{d condition} in this work. Instead, we rather consider the structure of three examples of solutions from which we identify the correct symmetry algebra, i.e. $\overline{\mathfrak{h}} \subset \mathfrak{h}$, necessary to construct a consistent quadratic contact 4d CS theory, generalizing the quadratic action originally introduced in \cite{NA loc CS}, via dualization of the moment map. Thus: 
\medskip

$\bullet \,$ \textbf{Solution (i)} $m=n=1$, or $\beta= \Omega+\kappa$. Symmetric under the exchange $\Omega \leftrightarrow  \kappa$.
\medskip

This solution is the only one originally considered in \cite{Yo}. We have that $\beta(Z)=r+r'\neq 0$ and
\begin{equation}
\beta(Z)d_{\mathbb{M}}\Omega  -r'd_{\mathbb{M}}\kappa=(r+r')d_{\text{M}}\omega-\left( (r+r')\frac{\zeta}{\zeta'}+r'\right)d_{\mathbb{M}}\kappa.
\end{equation}
Because of $\gamma_{\text{top}}  \sim d\tau \wedge \alpha \wedge d_{\text{M}} \alpha \neq 0$ and $\pounds_{\mathcal{R}}\lambda \neq 0$, $r\neq 0$ by assumption, we are forced to take $r'=0$. Hence,
\begin{equation}
d(\eta,\lambda, Z)=r\dint\nolimits_{\mathbb{M}}\Omega \wedge \kappa\wedge \left(d_{\text{M}}\omega  -\frac{\zeta}{\zeta'} d_{\mathbb{M}}\kappa \right)   \text{Tr}\left( \eta \pounds_{\mathcal{R}'} \lambda  \right)=0.
\end{equation}
For generic $\zeta$, the latter is satisfied if we require that $\pounds_{\mathcal{R}'} \eta=0$ and this condition can be solved by taking $d_{\mathbb{M}}\eta = \kappa \pounds_{\mathcal{R}}\eta + \gamma(\eta)$, where $\gamma(\eta) \in \Omega_{\mathbb{M}}^{1}\otimes \mathfrak{g}\cap \text{ker}_{\iota_{\mathcal{R}'}}$ is arbitrary. For this solution, we have that $Z=r\mathcal{R}$ and $\overline{\mathcal{H}}=U(1)_{\mathcal{R}}\ltimes \tilde{\mathcal{G}}$, with $\overline{\mathfrak{h}}=\mathbb{R}\oplus \tilde{\mathcal{G}}_{\text{Lie}}$. Notice that despite of the fact that $\Omega$ and $\kappa$ enter symmetrically in $\beta$, this is not the best choice for $\beta$, as it implies some gauge symmetry transformation issues, as mentioned in item I) in the introduction, leading to the lack of gauge covariance of some expressions \cite{Yo}. We will not consider this particular solution anymore because it was already explored in \cite{Yo}. 
\medskip

$\bullet \,$ \textbf{Solution (ii)} $m=1$, $n=0$, or $\beta=\Omega$. Not symmetric under the exchange $\Omega \leftrightarrow  \kappa$.
\medskip

We have that $\beta(Z)=r'\neq 0$ and
\begin{equation}
\beta(Z)d_{\mathbb{M}}\Omega  -r'd_{\mathbb{M}}\kappa=r'd_{\text{M}}\omega- r'\left(\frac{\zeta}{\zeta'}+1\right)d_{\mathbb{M}}\kappa.
\end{equation}
Because of $\gamma_{\text{top}}  \sim d\tau \wedge \alpha \wedge d_{\text{M}} \alpha \neq 0$, we are forced to take $r=r'$. Hence,
\begin{equation}
d(\eta,\lambda, Z)=r\dint\nolimits_{\mathbb{M}}\Omega \wedge \kappa\wedge \left[d_{\text{M}}\omega- r'\left(\frac{\zeta}{\zeta'}+1\right)d_{\mathbb{M}}\kappa    \right]   \text{Tr}\left( \eta \pounds_{\mathcal{R}'} \lambda  \right)=0. \label{sol ii}
\end{equation}
For generic $\zeta$ and gauge parameters that are not annihilated by the action of $\pounds_{\mathcal{R}'}$, the expression right above can be put to zero if we require the vanishing of the boundary term
\begin{equation}
\dint\nolimits_{\mathbb{M}}d\tau \wedge \alpha \wedge d_{\text{M}}\omega\, \text{Tr}\left( \eta \pounds_{\mathcal{R}'} \lambda  \right)=0 \label{def of G_0}
\end{equation}
and take $\zeta'=-\zeta$. In reaching \eqref{def of G_0}, we have replaced $\Omega \rightarrow \omega_{\zeta}$ in \eqref{sol ii}. By the integration formula \eqref{formula BC}, it is then natural to expect that the gauge parameters must obey certain restrictions in order to satisfy the condition \eqref{def of G_0}, which formally defines the Lie algebra $\mathcal{G}_{0\text{Lie}}$ of the gauge group $\mathcal{G}_{0}\subset \mathcal{G}$ introduced before. In this case, the admissible vector fields are of the form $Z=(r/\zeta')R$ and we have that $\overline{\mathcal{H}}=U(1)_{R}\ltimes \tilde{\mathcal{G}}_{0}$, with $\overline{\mathfrak{h}}=\mathbb{R}\oplus \tilde{\mathcal{G}}_{0\text{Lie}}$, where the meaning of the sub-index $0$ is now clear. We will not consider this solution either, as it relates the two deformations parameters and due to the fact that the $t$-scaling symmetry requires that $\zeta'\neq 0$, this would prevent taking the limit $\zeta\rightarrow 0$. For this reason, we will not continue the analysis of this solution in the rest of this work. Yet, \eqref{def of G_0} remains valid.

For future reference we use the condition \eqref{def of G_0} as an inspiration to introduce, up to a multiplicative non-zero constant, a non-degenerate, symmetric and ad-invariant bilinear form $\big\langle\! \! \big\langle \cdot, \cdot \big\rangle \! \! \big\rangle _{\omega}:\mathcal{G}_{\text{Lie}}\times \mathcal{G}_{\text{Lie}} \rightarrow \mathbb{R}$ on the pole-defect surface $\Sigma \times \mathfrak{p}$, defined by the $\omega$-dependent expression
\begin{equation}
\big\langle\! \! \big\langle \eta, \lambda \big\rangle \! \! \big\rangle _{\omega}:=-i\dint\nolimits_{\mathbb{M}}d\tau \wedge \alpha \wedge d_{\text{M}}\omega\, \text{Tr}\left( \eta  \lambda  \right).\label{gen gauge cond}
\end{equation}
Thus, the gauge algebra $\mathcal{G}_{0\text{Lie}}$ is now defined more elegantly by
\begin{equation}
\mathcal{G}_{0\text{Lie}}:=\bigl\{\eta,\lambda \in \mathcal{G}_{\text{Lie}}\, \big| \,\big\langle\! \! \big\langle \eta, \lambda \big\rangle \! \! \big\rangle _{\omega}=0 \bigr\}. \label{def of G_0''}
\end{equation}
Notice that the definition \eqref{gen gauge cond} does not include the Lie derivative $\pounds_{\mathcal{R}'}$ this time and the reason for this is because we are interested in looking for Lie-algebraic conditions to be imposed upon the gauge elements in $\mathcal{G}_{\text{Lie}}$, rather than in finding analytical conditions\footnote{It would be interesting to explore this type of solutions and their implications on the associated 2d IFT's.}, in the sense of imposing conditions of the form $\pounds_{\mathcal{R}'}\eta =0$, as in solution i). 

As a particular example of the bilinear form \eqref{gen gauge cond} we have that locally on $\Sigma \times \mathcal{U}$, when $C=\mathbb{CP}^{1}$ and $\mathcal{U}\subset \mathbb{CP}^{1}$, it becomes 
\begin{equation}
 \big\langle\! \! \big\langle \eta, \lambda \big\rangle \! \! \big\rangle _{\omega}=\frac{1}{2\pi i}\dint\nolimits_{\Sigma}d\tau \wedge d\sigma \left(\dint_{\mathcal{U}}d_{C}\omega_{C}\,\text{Tr}(\eta \lambda)\right)=\dint\nolimits_{\Sigma}d\tau \wedge d\sigma\left(\sum_{r=1}^{M}\sum_{p=0}^{m_{r}-1}\frac{k_{p}^{(r)}}{p!}\partial^{p}_{z}\,\text{Tr}(\eta \lambda) \big|_{z=z_{r}}\right),\label{def of G_0'}
\end{equation}
where we have used $\alpha \big|_{S^{1}}=\alpha_{\sigma}d\sigma $, with $\alpha_{\sigma}=\frac{1}{2 \pi}$ and the integration formula \eqref{formula BC} in order to reach the final form. Remarkably, the integrand in \eqref{def of G_0'} nicely relates to the bilinear forms \eqref{Inner R}, \eqref{Inner C1} and \eqref{Inner double}, that are behind the gauge field $\mathbb{A}$ boundary eom, when it is specialized to the pole structure of $\omega$ considered in $\S $\eqref{4.2} on a case by case basis.  
\medskip

$\bullet \,$ \textbf{Solution (iii)} $m=0$, $n=1$, or $\beta=\kappa$. Not symmetric under the exchange $\Omega \leftrightarrow  \kappa$.
\medskip

We have that $\beta(Z)=r\neq 0$ and
\begin{equation}
\beta(Z)d_{\mathbb{M}}\Omega  -r'd_{\mathbb{M}}\kappa=rd_{\text{M}}\omega- \left(r\frac{\zeta}{\zeta'}+r'\right)d_{\mathbb{M}}\kappa. \label{iii}
\end{equation}
By the same token, the invariance condition can be solved if we require that $\eta \in \mathcal{G}_{0 \text{Lie}}$ as defined in \eqref{def of G_0''} and set $r'=-(\zeta/\zeta')r$, for generic values of the parameters $\zeta, \zeta'$. In this case, the admissible vector fields are of the form $Z=(r/\zeta')R$ and we have that $\overline{\mathcal{H}}=U(1)_{R}\ltimes \tilde{\mathcal{G}}_{0}$, with $\overline{\mathfrak{h}}=\mathbb{R}\oplus \tilde{\mathcal{G}}_{0\text{Lie}}$. This solution provides a slight generalization of the inner product defined in \cite{Loop} (\S 4.9) and used in \cite{NA loc CS}, which is roughly obtained by taking $\Omega \rightarrow 1$, $\mathbb{M}\rightarrow \text{M}$ and $\kappa =\zeta' \alpha$ on the rhs of \eqref{inner product}, after putting the 1-form $\Omega$ to its leftmost position. This is the solution we will consider in the following in order to proceed with the rest of the construction. 

Solutions ii) and iii) only require the gauge parameters to be in $\mathcal{G}_{0 \text{Lie}}$ allowing to posses an arbitrary dependence on the fiber coordinates $(\tau,\sigma)$, a forbidden property in solution i). In all three cases, only one $U(1)$ subgroup of the two $U(1)^{\times 2}$ actions generated by $\mathcal{R},\mathcal{R}'$ is admitted if one is to have a well-defined invariant and non-degenerate inner product on $\overline{\mathfrak{h}}\subset \mathfrak{h}$. For solutions ii) and iii), the surviving $U(1)$ action is given precisely by the rigid translations along the $S^{1}$ fibers of M generated by the Reeb vector field $R$. In what follows, we will focus on solution iii) exclusively, as it does not impose any further constraint on the parameters $\zeta$ and $\zeta'$. 

Now, we proceed to use the new inner product \eqref{inner product} with $\beta=\kappa$ in order to dualize the moment map $\mu: \overline{\mathcal{A}}\rightarrow \overline{\mathfrak{h}}^{*}$, i.e. we want to get a map $\mu:\overline{\mathcal{A}}\rightarrow \overline{\mathfrak{h}}$, from the quotient space $\overline{\mathcal{A}}$ to the algebra $\overline{\mathfrak{h}}$, instead of its dual $\overline{\mathfrak{h}}^{\ast}$. This is easily done by solving
\begin{equation}
\big\langle \mu ,\left( Y,\lambda ,b\right) \big\rangle =\big( \mu
,\left( Y,\lambda ,b\right) \big) , \label{solving}
\end{equation}%
where
\begin{equation}
\big\langle \mu ,\left( Y,\lambda ,b\right) \big\rangle=-\frac{1}{2}%
\dint\nolimits_{\mathbb{M}}\omega_{\zeta} \wedge \kappa \wedge \text{Tr}\left( 
\pounds _{Y}\mathbb{A}\wedge \mathbb{A}\right) -\dint\nolimits_{%
\mathbb{M}}\omega_{\zeta} \wedge \kappa \wedge \text{Tr}\left( \lambda F_{\mathbb{A}%
}\right) +\dint\nolimits_{\mathbb{M}}\omega_{\zeta}\wedge d_{\mathbb{M}}
\kappa  \wedge \text{Tr}\left( \lambda \mathbb{A}\right) +b, \label{paired moment}
\end{equation}
for $(Y,\lambda, b)\in \overline{\mathfrak{h}}$ and some element $\mu =\left( X',\eta' ,a'\right)\in \overline{\mathfrak{h}} $ on the rhs, where $X'=(p/\zeta')R$ and $Y=(q/\zeta')R$. Because of $\iota_{X}\Omega=-(\zeta/\zeta')\neq 0$, we have exploited the invariance of \eqref{total moment} under the manifest symmetry $\Omega \rightarrow \Omega+f \kappa$, for some non-vanishing $f\in \Omega_{\mathbb{M}}^{0}$, in order to project $\Omega$ along $\text{ker}_{\iota_{X}}$, by replacing 
\begin{equation}
\Omega \longrightarrow \omega_{\zeta}= (1-\kappa\, \iota_{X})\Omega. \label{key shift}
\end{equation} 
Clearly, because of $\iota_{X}\kappa=1$ it obeys $\iota_{X}\omega_{\zeta}=0$ and from this we get that
\begin{equation}
\omega_{\zeta}=\omega+2\zeta\alpha_{\tau} d\tau. \label{nice}
\end{equation}
The latter explains the origin of the expression for the 1-form $\omega_{\zeta}$ introduced above in $\S$\eqref{2} and requested to satisfy the relations \eqref{Omega X conditions} and \eqref{zeta=0 limit}, recovering $\omega$ in the $\zeta \rightarrow 0$ limit. Furthermore, in \eqref{paired moment} (Cf. \eqref{total moment}), we have dropped the boundary term
\begin{equation}
\dint\nolimits_{\mathbb{M}}d_{\mathbb{M}} \omega_{\zeta}\wedge 
\kappa  \wedge \text{Tr}\left( \lambda \mathbb{A}\right)=0, \label{neglected}
\end{equation}
as $\lambda$ is now required to be an element of the gauge algebra $\mathcal{G}_{0\text{Lie}}$. We will come back to this condition, and also to the expression \eqref{def of G_0} later on.

Continuing with the dualization process and following \cite{Yo}, we find that $p=-1$ and write the dual moment solution in the form
\begin{equation}
\mu =(X',\eta',a')=\left( -X,\frac{\omega_{\zeta} \wedge \kappa \wedge F_{%
\mathbb{A}}-\omega_{\zeta}\wedge d_{\mathbb{M}} \kappa  \wedge 
\mathbb{A}}{\gamma_{\text{top}} },\frac{1}{2}%
\dint\nolimits_{\mathbb{M}}\omega_{\zeta} \wedge \kappa \wedge \text{Tr}\left( 
\pounds _{X}\mathbb{A}\wedge \mathbb{A}\right) \right) , \label{dual total moment}
\end{equation}
where now we have that $X=(1/\zeta')R$. Some properties of \eqref{dual total moment} are that $\eta'$ is a Lie algebra valued equivariant function, $a'$ is purely imaginary and\footnote{This is because $\omega_{\zeta}$-shifts are combinations of $\Omega$ and $\kappa$ shifts, as $\omega_{\zeta}=\Omega+\frac{\zeta}{\zeta'}\kappa$.} $\mu(^{\omega_{\zeta}}\mathbb{A})=\mu(\mathbb{A})$. Note in passing that the co-cycle \eqref{co-cycle} now reduces to the expression
\begin{equation}
c(\eta,\lambda)=-\dint\nolimits_{\mathbb{M}}\gamma_{\text{top}} \text{Tr}\left( 
\eta \pounds _{X}\lambda \right)=2\alpha_{\tau}\zeta \zeta' \dint\nolimits_{\mathbb{R}} d\tau \left(-\dint\nolimits_{\text{M}}\alpha \wedge d_{\text{M}}\alpha \, \text{Tr}(\eta\pounds_{R}\lambda)    \right).
\end{equation}
The quantity in parenthesis is precisely the co-cycle first discovered in \cite{NA loc CS} in the 3d CS theory case.

Similar to the results derived from solution i) and considered in \cite{Yo}, the dualized moment map can be written as well, in the compact form
\begin{equation}
\mu=(-X,\Phi_{\text{on}}-\iota_{X}\mathbb{A}, a').
\end{equation}
Right above, we have that $\Phi_{\text{on}}$ is as given by \eqref{Phi eom} and that
\begin{equation}
\iota_{X}\mathbb{A}=\frac{\omega_{\zeta} \wedge d_{\mathbb{M}}\kappa \wedge 
\mathbb{A}}{\gamma_{\text{top}} }. \label{Phi and con}
\end{equation}%
Furthermore, we have that \cite{Yo}
\begin{equation}
\begin{aligned}
2a'&=\dint\nolimits_{\mathbb{M}}\omega_{\zeta} \wedge \kappa \wedge \text{Tr}\left( 
\pounds _{X}\mathbb{A}\wedge \mathbb{A}\right) \\
&=\dint\nolimits_{\mathbb{M}}\omega_{\zeta} \wedge CS\left( \mathbb{A}\right)
-\dint\nolimits_{\mathbb{M}}\gamma_{\text{top}} \text{Tr}\Big[ \Big( 2\Phi_{\text{on}}-B(\mathbb{A})-\iota_{X}\mathbb{A}\Big) \iota_{X}\mathbb{A}\Big],
\end{aligned}
\end{equation}
where we have defined the boundary term contribution
\begin{equation}
B(\mathbb{A})=\frac{d_{\mathbb{M}}\omega_{\zeta} \wedge \kappa
\wedge \mathbb{A}}{\gamma_{\text{top}}}.
\end{equation}
The quantities $\Phi_{\text{on}}$, $\iota_{X}\mathbb{A}$ and $B(\mathbb{A})$ all belong to $\Omega_{\mathbb{M}}^{0}\otimes\mathfrak{g}$. They transform differently under gauge transformations, as can be easily verified explicitly. 

\subsubsection{Gauge invariance II} \label{5.1.3}

In $\S \eqref{4}$, we studied solutions for the fields $(\mathbb{A},\Phi)$ boundary eom and showed, in the case of $\mathbb{A}$, their consistency with the usual 4d CS theory boundary eom solutions. Recall that this was possible only after imposing an extra set of conditions to be satisfied by the 1-form $\kappa$. At this stage, we can safely restrict our 4d theories to the space of fields subordinated to those solutions and, in practice, this means that the boundary contributions to the actions \eqref{Omega shift inv} and \eqref{double shift inv} given, respectively, by
\begin{equation}
\dint\nolimits_{\mathbb{M}}d_{\mathbb{M}}\omega_{\zeta}\wedge \kappa\wedge \text{Tr}\left( \mathbb{A}\iota_{X}\mathbb{A}  \right)=0, \text{\qquad \qquad}\dint\nolimits_{\mathbb{M}}d_{\mathbb{M}}\omega_{\zeta}\wedge \kappa\wedge \text{Tr}\left( \mathbb{A}\Phi \right)=0, \label{two term}
\end{equation}
vanish. This fact, in principle, simplifies the proof of the claim considered above in \eqref{general gauge transf}, that $^{g}S=S$. Indeed, all boils down to showing that the regularized action, now taking the more familiar form
\begin{equation}
S(\mathbb{A})_{\text{reg}}=ic\dint\nolimits_{\mathbb{M}}\omega_{\zeta} \wedge CS\left( 
\mathbb{A}\right), \label{reg reg CS}
\end{equation} 
is gauge invariant. Thus, from the identity \eqref{CS gauge}, we end up with the expression
\begin{equation}
S\big(\! \,^{g}\mathbb{A}  \big)_{\text{reg}}=S\left(\mathbb{A}  \right)_{\text{reg}}+ic\dint\nolimits_{\mathbb{M}}d_{\mathbb{M}}\omega_{\zeta}\wedge \text{Tr}\big(\mathbb{A}\wedge\mathbb{J}\big)+ ic\dint\nolimits_{\mathbb{M}}\omega_{\zeta}\wedge \chi(g). \label{reg bc gauge var}
\end{equation}
 
Before starting with the proof, let us consider first two more boundary terms we still need to verify they vanish. The first one is a constraint formally defining the gauge algebra $\mathcal{G}_{0\text{Lie}}$ via equation \eqref{def of G_0} and the second one corresponding to the contribution \eqref{neglected} to the moment map \eqref{paired moment}, that was neglected in the process of dualization of $\mu$.

The gauge group Lie algebra defining condition \eqref{def of G_0} is equivalent, see \eqref{def of G_0'}, to\footnote{In \cite{Homotopical, E models}, the gauge group $\mathcal{G}_{0}$ is elegantly determined via the condition $^{g}S=S$. Here, we combine, in a more pedestrian way, the boundary eom solutions space algebraic structure with the condition \eqref{def of G_0} to do so.  }  
\begin{equation}
\big\langle\! \! \big\langle \eta, \lambda' \big\rangle \! \! \big\rangle _{\omega}=0,\label{key integral}
\end{equation}
where we have written $\lambda'=\pounds_{\mathcal{R}'}\lambda \in \Omega_{\mathbb{M}}^{0}\otimes \mathfrak{g}$. The fact that $\pounds_{\mathcal{R}'}$ is acting on a gauge parameter, does not affect the analytic structure thereof along the $z$ coordinate, as it only involves $\tau$ and $\sigma$ partial derivatives. When restricted to the PCM-type theories as the ones considered in $\S \eqref{4}$, we get that for a real double pole, a pair of real simple poles and a pair of complex conjugate simple poles, \eqref{key integral} is, respectively, equivalent to the following conditions
\begin{equation}
\begin{aligned}
\Big\langle\! \! \Big\langle \Big(\eta\big|_{z=z_{r}},  \partial_{z}\eta\big|_{z=z_{r}}   \Big),\Big( \lambda '\big|_{z=z_{r}},\partial_{z}\lambda '\big|_{z=z_{r}}   \Big) \Big\rangle\! \! \Big\rangle _{\mathfrak{d}_{\text{ab}}}^{(r)}&=0,\\
\Big\langle\! \! \Big\langle \Big(\eta\big|_{z=z_{1}},  \eta\big|_{z=z_{2}}   \Big),\Big( \lambda '\big|_{z=z_{1}},\lambda '\big|_{z=z_{2}}   \Big) \Big\rangle\! \! \Big\rangle _{\mathfrak{d}}&=0,\\
\Big\langle\! \! \Big\langle \eta\big|_{z=z_{1}},\lambda '\big|_{z=z_{1}} \Big\rangle\! \! \Big\rangle _{\mathfrak{g}}&=0,
\end{aligned}
\end{equation}
where we have used the algebraic data introduced at the beginning of $\S \eqref{4.2}$. Thus, the gauge algebra $\mathcal{G}_{0\text{Lie}}$, also satisfying \eqref{def of G_0''} is finally identified, for the PCM-type theories, to be of the following form
\begin{equation}
\mathcal{G}_{0\text{Lie}}:=\bigl\{\eta\in \mathcal{G}_{\text{Lie}}\, \Big| \,\eta \big|_{\Sigma \times\mathfrak{p}}\in \Omega_{\Sigma }^{0}\otimes \underline{\mathfrak{l}} \bigr\}.
\end{equation}
A consequence of this is that the boundary eom space of solutions is formed by the $\mathcal{G}_{0}$-orbits of the fields $(\mathbb{A},\Phi)$ restricted to the pole-defect surface $\Sigma \times \mathfrak{p}$. Also notice that \eqref{BC Phi off} and \eqref{kappa gauge BC} are preserved by the action of $\mathcal{G}_{0}$. In a similar way, the condition \eqref{neglected}, reduces to
\begin{equation}
\dint_{\Sigma}d\tau \wedge d\sigma \left( \dint_{C}d_{C}\omega_{C}\epsilon^{\mu \nu}\kappa_{\mu}\text{Tr}\big( A_{\nu}\eta   \big)\right)=0 \label{2 integral}
\end{equation}
and vanishes as well. To show this, we notice that the second term on the rhs of \eqref{Again} and \eqref{2 integral} match if we map $\delta \Phi \rightarrow \eta$, as the restrictions of $\Phi$ and $\eta$ to $\Sigma \times \mathfrak{p}$ are valued on the same space the result follows. As a by product, the second term on the rhs of \eqref{reg bc gauge var} also vanishes after writing it in the form
\begin{equation}
\dint\nolimits_{\mathbb{M}}d_{\mathbb{M}}\omega_{\zeta}\wedge \text{Tr}\big(\mathbb{A}\wedge\mathbb{J}\big)=\dint\nolimits_{\mathbb{M}}d_{\mathbb{M}}\omega_{\zeta}\wedge \kappa \wedge \text{Tr}\big(\mathbb{J}\iota_{R}\mathbb{A}-\mathbb{A}\iota_{R}\mathbb{J}\big),
\end{equation}
consequence of the trivial 5-form contraction
\begin{equation}
0=\iota_{R}\Big( d_{\mathbb{M}}\omega_{\zeta}\wedge \kappa \wedge \text{Tr}\big(\mathbb{A}\wedge\mathbb{J}\big)  \Big).
\end{equation}
Using the fact that the restrictions of $(\mathbb{J},\mathbb{A})$ and $(\iota_{R}\mathbb{J},\iota_{R}\mathbb{A})$ to the pole-defect surface $\Sigma \times \mathfrak{p}$ belong, respectively, to $\Omega_{\Sigma}^{1}\otimes \underline{\mathfrak{l}}$ and $\Omega_{\Sigma}^{0}\otimes \underline{\mathfrak{l}}$, we get the result.

The proof that $^{g}S=S$, then reduces to show that the last contribution in \eqref{reg bc gauge var} drops out. More explicitly, we write
\begin{equation}
\dint\nolimits_{\mathbb{M}}\omega_{\zeta}\wedge \chi(g)=\dint\nolimits_{\mathbb{M}}\omega\wedge \chi(g)+2\zeta\alpha_{\tau}\dint\nolimits_{\mathbb{M}}d\tau \wedge \chi(g). \label{WZ term}
\end{equation}
The last term on the rhs right above vanishes because of the WZ 3-form $\chi(g)$ is closed and the 4d manifold $\mathbb{M}$ is such that $\partial \mathbb{M}=0$, i.e. we write $d \tau \wedge \chi(g)=d_{\mathbb{M}}\big( \tau \chi(g)  \big)$ and use Stokes theorem. The first term on the rhs in \eqref{WZ term} is known to vanish for $\mathbb{M}=\mathbb{R}^{2}\times \mathbb{CP}^{1}$ by the arguments\footnote{I thank B. Vicedo for patiently commenting on this.} considered in \cite{Homotopical}, based on a Homotopical analysis of the 4d CS theory. However, in the more general situation $\mathbb{M}=\mathbb{R} \times \text{M}$, with M as defined in \eqref{n bundle}, the vanishing of that term is still unclear at the time and remains as an important open problem. 

In summary, up to the remaining WZ term contribution in \eqref{WZ term}, all actions functionals considered so far are invariant under the action of the gauge group $\mathcal{G}_{0}$, when supplemented with an equivariant 1-form $\kappa$  having fiber (vertical) components that are real constants \eqref{result equi}. In what follows, we will not consider boundary contribution to the action functionals anymore. They can now be safely discarded.

\subsection{Defect data }

This time, we introduce the 1d CS theory defect symplectic data associated to the defect contribution \eqref{1d CS}. We follow \cite{Wilson NA loc} closely on this matter.

\subsubsection{Adding coadjoint orbits}

Consider the following closed, invariant and non-degenerate symplectic form on $L\mathcal{O}:=L\mathcal{O}_{\lambda_{1}}\times ... \times L\mathcal{O}_{\lambda_{N}}$, associated to the (co)-adjoint orbit defects inserted at the points $z_{j}\in C$, $j=1,...N$, given by
\begin{equation}
\hat{\Omega}'=\sum_{j=1}^{N}\hat{\Omega}'_{j},\label{orbit symp form}
\end{equation}
where 
\begin{equation}
\hat{\Omega}'_{j}:=il \oint\nolimits_{S^{1}}\kappa \wedge \text{Tr}\left(\lambda_{j}f^{-1}\hat{\delta}f \wedge f^{-1}\hat{\delta}f   \right). \label{orbit symp form'}
\end{equation}
The symbol $\hat{\delta}$ denotes now the exterior differential on $LG$. 

The action of $\overline{\mathfrak{h}}$ on any $L\mathcal{O}_{\lambda_{j}}$, induces the vector field
\begin{equation}
\delta f =V'(X,\eta,a)=-\eta|_{S^{1}}\cdot f -\pounds_{X}f, \label{vector field on LO}
\end{equation}
where we have assumed that the central element $a$ acts trivially and where $X=(p/\zeta')R$, as discussed in the paragraph below \eqref{iii}. The first term on the rhs of \eqref{vector field on LO} comes from an infinitesimal expansion of the gauge transformation for $f$ given in \eqref{gauge defects}, while the second term reflects the action of the generator of rigid translation along $S^{1}$. In what follows we will suppress the $|_{S^{1}}$ restriction in the gauge parameter $\eta$ in order to avoid clutter.

The moment map $\mu_{j}' : \mathcal{O}_{\lambda_{j}}\rightarrow \overline{\mathfrak{h}}$ associated to the action of this vector field is given by the contraction
\begin{equation}
-\iota_{V'(X,\eta,a )}\hat{\Omega }_{j}'=\hat{\delta }\big\langle \mu'_{j} ,(X,\eta,a) \big\rangle. \label{def moment'}
\end{equation}%
We find that
\begin{equation}
\big\langle \mu'_{j} ,(X,\eta,a) \big\rangle=il \oint\nolimits_{S^{1}}\kappa \wedge \text{Tr} \Big( \lambda_{j} \left( f^{-1}\eta f+f^{-1}\pounds_{X}f   \right)  \Big),
\end{equation}
directly implying the invariance relations, Cf. \eqref{Inv},
\begin{equation}
\pounds_{V'(\mathbf{\phi})} \hat{\Omega}'_{j}=0, \text{ \qquad \qquad }\pounds_{V'(\mathbf{\phi)}} \mu'_{j}=0. \label{Inv'}
\end{equation}

The Poisson algebra is given by
\begin{equation}
\Big\{ \big\langle \mu'_{j} ,\left( X,\eta ,a\right) \big\rangle,\big\langle \mu'_{k} ,\left(
Y,\lambda ,b\right) \big\rangle  \Big\} =\delta_{jk}\Big\langle \mu'_{j} ,
\Big(0,\,%
\left[ \eta ,\lambda \right] +\pounds _{X}\lambda -\pounds _{%
Y}\eta ,\,0\Big) 
\Big\rangle ,
\end{equation}
where we have used the fact that $[X,Y]=0$. Also, any possible constant added to $\mu'_{j}$ is required to be zero in order for the action to be Hamiltonian. Furthermore, notice that the central term $c(\eta,\lambda)$ is absent and this means that effectively, the symmetry algebra $\overline{\mathfrak{h}}$ acting on the $N$ defects insertions changes from $\mathbb{R}\oplus \tilde{\mathcal{G}}_{0\text{Lie}}$ to $\mathbb{R}\oplus \mathcal{G}_{0\text{Lie}} $, as the central term is completely irrelevant and $\langle \mu'_{j},(0,0,a)   \rangle=0$. Actually, from \eqref{gauge defects} we notice that, at the level of the defects, the symmetry algebra is slightly bigger being $\mathbb{R}\oplus \mathcal{G}_{\text{Lie}}$, where the first factor corresponds to the rigid translations along $S^{1}$ and the second one corresponds to the gauge parameters of the formal gauge group. This was already noticed when we introduced the 1d CS theory defect action in \S \eqref{2.2}.

Similarly as done in \eqref{solving}, we now proceed to dualize the moment map $\mu'_{j}$. For this to be accomplished, we need to solve the following equation
\begin{equation}
\big\langle \mu'_{j},\left( Y,\lambda ,b\right) \big\rangle =\big( \mu'_{j}
,\left( Y,\lambda ,b\right) \big),\text{\qquad \qquad}j=1,...N ,
\end{equation}%
for some element $\mu'_{j} =\big( \tilde{X}_{j}, \tilde{\eta}_{j} ,\tilde{a}_{j}\big)\in \overline{\mathfrak{h}} $ on the rhs and arbitrary $(Y,\lambda, b)$, where $\tilde{X}_{j}=(p_{j}/\zeta')R$ and $Y=(q/\zeta')R$. We find that $p_{j}=0$ and write the dual moment solution in the form
\begin{equation}
\mu'_{j} =(\tilde{X}_{j}, \tilde{\eta}_{j} ,\tilde{a}_{j})=-il\left( 0,\frac{\delta_{S^{1}}\wedge \kappa\, U_{j}}{\gamma_{\text{top}} },
\dint\nolimits_{\mathbb{M}}\delta_{S^{1}}\wedge \kappa\, \text{Tr}\left( 
\lambda_{j} f^{-1}\pounds _{X}f\right) \right) , \label{dual prime}
\end{equation}
where as before, we have set $X=(1/\zeta')R$. The total moment map for the $N$ defects is then given by
\begin{equation}
\mu'=(\tilde{X},\tilde{\eta},\tilde{a})=\sum_{j=1}^{N}\mu'_{j}.\label{dual prime moment}
\end{equation}
For further reference, the third entry on the rhs of \eqref{dual prime} can, alternatively, be written in the form
\begin{equation}
\dint\nolimits_{\mathbb{M}}\delta_{S^{1}}\wedge \kappa\, \text{Tr}\left( 
\lambda_{j} f^{-1}\pounds _{X}f\right)=\oint_{S^{1}}\text{Tr} \left(  \lambda_{j} f^{-1}d_{\mathbb{M}}f \right), \label{atilde}
\end{equation}
where we have used both integration formulas in \eqref{integral formula}. From the time being, notice that \eqref{atilde} and the second relation in \eqref{defect data'}, imply that $\tilde{a}$ in \eqref{dual prime moment} is a purely imaginary number. We further assume that $\delta_{S^{1}}$ possesses, just the right properties, to make $\tilde{\eta}$ in \eqref{dual prime moment}, a Lie algebra valued equivariant function.  

\subsection{Quadratic action with defects}

After dualizing the moment maps $\mu: \overline{\mathcal{A}}\rightarrow \overline{\mathfrak{h}}^{\ast}$, and $\mu': L\mathcal{O}_{\lambda_{1}}\times ... \times L\mathcal{O}_{\lambda_{N}}  \rightarrow \overline{\mathfrak{h}}^{\ast}$,
now we compute the square of $\mathbf{\mu}:=\mu+\mu' \in \overline{\mathfrak{h}}$, given by the sum of \eqref{dual total moment} and \eqref{dual prime moment}. We find, from 
\begin{equation}
\mathbf{\mu}=\big(X',\eta'+\tilde{\eta},a'+\tilde{a}\big)=\big(-X,\tilde{\Phi}_{\text{on}}-\iota_{X}\mathbb{A},a'+\tilde{a}\big)
\end{equation}
and \eqref{integral formula}, that
\begin{equation}
\left( \mathbf{\mu} ,\mathbf{\mu} \right) =-\dint\nolimits_{\mathbb{M}}\gamma_{\text{top}} \text{Tr}\Big( (\eta'+\tilde{\eta})^{2}\Big) -2\kappa(X')(a'+\tilde{a}),
\end{equation}%
equals
\begin{equation}
\left( \mathbf{\mu} ,\mathbf{\mu} \right) =\dint\nolimits_{\mathbb{M}}\omega_{\zeta} \wedge CS\left( 
\mathbb{A}\right)-\dint\nolimits_{\mathbb{M}}\gamma_{\text{top}} \text{Tr}\left( \tilde{\Phi}_{\text{on}} ^{2}%
\right)-2il \sum_{j=1}^{N}\oint_{S^{1}}\text{Tr}\left( \lambda_{j}f^{-1}d_{\mathbb{A}}f   \right), \label{mu mu}
\end{equation}%
with $\tilde{\Phi}_{\text{on}} $ as given in \eqref{Phi defect  eom}. The contact 4d Chern-Simons theory with $N$ defects is then defined by the quadratic, real and $\mathcal{S}$-invariant action functional on $\overline{\mathcal{A}}\times L\mathcal{O}_{\lambda_{1}}\times ... \times L\mathcal{O}_{\lambda_{N}}$, defined by Cf. \eqref{facto 1} 
\begin{equation}
S(\mathbb{A},U)_{\text{c-}4d\text{-CS}}:=ic\left( \mathbf{\mu} ,\mathbf{\mu}\right), \label{Generalized 4CS action}
\end{equation}
which is nothing but the dual action with $N$ defects constructed above in \eqref{defect dual}. Notice that the dual theory can be put in quadratic form only because of the Hamiltonian symmetry algebra acting on the quotient space $\overline{\mathcal{A}}$, was reduced from $\mathfrak{h}$ to $\overline{\mathfrak{h}}\subset \mathfrak{h}$, where a well-defined inner product can, actually, be defined. See solution iii) in $\S \eqref{5.1.2}$. 

In order to keep simplicity in the following discussion we will suppress, without loss of generality, all the (co)-adjoint orbit defects. Consider now the extended action \eqref{double shift inv}. By fixing the $\kappa$-shift symmetry by means of the gauge fixing condition $\Phi=0$, we obtain the regularized 4d CS theory
\begin{equation}
S(\mathbb{A})_{\text{reg}}=ic\dint\nolimits_{\mathbb{M}}\omega_{\zeta} \wedge CS\left( 
\mathbb{A}\right). \label{reg again}
\end{equation}
The latter action being invariant under $\omega_{\zeta}$-shifts only. 

We can now consider two different limits of the action \eqref{reg again}. The first limit is when we take $\zeta \rightarrow 0$. The resulting action is given by
\begin{equation}
S(\mathbb{A})_{4d\text{-CS}}=ic\dint\nolimits_{\mathbb{M}}\omega \wedge CS\left( 
\mathbb{A}\right), \label{zeta''}
\end{equation}
which is formally equal to the usual 4d CS theory but now defined on $\mathbb{M}=\mathbb{R} \times \text{M}$. In this limit the theory also preserves, using local coordinates $(\tau, \sigma, z, \overline{z})$, the original $(1,0)$-shift symmetry \eqref{chi} and it is independent of the gauge field component $A_{z}$ as well. In this sense, the original and the regularized 4d CS theories are virtually the same, as both reproduce the same sample of 2d IFT's considered in $\S \eqref{4}$. The second limit is when we take\footnote{Alternative, we can interpret this limit as if $\zeta \rightarrow \infty$. Thus, we have that $\zeta \in [0, \infty)$.} $\omega =0$. In this case, we get, after decomposing $\mathbb{A}=A_{\tau}d\tau+A$, that
\begin{equation}
\tilde{S}(\mathbb{A})_{3d\text{-CS}}=2ic\alpha_{\tau}\zeta \int\nolimits_{\mathbb{R}}d\tau \bigg( \int\nolimits_{\text{M}} CS\left( 
A\right)   \bigg).
\end{equation}
The expression inside the parentheses is the ordinary 3d-CS theory on M originally considered in \cite{NA loc CS} and in this limit, the time component of the gauge field completely decouples. Thus, the regularized action \eqref{reg again} interpolates between the 4d-CS theory on $\mathbb{R}\times \text{M}$ and the 3d-CS theory on M trivially embedded in $\mathbb{M}$, denoted here as $\tilde{S}(A)_{3d\text{-CS}}$. In conclusion, the action \eqref{reg again} is defined on the $\zeta$-dependent quotient space $\mathcal{A}/s\omega_{\zeta}$ and roughly, we have that
\begin{equation}
\underset{\mathcal{A}/sd\tau}{\tilde{S}(\mathbb{A})_{3d\text{-CS}}} \overset{\omega \rightarrow 0}{\xleftarrow{\hspace*{2.0cm}}} \underset{\mathcal{A}/s\omega_{\zeta}}{S(\mathbb{A})_{\text{reg}}}\overset{\zeta \rightarrow 0}{\xrightarrow{\hspace*{2.0cm}}} \underset{\mathcal{A}/s\chi}{S(\mathbb{A})_{4d\text{-CS}}}, \label{two coset limits}
\end{equation}
where below all three actions we have displayed the quotient spaces where the corresponding action functional is naturally defined. 

If now we insert the adjoint scalar field $\Phi$ bulk eom back into the action \eqref{double shift inv}, we obtain, as shown above, the dual or contact 4d CS theory action functional \eqref{dual}, now being expressed in the quadratic form
\begin{equation}
S(\mathbb{A})_{\text{dual}}=ic(\mu,\mu), \label{the result}
\end{equation}
see \eqref{Generalized 4CS action} above, where $\mu$ denotes ${\mathbf\mu}$ with all the (co)-adjoint orbit defects being turned off. 

Two natural limits of the extended action \eqref{double shift inv} to be considered as well are $\zeta \rightarrow 0$ and $\omega=0$. In the first limit, the extended action is linear in $\Phi$ and no notion of duality is available, Cf. \eqref{gamma Top}. This is a problematic limit rendering the original 4d-CS theory not dualizable and the main reason why we had to regularize it in the first place, by relaxing the condition that $\mathbb{M}$ should look like the product $\Sigma \times C$ globally. For the second limit we get, after decomposing $\mathbb{A}=A_{\tau}d\tau + A$, that
\begin{equation}
\tilde{S}(\mathbb{A},\Phi)_{\text{ext}}=2ic\alpha_{\tau}\zeta \dint\nolimits_{\mathbb{R}}d\tau \left( \dint\nolimits_{\text{M}}CS(A)-2\dint\nolimits_{\text{M}}\alpha \wedge \text{Tr} \big(\Phi F_{\text{A}}    \big)+\dint\nolimits_{\text{M}}\alpha \wedge d_{\text{M}}\alpha \, \text{Tr}(\Phi^{2})\right).
\end{equation}
The quantity inside the parenthesis is precisely the extended 3d action functional $S(A,\Phi)_{\text{ext}}$ introduced in \cite{NA loc CS}, but now trivially embedded in $\mathbb{M}$ and denoted here as $\tilde{S}(\mathbb{A},\Phi)_{\text{ext}}$. We have used the $t$-rescaling symmetry with $t=1/\zeta'$ in order to absorb $\zeta'$. As before, when $\omega=0$, the time component of $\mathbb{A}$ completely decouples from the theory. Thus, in the updated classical round-trip diagram introduced at the end of section $\S \eqref{2.1.3}$, all boundary contributions are absent and now we have that:
\begin{equation*}
\begin{array}{ccc}
S(\mathbb{A})_{4d\text{-CS}}=\small{\text{action\,}} \eqref{zeta''} & \qquad {\overset{\omega_{\zeta} \text{-shift
extension}}{\xrightarrow{\hspace*{2.5cm}}  } }& \qquad  \boxed{S(\mathbb{A})_{\text{reg}}=\small\text{action\,} \eqref{reg again}}
\\ 
&  &  \\ 
\; \;  \Bigg\uparrow
\begin{array}{c}
\footnotesize\text{step I: }\, \scriptstyle{\Phi_{\text{on}}=0}  \\ 
\! \! \! \! \! \! \! \! \!  \quad  \footnotesize\text{step II: } \, \scriptstyle{\zeta \rightarrow 0}%
\end{array}
&  & \qquad \Bigg\downarrow \, \kappa \footnotesize\text{-shift extension} \\ 
&  &  \\ 
\boxed{S(\mathbb{A})_{\text{con}}=\small\text{action\,}\eqref{the result}}
& \qquad  \overset{\Phi \text{-integration}}{\xleftarrow{\hspace*{2.5cm}} } & \qquad  S(\mathbb{A},\Phi )_{\text{ext}}=%
\small\text{action\,} \eqref{double shift inv}%
\end{array}
\end{equation*} 
 
Finally, we comment on the condition imposed upon the 1-form $\kappa$ found in \S \eqref{4}, namely that the fiber components of $\kappa_{\mu}$, $\mu=\tau ,\sigma$ are constants. For $\mu=\tau$, this is trivially satisfied because we have already chosen that $\alpha_{\tau}\neq 0 \in \mathbb{R}$, while for $\mu=\sigma$ the same is valid, i.e. $\alpha_{\sigma}\neq 0 \in \mathbb{R}$, by the Seifert condition imposed on the 3-manifold M. It is indeed satisfactory to see how the Seifert condition on M is not only necessary for the inner product \eqref{inner product} to be well-defined, allowing to put the dual action with (co)-adjoint orbit defects insertions \eqref{defect dual} in the quadratic form \eqref{mu mu}, \eqref{Generalized 4CS action}, but also to nicely relate, in the $\zeta \rightarrow 0$ limit, the regularized 4d CS theory with the integrable sigma models living on the pole-defect surface $\Sigma \times \mathfrak{p}$, at least, for the sample of standard integrable $\sigma$-models we just considered. 

\section{Path integral symplectic measures}\label{6}

An important property of the quotient space $\overline{\mathcal{A}}$ is that it can be endowed with a K\"ahler-type metric. As a consequence, the conventional path integral volume form on $\overline{\mathcal{A}}$ can, formally, be defined in terms of the Liouville volume form associated to the K\"ahler-type form. See \cite{NA loc CS} for the 3d case on M. In this section, we review and adapt such a result for the contact 4d CS theory, but in a more general way than the proof provided in \cite{Yo}, where only the case $\text{M}=S^{3}$ was considered as the main working example and where some issues concerning reality conditions were not covered properly. Essentially, the main goal of this section is to provide an equivariant version of the calculation of \cite{NA loc CS} that includes the twist form $\omega$ as well. We also consider, based on a discussion presented in \cite{Wilson NA loc} on the same topic, the symplectic measure on the loopspace $L\mathcal{O}$ corresponding to the (co)-adjoint orbit defect insertions. The new ingredient that emerges when dealing with complex Lie groups, in comparison with the real ones, is the presence of pseudo-K\"ahler structures, where their associated metrics are not necessarily positive definite.

\subsection{Measures on $\overline{\mathcal{A}}$ and $L\mathcal{O}$}

\subsubsection{Complex structure on $\Omega_{{\text{Hor,}}\mathbb{M}}^{1}$}

The K\"ahler metric structure on $\overline{\mathcal{A}}$ crucially depends on the existence of a complex structure $J$, such that $J^{2}=-1$, when acting on the space of horizontal 1-forms of the 4d manifold $\mathbb{M}$, denoted by $\Omega_{{\text{Hor,}}\mathbb{M}}^{1}\subset \Omega_{\mathbb{M}}^{1}$ and defined as the subspace of $\Omega_{\mathbb{M}}^{1}$ that is annihilated by the contractions against the vector fields $\partial_{\tau}$ and $R$, namely\footnote{We denote by $\text{ker}_{\textit{O}}$ and $\text{im}_{\textit{O}}$, the kernel and images associated to the action of the operator $\textit{O}$.}
\begin{equation}
\Omega_{{\text{Hor,}}\mathbb{M}}^{1}=\text{ker}_{\iota_{\partial_{\tau}}}\cap \text{ker}_{\iota_{R}}. \label{Hor MM}
\end{equation} 
Recall that the global $\mathbb{R}$ factor and the circle fibers $S^{1}$ define, respectively, trivial and non-trivial $U(1)$ bundles over the base manifold $C$. This is why we referred to $(\tau,\sigma)$ in \S \eqref{4}, as the fiber coordinates. In this section we assume that all differential forms are real. 

The logic behind the construction of $J$ is the following: the Hodge duality operator $\star_{C}$ on $C$ acting on the space of differential forms $\Omega_{C}^{\bullet}$, can be defined, but not in a mandatory way\footnote{Actually, a metric is not required and $\star_{C}$ can be defined simply as a duality operator relating 2-forms and 0-forms on $C$ in terms of the defining relation $\star_{C}1=\sigma_{C}$, i.e. we only need a symplectic structure on $C$.}, in terms of the K\"ahler metric $g_{C}$ associated to the symplectic form $\sigma_{C}$ defining the contact form $\alpha_{r}$, as given by the third relation in \eqref{contact defs}. When acting on the subspace of 1-forms $\Omega_{C}^{1}$, it obeys $\star_{C}^{2}=-1$, hence defining an integral complex structure. From \eqref{n bundle}, we infer that the projection map $\underline{\pi}$ induces an effective 2d Hodge duality operator $\star_{2}$ on $\text{M}$, via the relation $\star_{2}\circ \underline{\pi}^{\ast}=\underline{\pi}^{\ast}\circ \star_{C}$, that satisfy $\star_{2}^{2}=-1$ as well, when acting on the subspace of horizontal 1-forms on $\text{M}$, denoted by $\Omega_{\text{Hor,M}}^{1}\subset \Omega_{\text{M}}^{1}$ and defined as the subspace of $\Omega_{\text{M}}^{1}$ that is annihilated by the contraction against the Reeb field $R$. This 2d effective Hodge duality operator was introduced in \cite{NA loc CS} in terms of the Hodge duality operator $\star_{\text{M}}$ on M and the contraction $\iota_{R}$. Because of $\Omega_{\text{Hor,M}}^{1}\subset \Omega_{{\text{Hor,}}\mathbb{M}}^{1} $ and $\mathbb{M}=\mathbb{R}\times \text{M}$, one is inevitably forced to conclude that $J$ must be constructed in a similar manner in terms of $\star_{\mathbb{M}}$, which is the Hodge duality operator on $\mathbb{M}$, and the contractions $\iota_{\partial_{\tau}}$ and $\iota_{R}$. The key to show this is to relate $\star_{\mathbb{M}}$ and $\star_{\text{M}}$ and exploit what is known in the 3d case. Such a relation can be easily found because of the trivial relation that exists between the metrics $g_{\text{M}}$ and $g_{\mathbb{M}}$ on M and $\mathbb{M}$, respectively.

Before we continue the construction of $J$, it is convenient first to invoke some known results, taken from \cite{NA loc CS}, relating $\star_{\text{M}}$ and $\alpha_{r}$. We assume that the vector field $R$ acts on M as an isometry, so that the metric $g_{\text{M}}$ on M takes the form
\begin{equation}
ds_{\text{M}}^{2}=\underline{\pi}^{\ast}(ds_{C}^{2})+\alpha_{r} \otimes \alpha_{r}, \label{metric on M}
\end{equation} 
where $ds_{C}^{2}$ is defined in terms of the metric $g_{C}$ with an associated K\"ahler form $\sigma_{C}$ normalized as in \eqref{contact defs}, \eqref{normalizations}. As a consequence of this normalization, the Hodge duality operator defined by the metric \eqref{metric on M}, satisfies 
\begin{equation}
\star_{\text{M}}1=\alpha_{r} \wedge d_{\text{M}}\alpha_{r},\text{ \qquad \qquad }\star_{\text{M}} \alpha_{r} =d_{\text{M}}\alpha_{r}. \label{alpha key rel}
\end{equation}
To see this, we can use local bundle coordinates $x=(z,\overline{z},\sigma)$, where $\alpha_{r}$ is written as in \eqref{alpha real 1}, \eqref{components} and $ds_{C}^{2}$  takes the explicit form
\begin{equation}
ds_{C}^{2}=g_{z\overline{z}}dz\otimes d\overline{z}+g_{\overline{z}z}d\overline{z}\otimes dz,
\end{equation}
with $g_{z\overline{z}}=\partial_{z}\partial_{\overline{z}}\mathcal{K}$. In matrix form the metric and its inverse are given, respectively, by
\begin{equation}
\lbrack g_{\text{M}}]=\left( 
\begin{array}{ccc}
\alpha _{rz}^{2} & g_{z\overline{z}}+\alpha _{rz}\alpha _{r\overline{z}} & 
\alpha _{rz}\alpha _{\sigma } \\ 
g_{z\overline{z}}+\alpha _{rz}\alpha _{r\overline{z}} & \alpha _{r\overline{z}%
}^{2} & \alpha _{r\overline{z}}\alpha _{\sigma } \\ 
\alpha _{rz}\alpha _{\sigma } & \alpha _{r\overline{z}}\alpha _{\sigma } & 
\alpha _{\sigma }^{2}%
\end{array}%
\text{ }\right)
\end{equation}
and
\begin{equation}
[g_{\text{M}}^{-1}]=\frac{1}{\alpha
_{\sigma }g_{z\overline{z}}}\left( 
\begin{array}{ccc}
\text{\ }0 & \alpha _{\sigma } & -\alpha _{r\overline{z}} \\ 
\alpha _{\sigma } & 0 & -\alpha _{rz} \\ 
-\alpha _{r\overline{z}} & -\alpha _{rz} & \frac{1}{\alpha _{\sigma }}\big(
g_{z\overline{z}}+2\alpha _{rz}\alpha _{r\overline{z}}\big) 
\end{array}%
\right) .
\end{equation}
Recall the action of $\ast_{\text{M}}$ on the base space of differential 1-forms on M, i.e.
\begin{equation}
\star_{\text{M}}\left(dx^{i}   \right)=\frac{1}{2!}\sqrt{|g_{\text{M}}|}(g_{\text{M}})^{ij}\epsilon_{jkl}dx^{k}\wedge dx^{l}, \label{Hodge M}
\end{equation} 
where $x^{i}$, $i=1,2,3$ are coordinates on M, $(g_{\text{M}})^{ij}$ is the inverse of $(g_{\text{M}})_{ij}$ and $\epsilon_{ijk}$ is the 3d completely antisymmetric symbol. In the coordinates $(z,\overline{z},\sigma)$, with $\epsilon_{z\overline{z}\sigma}=i$ and $\sqrt{|g_{\text{M}}|}=\alpha_{\sigma}g_{z\overline{z}}$, we find that
\begin{equation}
\star_{\text{M}}\left(dz   \right)=i\alpha_{r}\wedge dz,\text{\qquad}\star_{\text{M}}\left(d\overline{z}   \right)=-i\alpha_{r}\wedge d\overline{z},\text{\qquad}\star_{\text{M}}\left(d\sigma   \right)=\frac{1}{\alpha_{\sigma}}\Big(d_{\text{M}}\alpha_{r}-\star_{\text{M}} \big(\alpha_{r}\big|_{\text{Hor}}\big)  \Big).\label{star M on basis}
\end{equation}
From this follows the second relation in \eqref{alpha key rel}. The first one is straightforward.

Consider now a metric on the 4d manifold $\mathbb{M}$ of the form
\begin{equation}
ds_{\mathbb{M}}^{2}=d\tau \otimes d\tau -ds_{\text{M}}^{2}.
\end{equation}
In the coordinates $x^{\tilde{\mu}}=(x^{0}=\tau, x^{i})$, with $\tilde{\mu}=0,1,2,3$, we have that
\begin{equation}
(g_{\mathbb{M}})_{\tilde{\mu} \tilde{\nu}}=\text{diag}\Big(1, -(g_{\text{M}})_{ij}\Big). \label{metric on MM}
\end{equation}
The action of $\ast_{\mathbb{M}}$ on the base space of differential 1-forms on $\mathbb{M}$ is given by
\begin{equation}
\star_{\mathbb{M}}\left(dx^{\tilde{\mu}}   \right)=\frac{1}{3!}\sqrt{|g_{\mathbb{M}}|}(g_{\mathbb{M}})^{\tilde{\mu} \tilde{\nu}}\epsilon_{\tilde{\nu} \tilde{\rho} \tilde{\lambda} \tilde{\delta}}dx^{\tilde{\rho}}\wedge dx^{\tilde{\lambda}}\wedge dx^{\tilde{\delta}}, \label{Hodge MM}
\end{equation}
where $\epsilon_{\tilde{\mu} \tilde{\nu}\tilde{ \rho}\tilde{ \lambda}}$ is the 4d completely antisymmetric symbol. From the metric expression \eqref{metric on MM} and the definitions \eqref{Hodge M}, \eqref{Hodge MM}, we find that
\begin{equation}
\star_{\mathbb{M}}\left( d\tau   \right)=\star_{\text{M}}1=\alpha_{r} \wedge d_{\text{M}}\alpha_{r}, \text{ \qquad \qquad }\star_{\mathbb{M}}\left( dx^{i}   \right)= d\tau \wedge \star_{\text{M}}\left(dx^{i}   \right).\label{Imp result}
\end{equation}

It is important to understand how the Hodge operator $\star_{\mathbb{M}}$ acts on the space of horizontal forms on M, i.e. $\Omega_{\text{Hor,M}}$. A basis of differential forms of this space, satisfy $dx_{\text{Hor}}^{i}=\underline{\pi}^{\ast}(dy^{i})$, where $y^{i}$ are coordinates on the base manifold $C$ and $dy^{i}$ forms a basis of the space of differential 1-forms on the base manifold $C$, i.e. $\Omega_{C}^{1}$. Notice that the index $i$ is now restricted to take only two possible values. Thus, from the second result in \eqref{Imp result}, we get
\begin{equation}
\star_{\mathbb{M}}\left( dx_{\text{Hor}}^{i}   \right)=d\tau \wedge \star_{\text{M}}\big(\underline{\pi}^{\ast}(dy^{i}) \big). \label{pre CS}
\end{equation}
In order to find the complex structure $J$, one projects \eqref{pre CS} along \eqref{Hor MM} by acting with $\iota_{R}\circ \iota_{\partial_{\tau}}$ as any new contraction against $\partial_{\tau}$ or $R$ will vanish by the nilpotency of the contraction operation. Thus, we get the following chain of results, valid for horizontal 1-forms, i.e.
\begin{equation}
J\cdot dx_{\text{Hor}}^{i}:= \iota_{R}\circ \iota_{\partial_{\tau}}\circ \star_{\mathbb{M}}\left( dx_{\text{Hor}}^{i}   \right)=\star_{2}\big(\underline{\pi}^{\ast}(dy^{i}) \big)=\underline{\pi}^{\ast}\big( \star_{C}(dy^{i})  \big)=\lambda\, dx_{\text{Hor}}^{i},
\end{equation}
where $\lambda=\pm i$, $i=\sqrt{-1}$ is given by the eigenvalue equation $\star_{C}(dy^{i})=\lambda dy^{i}$, assuming $dy^{i}$ has a well-defined eigenvalue, and where we have identified, up to a sign, the effective 2d Hodge duality operator first introduced in \cite{NA loc CS}
\begin{equation}
\star_{2}=\iota_{R}\circ \star_{\text{M}}.
\end{equation}
Contracting the first two expressions in \eqref{star M on basis} against $R$, we find the expected results
\begin{equation}
J\cdot dz=idz\text{\qquad \qquad} J\cdot d\overline{z}=-id\overline{z}. \label{expectedd}
\end{equation}
Thus, when acting on the space \eqref{Hor MM}, we have that
\begin{equation}
J=\iota_{R}\circ \iota_{\partial_{\tau}}\circ \star_{\mathbb{M}}=\star_{2}.
\end{equation}
This result is precisely what we were looking for.

We know proceed to study the structure of $\Omega_{\mathbb{M}}^{1}$, in terms of its horizontal and vertical subspaces.

Let us begin with its horizontal subspace. An important result that follows from \eqref{pre CS} is given by
\begin{equation}
dx^{i}_{\text{Hor}}\wedge \star_{\mathbb{M}}\big( dx_{\text{Hor}}^{j}   \big)= d\tau \wedge \alpha_{r} \wedge dx^{i}_{\text{Hor}}\wedge J\big( dx_{\text{Hor}}^{j}   \big).\label{result 2}
\end{equation}
In showing this, we have used the trivial contraction
\begin{equation}
0=\iota_{R}\Big(d\tau \wedge \alpha_{r} \wedge dx^{i}_{\text{Hor}} \wedge \star_{\text{M}}\big( dx_{\text{Hor}}^{j}   \big) \Big),
\end{equation}
implying that
\begin{equation}
d\tau \wedge dx^{i}_{\text{Hor}} \wedge \star_{\text{M}}\big( dx_{\text{Hor}}^{j}   \big)=-d\tau \wedge \alpha_{r} \wedge dx^{i}_{\text{Hor}} \wedge \star_{2}\big( dx_{\text{Hor}}^{j}   \big).
\end{equation}
In order to express $dx^{i}_{\text{Hor}}\in \Omega_{\text{Hor,M}}^{1}$ in terms of $dx^{\tilde{\mu}}\in\Omega_{\mathbb{M}}^{1}$, we introduce the following projector operator $\Pi:\Omega_{\mathbb{M}}^{1}\rightarrow \Omega_{\text{Hor,M}}^{1}$, defined by
\begin{equation}
\Pi=1-d\tau \iota_{\partial_{\tau}}-\alpha_{r} \iota_{R}. \label{ZZZ}
\end{equation}
This projector is defined in terms of two globally-defined and nowhere vanishing vector fields and differential forms and satisfies 
\begin{equation}
\iota_{\partial_{\tau}}\circ \Pi=\iota_{R}\circ \Pi=0.
\end{equation}
Essentially, it removes the fiber directions of any differential 1-form. Thus, we have that
\begin{equation}
\gamma \big|_{\text{Hor}}=\Pi \gamma,
\end{equation} 
for any $\gamma \in \Omega_{\mathbb{M}}^{1}$ and now \eqref{result 2} becomes
\begin{equation}
\Pi \gamma \wedge \star_{\mathbb{M}}\Pi \gamma'=d\tau \wedge \alpha_{r} \wedge \Pi \gamma \wedge J\Pi \gamma'. \label{hor equality}
\end{equation}
 
Consider now the vertical subspace. Notice that the kernel of $\Pi$, consists of all vertical 1-forms obtained by acting with the projector operator $\Pi^{\perp}: \Omega_{\mathbb{M}}^{1}\rightarrow \Omega_{\text{Ver,M}}^{1}$, defined by
\begin{equation}
\Pi^{\perp}=d\tau \iota_{\partial_{\tau}}+\alpha_{r} \iota_{R}
\end{equation} 
and this means that any 1-form can be decomposed trivially in the form
\begin{equation}
\gamma=\Pi\gamma +\Pi^{\perp}\gamma,\text{ \qquad \qquad}\Pi \circ \Pi^{\perp}=\Pi^{\perp} \circ \Pi=0.
\end{equation}

In terms of these operators the inner product of any pair of 1-forms is defined to be proportional to the integral over $\mathbb{M}$ of
\begin{equation}
\gamma \wedge \star_{\mathbb{M}}\gamma'=\Pi\gamma \wedge \star_{\mathbb{M}}\Pi\gamma' +\Pi\gamma \wedge \star_{\mathbb{M}}\Pi^{\perp}\gamma'+\Pi\gamma' \wedge \star_{\mathbb{M}}\Pi^{\perp}\gamma +\Pi^{\perp}\gamma \wedge  \star_{\mathbb{M}}  \Pi^{\perp}\gamma'. \label{norm gamma}
\end{equation}
The first term on the rhs can be rewritten as in \eqref{hor equality}, the second and third contributions vanish because of from, say
\begin{equation}
\star_{\mathbb{M}}\Pi^{\perp}\gamma'=\big(\alpha_{r}\iota_{\partial_{\tau}}\gamma'+ d\tau \iota_{R}\gamma'\big)\wedge d_{\text{M}}\alpha_{r},
\end{equation}
the third relation in \eqref{contact defs} and $\Pi \gamma=\underline{\pi}^{\ast}\gamma_{C}$, for some $\gamma_{C}\in \Omega_{C}^{1}$, we find that
\begin{equation}
\Pi\gamma \wedge d_{\text{M}}\alpha_{r}\sim \underline{\pi}^{\ast}\big( \gamma_{C} \wedge \sigma_{C} \big)=0
\end{equation}
vanishes for dimensional reasons. The last term on the rhs in \eqref{norm gamma} thus gives
\begin{equation}
\Pi^{\perp}\gamma \wedge  \star_{\mathbb{M}}  \Pi^{\perp}\gamma'=d\tau \wedge \alpha_{r}\wedge d_{\text{M}}\alpha_{r}\Big[ \big( \iota_{\partial_{\tau}}\gamma \big)\big( \iota_{\partial_{\tau}}\gamma' \big)- \big( \iota_{R}\gamma \big)\big( \iota_{R}\gamma' \big)\Big].
\end{equation}
The relative sign on the rhs right above comes from the signature of the metric in \eqref{metric on MM}.

Finally, the lhs in \eqref{norm gamma}, then induces the following orthogonal decomposition
\begin{equation}
\Omega_{\mathbb{M}}^{1}=\Omega_{{\text{Hor,}}\mathbb{M}}^{1}\oplus \Omega_{{\text{Ver,}}\mathbb{M}}^{1}. \label{dec hor ver}
\end{equation}
On $\Omega_{{\text{Hor,}}\mathbb{M}}^{1}$, we define the inner product as being proportional to
\begin{equation}
\big( \gamma,\gamma'  \big)_{\text{Hor}}=\dint_{\mathbb{M}}d\tau \wedge \alpha_{r} \wedge \Pi \gamma \wedge J\Pi \gamma', \label{inner hor}
\end{equation}
while on its orthocomplement, we define it to be proportional to
\begin{equation}
\big( \gamma,\gamma'  \big)_{\text{Ver}}= \dint_{\mathbb{M}} d\tau \wedge \alpha_{r}\wedge d_{\text{M}}\alpha_{r}\Big[ \big( \iota_{\partial_{\tau}}\gamma \big)\big( \iota_{\partial_{\tau}}\gamma' \big)-\big( \iota_{R}\gamma \big)\big( \iota_{R}\gamma' \big)\Big]. \label{inner ver}
\end{equation}
Both expression being real and metric-independent. The only reminiscent data coming from the metric is the signature in \eqref{inner ver}, which is a topological invariant anyway. Notice that, because of $\Pi(d\tau)=\Pi(\alpha_{r})=0$, the inner product \eqref{inner hor} is invariant under the shift symmetry
\begin{equation}
\gamma\rightarrow \gamma+sd\tau + s'\alpha_{r}, \label{shift t alpha}
\end{equation}
where $s,s'\in \Omega_{\mathbb{M}}^{0}$ are arbitrary functions on $\mathbb{M}$. The same holds for $\gamma'$. 

In the next section, we will use this construction as an inspiration for introducing appropriate norms in the decomposition $\mathcal{A}=\overline{\mathcal{A}}\oplus \mathcal{S}$, that are relevant for the construction of the contact 4d CS theory path integral localization formula. The new ingredient being the presence of equivariant differential forms. Unfortunately, the Hodge duality operator $\star_{\mathbb{M}}$, does not map equivariant differential forms into equivariant differential forms. Fortunately, as we saw, the inner products \eqref{inner hor} and \eqref{inner ver} associated to the decomposition \eqref{dec hor ver} are metric independent and the strategy to be addressed below is to work with real connections and differential forms in order to infer and introduce later well-defined norms on $\overline{\mathcal{A}}$ and $\mathcal{S}$ that respect the required equivariance properties when complex-valued objects are considered. All this, without invoking any metric defined on $\mathbb{M}$.

\subsubsection{Measure on $\overline{\mathcal{A}}$}

Let us first consider the situation where $\omega \rightarrow 0$ and $\zeta \neq 0$, i.e. we start with the regularized 4d CS theory defined on the quotient space $\mathcal{A}/sd\tau$, corresponding to the leftmost position in the diagram \eqref{two coset limits}. This is precisely the situation where the usual 3d CS theory defined on M is trivially embedded in $\mathbb{M}=\mathbb{R}\times \text{M}$. In this case, the results found in the last section apply straightforwardly with the projectors $\Pi$ and $\Pi^{\perp}$ now acting, by extension, on the space of connections $\mathcal{A}$. In what follows, connections and differential forms are still considered as real unless otherwise specified.

The (pseudo)-Riemannian norm of a connection $\mathbb{A}$ is defined, up to a multiplicative constant, by the integral over $\mathbb{M}$ of
\begin{equation}
\text{Tr} \big( \mathbb{A}\wedge \star_{\mathbb{M}}\mathbb{A} \big)=\text{Tr}\left(A_{\tilde{\mu}}A_{\tilde{\nu}} \right)dx^{\tilde{\mu}}\wedge \star_{\mathbb{M}}\big( dx^{\tilde{\nu}}\big),\label{core inner}
\end{equation} 
where we have used the decomposition $\mathbb{A}=A_{\tilde{\mu}}dx^{\tilde{\mu}}$. From the results found above applied now to $dx^{\tilde{\mu}}\wedge \star_{\mathbb{M}}\big( dx^{\tilde{\nu}}\big)$ in \eqref{core inner}, we get, up to multiplicative constants that
\begin{equation}
\big( \mathbb{A},\mathbb{A} \big)_{\text{Hor}}=\dint_{\mathbb{M}}d\tau \wedge \alpha_{r} \wedge \text{Tr} \Big( \Pi (\mathbb{A}) \wedge J\Pi (\mathbb{A}) \Big)\label{quotient}
\end{equation}
and
\begin{equation}
\big( \mathbb{A},\mathbb{A}  \big)_{\text{Ver}}=\dint_{\mathbb{M}} d\tau \wedge \alpha_{r}\wedge d_{\text{M}}\alpha_{r}\, \text{Tr}\Big[\big(\iota_{\partial_{\tau}}\mathbb{A}  \big)\big(\iota_{\partial_{\tau}}\mathbb{A} \big)- \big( \iota_{R}\mathbb{A} \big)\big( \iota_{R}\mathbb{A} \big)  \Big]. \label{orto inner}
\end{equation}
The norm \eqref{quotient} is defined on the quotient space formed by the space $\mathcal{A}$ modulo connections of the form $sd\tau +s'\alpha_{r}$, for $s,s'\in \Omega_{\mathbb{M}}^{0}\otimes\mathfrak{g}^{\mathbb{R}}$, C.f \eqref{shift t alpha}, and this follows from the fact that $\Pi(d\tau)=\Pi(\alpha_{r})=0$.
Also notice the resemblance of the norm \eqref{orto inner}, with the invariant and non-degenerate inner product on the gauge algebra $\mathcal{G}_{\text{Lie}}$, defined before by the relation
\begin{equation}
(\eta ,\lambda):=-\dint\nolimits_{\mathbb{M}}\gamma_{\text{top}} \text{Tr}\left( \eta \lambda \right), \label{inner on S}
\end{equation} 
for any pair $\eta,\lambda\in \Omega_{\mathbb{M}}^{0}\otimes \mathfrak{g}$ of Lie algebra valued functions on $\mathbb{M}$. See the first contribution to the rhs of \eqref{inner product}. Thus, if the inner product is positive definite or not, depends on the signature's bilinear form $\text{Tr}$ real slice. This will be important later, mainly because of the Lie algebras involved in the formulation of the 4d CS theory are, by definition, always complex.  

We now proceed to turn on the equivariant twist form $\omega$, still with $\zeta\neq0$. In this case, we start now with the regularized 4d CS theory defined on the quotient space $\mathcal{A}/s\omega_{\zeta}$, corresponding to the general case, i.e. the middle position in the diagram \eqref{two coset limits}.     

We begin by introducing operators $P$ and $P^{\perp}$, acting on the space $\mathcal{A}$, such that $P: \mathcal{A}\rightarrow \text{im}P$ and $P^{\perp}:\mathcal{A}\rightarrow \text{im}P^{\perp}$, induce an orthogonal decomposition $\mathcal{A}=\text{im}P\oplus \text{im}P^{\perp}$ with respect to the norm \eqref{core inner}. Propose that
\begin{equation}
\mathbb{A}=P (\mathbb{A})+P^{\perp}(\mathbb{A}),\text{ \qquad}P(\mathbb{A})=\mathbb{A}-p\Omega_{r}-q \kappa_{r},\text{\qquad}P^{\perp}(\mathbb{A})=p\Omega_{r}+q \kappa_{r},
\end{equation}
where $p,q:\mathcal{A}\rightarrow \Omega_{\mathbb{M}}^{0}\otimes \mathfrak{g}^{\mathbb{R}}$ are to be determined by an orthogonality condition and $\Omega_{r}, \kappa_{r}$ are of the same form as in \eqref{solutions omega, kappa} but with\footnote{Recall the assumption that all quantities are considered as real valued. } $\omega \rightarrow \omega_{r}$ and $\alpha \rightarrow \alpha_{r}$. We will also replace momentarily the constant parameter $\alpha_{\tau}\rightarrow \alpha_{\tau}'$ in the definition of $\kappa_{r}$. Then, \eqref{core inner} gives
\begin{equation}
\text{Tr} \big( \mathbb{A}\wedge \star_{\mathbb{M}}\mathbb{A} \big)=\text{Tr} \Big( P(\mathbb{A})\wedge \star_{\mathbb{M}}P(\mathbb{A}) \Big)+\text{Tr} \Big( P^{\perp}(\mathbb{A})\wedge \star_{\mathbb{M}}P^{\perp}(\mathbb{A}) \Big)+2\text{Tr} \Big( P(\mathbb{A})\wedge \star_{\mathbb{M}}P^{\perp}(\mathbb{A}) \Big).
\end{equation}

The orthogonality condition requires the vanishing of the last contribution on the rhs right above and this imposition imply that
\begin{equation}
0=\text{Tr}\Big( p\mathbb{A}\wedge \star_{\mathbb{M}}\Omega_{r} +q\mathbb{A}\wedge \star_{\mathbb{M}}\kappa_{r}-p^{2}\Omega_{r} \star_{\mathbb{M}}\Omega_{r}-q^{2} \kappa_{r} \star_{\mathbb{M}}\kappa_{r}-2pq\Omega_{r}\wedge \star_{\mathbb{M}}\kappa_{r}\Big).\label{condition ort}
\end{equation}
Let us consider the following results
\begin{equation}
\begin{aligned}
\kappa_{r} \star_{\mathbb{M}}\kappa_{r}&=\zeta'^{2}\big(\alpha_{\tau}'^{2}-1\big)d\tau \wedge \alpha_{r}\wedge d_{\text{M}}\alpha_{r},\\
\Omega_{r} \star_{\mathbb{M}}\Omega_{r}&=\omega_{r}\wedge \star_{\mathbb{M}}\omega_{r}+\zeta^{2}\big(\alpha_{\tau}^{2}-1\big)d\tau \wedge \alpha_{r}\wedge d_{\text{M}}\alpha_{r},\\
\Omega_{r} \star_{\mathbb{M}}\kappa_{r}&=\zeta \zeta' \big(\alpha_{\tau}\alpha_{\tau}'+1 \big)d\tau \wedge \alpha_{r}\wedge d_{\text{M}}\alpha_{r},\\
\Omega_{r}\wedge \kappa_{r}\wedge d_{\mathbb{M}}\kappa_{r}&=\zeta \zeta'^{2}\big(\alpha_{\tau}+ \alpha_{\tau}'  \big)d\tau \wedge \alpha_{r}\wedge d_{\text{M}}\alpha_{r}. \label{123}
\end{aligned}
\end{equation}
The last equation in \eqref{123} is, as we saw before, necessary for the construction of the regularized 4d CS theory in the general case, hence we require that $\alpha_{\tau}+ \alpha_{\tau}' \neq 0$. The expression \eqref{condition ort} nicely simplifies if we set $\alpha_{\tau}=\alpha_{\tau}'=1$ and because of the $t$-rescaling invariance we can further fix $\zeta'=1$. The only genuine free parameter turns out to be $\zeta$. The contribution $\omega_{r}\wedge \star_{\mathbb{M}}\omega_{r}$ vanishes when we promote $\omega_{r}\rightarrow \omega$ from being a real 1-form to be the equivariant twist form, so this contribution will be ignored it in the rest of the analysis\footnote{This is because $\omega \wedge \star_{\mathbb{M}}\omega\sim \omega \wedge \omega=0.$}. Thus, \eqref{condition ort} requires that
\begin{equation}
p=\frac{\mathbb{A}\wedge \star_{\mathbb{M}}\kappa_{r}}{\Omega_{r}\wedge \kappa_{r}\wedge d_{\mathbb{M}}\kappa_{r}},\text{\qquad \qquad}q=\frac{\mathbb{A}\wedge \star_{\mathbb{M}}\Omega_{r}}{\Omega_{r}\wedge \kappa_{r}\wedge d_{\mathbb{M}}\kappa_{r}}.
\end{equation}
Now, using
\begin{equation}
\star_{\mathbb{M}}\kappa_{r}=\kappa_{r}\wedge d_{\mathbb{M}}\kappa_{r}, \text{\quad}\star_{\mathbb{M}}\Omega_{r}=\star_{\mathbb{M}}\omega_{r}-\Omega_{r}\wedge d_{\mathbb{M}}\kappa_{r},\text{\quad}\star_{\mathbb{M}}\omega_{r}=d\tau \wedge \alpha_{r}\wedge \star_{2}\omega_{r},
\end{equation}
we find that
\begin{equation}
p=\iota_{\mathcal{R}'}\mathbb{A},\text{\qquad \qquad}q=\iota_{\mathcal{R}}\mathbb{A}-\theta(\mathbb{A}),
\end{equation}
where the map $\theta: \Omega_{\mathbb{M}}^{1}\rightarrow \Omega_{\mathbb{M}}^{0}$ is defined by
\begin{equation}
\theta(\cdot):=\frac{d\tau \wedge \alpha_{r}\wedge \star_{2}\omega_{r}\wedge (\cdot)}{\Omega_{r}\wedge \kappa_{r}\wedge d_{\mathbb{M}}\kappa_{r}}.
\end{equation}

After the orthogonal decomposition is ensured, we consider now the norms on the spaces $\text{im}P$ and $\text{im}P^{\perp}$. Let us first begin with $\text{im}P$ and write
\begin{equation}
P(\mathbb{A})=\mathscr{P}_{r}(\mathbb{A})+\theta(\mathbb{A})\kappa_{r},
\end{equation} 
where $\mathscr{P}_{r}:\Omega_{\mathbb{M}}^{1}\rightarrow \Omega_{\text{Hor,M}}^{1}$ is a projector operator defined by
\begin{equation}
\mathscr{P}_{r}:=1-\Omega_{r}\iota_{\mathcal{R}'}-\kappa_{r}\iota_{\mathcal{R}}.
\end{equation}
Following \cite{Yo, NA loc CS}, we consider the trivial contraction
\begin{equation}
0=\iota_{\mathcal{R}'}\Big( \Omega_{r}\wedge P(\mathbb{A})\wedge \star_{\mathbb{M}}P(\mathbb{A})  \Big)
\end{equation}
to show that
\begin{equation}
\text{Tr} \Big( P(\mathbb{A})\wedge \star_{\mathbb{M}}P(\mathbb{A}) \Big)=-\Omega_{r}\wedge \text{Tr} \Big( P(\mathbb{A})\wedge \iota_{\mathcal{R}'}\circ \star_{\mathbb{M}}P(\mathbb{A}) \Big). \label{contrac 1}
\end{equation}
In a similar manner, combining the trivial contraction
\begin{equation}
0=\iota_{\mathcal{R}}\Big( \Omega_{r}\wedge \kappa_{r}\wedge P(\mathbb{A})\wedge \iota_{\mathcal{R}'}\circ\star_{\mathbb{M}}P(\mathbb{A})  \Big),
\end{equation}
with \eqref{contrac 1}, gives
\begin{equation}
\text{Tr} \Big( P(\mathbb{A})\wedge \star_{\mathbb{M}}P(\mathbb{A}) \Big)=\frac{1}{2\zeta}\Omega_{r}\wedge \kappa_{r}\wedge \text{Tr} \Big( \mathscr{P}_{r}(\mathbb{A})\wedge J\mathscr{P}_{r}(\mathbb{A}) \Big), \label{real coset norm}
\end{equation}
where we have used
\begin{equation}
\star_{\mathbb{M}}P(\mathbb{A})=d\tau \wedge \kappa_{r}\wedge \star_{2}\mathscr{P}_{r}(\mathbb{A})+\theta(\mathbb{A})\kappa_{r}\wedge d_{\mathbb{M}}\kappa_{r},
\end{equation}
in order to write the final expression in terms of $\mathscr{P}_{r}$. The rhs in \eqref{real coset norm} is invariant under the shift symmetry
\begin{equation}
\mathbb{A}\rightarrow \mathbb{A}+s\Omega_{r}+s'\kappa_{r}
\end{equation}
and, from the norm point of view, this identifies $\text{Im}P$ with the quotient space formed by the space of connections $\mathcal{A}$ modulo connections of the form $s\Omega_{r}+s'\kappa_{r}$, i.e. the real version of $\overline{\mathcal{A}}$. On the space $\text{im}P^{\perp}$, we quickly find that
\begin{equation}
\text{Tr} \Big( P^{\perp}(\mathbb{A})\wedge \star_{\mathbb{M}}P^{\perp}(\mathbb{A}) \Big)=2\Omega_{r}\wedge \kappa_{r}\wedge d_{\mathbb{M}}\kappa_{r}\, \text{Tr} \Big[\big(\iota_{\mathcal{R}'}\mathbb{A}  \big)\big( \iota_{\mathcal{R}}\mathbb{A}-\theta(\mathbb{A}) \big)   \Big],\label{real complement}
\end{equation}
hence providing a norm on the shift group $\mathcal{S}$. From the fact that $\theta(\kappa_{r})=\theta(\Omega_{r})=0$ and after neglecting the contribution $\omega_{r}\wedge \star_{2}\omega_{r}$, we conclude that $\theta(\mathbb{A})=\theta \big( \mathscr{P}_{r}(\mathbb{A})  \big)$. 

The norm \eqref{core inner}, then induces the following orthogonal decomposition
\begin{equation}
\mathcal{A}=\overline{\mathcal{A}}\oplus \mathcal{S}. \label{real decomposition}
\end{equation}
On the quotient space $\overline{\mathcal{A}}$, we define the inner product as being proportional to
\begin{equation}
\big( \mathbb{A},\mathbb{A} \big)_{\overline{\mathcal{A}}}=\frac{1}{2\zeta}\dint_{\mathbb{M}}\Omega_{r}\wedge \kappa_{r}\wedge \text{Tr} \Big( \mathscr{P}_{r}(\mathbb{A})\wedge J\mathscr{P}_{r}(\mathbb{A}) \Big), \label{inner coset real}
\end{equation}
while on its ortho-complement $\mathcal{S}$, we define it to be proportional to
\begin{equation}
\big( \mathbb{A},\mathbb{A} \big)_{\mathcal{S}}= 2\dint_{\mathbb{M}} \Omega_{r}\wedge \kappa_{r}\wedge d_{\mathbb{M}}\kappa_{r}\, \text{Tr} \Big[\big(\iota_{\mathcal{R}'}\mathbb{A}  \big)\Big( \iota_{\mathcal{R}}\mathbb{A}-\theta \big( \mathscr{P}_{r}(\mathbb{A})  \big) \Big)   \Big]. \label{real complement}
\end{equation}
Path integral measures are to be defined in terms of the norms \eqref{inner coset real} and \eqref{real complement}. In the latter case, the norm on $\mathcal{S}$ seems to explicitly depend on $\overline{\mathcal{A}}$ via the action of $\theta$ on $\mathscr{P}_{r}({\mathbb{A}})$, but this is actually not an issue as path integral measures are invariant under translations in field space and this term can be absorbed by a field redefinition. In the limit $\omega_{r} \rightarrow 0$, we recover \eqref{quotient} and \eqref{orto inner} and if we further take $A_{\tau}=0$, we recover, inside the parentheses, the expressions first introduced in \cite{NA loc CS} for the 3d CS theory case, namely
\begin{equation}
\big( A,A \big)_{\overline{\mathcal{A}}}=\dint_{\mathbb{R}}d\tau\left(\dint_{\text{M}}\alpha_{r}\wedge \text{Tr} \Big[ \Pi(A)\wedge J\Pi(A) \Big]\right),
\end{equation}
and
\begin{equation}
\big( A,A \big)_{\mathcal{S}}=\dint_{\mathbb{R}}d\tau\left(-\dint_{\text{M}}\alpha_{r}\wedge d_{\text{M}}\alpha_{r}\, \text{Tr} \Big[ \big( \iota_{R}A \big)^{2} \Big]\right).
\end{equation}

The pattern is now clear and we are in the position to `equivariantize' the results we just found in order to handle equivariant gauge connections and equivariant differential forms. We have gathered substantial information concerning the metric independence of the norms as both depend solely on $\alpha_{r}$ and $J$. In the decomposition $\mathcal{A}=\overline{\mathcal{A}}\oplus \mathcal{S}$, where gauge connections and differential forms are now considered to be equivariant, the norm on $\mathcal{A}$ is defined by
\begin{equation}
\big( \mathbb{A},\mathbb{A} \big)_{\mathcal{A}}:=c_{1}\big( \mathbb{A},\mathbb{A} \big)_{\overline{\mathcal{A}}}+c_{2}\big( \mathbb{A},\mathbb{A} \big)_{\mathcal{S}},\label{orto decom}
\end{equation}
where $c_{1},c_{2} \in \mathbb{R}$ are real numbers and where
\begin{equation}
\begin{aligned}
\big( \mathbb{A},\mathbb{A} \big)_{\overline{\mathcal{A}}}&:=\hat{\Omega}\Big(\mathscr{P}(\mathbb{A}),J\mathscr{P}(\mathbb{A})  \Big)=-\frac{1}{2}\dint_{\mathbb{M}}\Omega\wedge \kappa\wedge \text{Tr} \Big( \mathscr{P}(\mathbb{A})\wedge J\mathscr{P}(\mathbb{A}) \Big), \\
\big( \mathbb{A},\mathbb{A} \big)_{\mathcal{S}}&:= -i\dint_{\mathbb{M}} \gamma_{\text{top}}\, \text{Tr} \Big[\big(\iota_{\mathcal{R}'}\mathbb{A}  \big)\big( \iota_{\mathcal{R}}\mathbb{A} \big)   \Big]. \label{explicit equi norm}
\end{aligned}
\end{equation}
Right above we have dropped the $\theta$ term contribution for the reason explained before, introduced the projector $\mathscr{P}: \mathcal{A}\rightarrow \overline{\mathcal{A}}$, with
\begin{equation}
\mathscr{P}:=1-\Omega \iota_{\mathcal{R}'}-\kappa \iota_{\mathcal{R}}\label{Abar projector}
\end{equation} 
and used \eqref{pre-symplectic} to put explicitly the norm on $\overline{\mathcal{A}}$ in the K\"ahler-type form. Notice that $\mathscr{P}$ preserves the equivariance properties of $\mathbb{A}$ and that the reality of the first line in \eqref{explicit equi norm} follows from the fact that $\hat{\tau}\circ J=-J \circ \hat{\tau}$, when acting on Lie algebra valued horizontal 1-forms. Now, to verify that \eqref{orto decom} is an orthogonal decomposition, we simply decompose any gauge connection in the form
\begin{equation}
\mathbb{A}=\mathscr{P}(\mathbb{A})+\mathscr{P}^{\perp}(\mathbb{A}),\qquad \qquad \mathscr{P}^{\perp}=\Omega\iota_{\mathcal{R}'}+\kappa \iota_{\mathcal{R}} \label{dec A}
\end{equation}
and use $\mathscr{P}(\Omega)=\mathscr{P}(\kappa)=0$ to show that $\mathscr{P}\circ \mathscr{P}^{\perp}=\mathscr{P}^{\perp} \circ \mathscr{P}=0 $. The result then follows. 

It is important the identify the signature of $\big( \mathbb{A},\mathbb{A} \big)_{\overline{\mathcal{A}}}$. To see this, it is useful to use local coordinates $(\tau, \sigma, z, \overline{z})$, where $J$ acts as in \eqref{expectedd}. Then, the first line in \eqref{explicit equi norm} takes the explicit form
\begin{equation}
\big( \mathbb{A},\mathbb{A} \big)_{\overline{\mathcal{A}}}=2i \zeta \zeta' \alpha_{\tau} \alpha_{\sigma}\dint_{\mathbb{M}}d\tau \wedge d\sigma \wedge dz \wedge d\overline{z}\, \text{Re}\Big[\text{Tr} \big( \mathscr{P}(\mathbb{A})_{z}\mathscr{P}(\mathbb{A})_{\overline{z}} \big)\Big], \label{signature'}
\end{equation}
where we have used the decomposition \eqref{ReIm parts} in order to select the real part of the bilinear form $\text{Tr}$ defined on the complex Lie algebra $\mathfrak{g}$. For example, if in the decomposition $\mathfrak{g}=\mathfrak{g}^{\mathbb{R}}\oplus i \mathfrak{g}^{\mathbb{R}}$, $\mathfrak{g}^{\mathbb{R}}$ is the Lie algebra of a compact lie group $G^{\mathbb{R}}$, where $\text{Tr}_{\mathbb{R}}$ is considered as positive definite, the norm on $\overline{\mathcal{A}}$ is clearly not positive definite, reflecting the pseudo-K\"ahler nature of the manifold $\overline{\mathcal{A}}$. The signature being $(n,n)$, where $n=\text{dim}_{\mathbb{R}}\mathfrak{g}^{\mathbb{R}}$. This is why we have referred to the metric on $\overline{\mathcal{A}}$ as being of K\"ahler-type. 

In terms of the decomposition \eqref{dec A}, the action of the gauge group $\mathcal{G}_{0}$
\begin{eqnarray}
^{g}\mathbb{A}=\mathscr{P}\big(^{g}\mathbb{A}\big)+\mathscr{P}^{\perp}\big(^{g}\mathbb{A}\big)
\end{eqnarray}
induces the actions
\begin{equation}
\begin{aligned}
\mathscr{P}\big(^{g}\mathbb{A}\big)&= g^{-1}\mathscr{P}(\mathbb{A})g+g^{-1}\big(d_{\mathbb{M}}-\Omega \pounds_{\mathcal{R}'} -\kappa \pounds_{\mathcal{R}}  \big)g,\\
\mathscr{P}^{\perp}\big(^{g}\mathbb{A}\big)&=g^{-1}\mathscr{P}^{\perp}(\mathbb{A})g+g^{-1}\big(\Omega \pounds_{\mathcal{R}'} +\kappa \pounds_{\mathcal{R}}  \big)g,\label{deco key}
\end{aligned}
\end{equation}
along $\overline{\mathcal{A}}$ and $\mathcal{S}$, respectively. Now, we recast the two gauge fixing conditions \eqref{tau gauge fixing} and \eqref{sigma gauge fixing}, for the $\omega_{\zeta}$-shift and $\kappa$-shift symmetries, respectively, in the compact form (recall that $X=(1/\zeta')R$)
\begin{equation}
\mathscr{P}^{\perp}(\mathbb{A})=0.
\end{equation}
As the contact 4d CS theory is defined on the quotient space formed by the space of gauge fields on $\mathbb{M}$ modulo the space of gauge fields of the form $s\kappa + s' \Omega$, i.e. $\overline{\mathcal{A}}$, the second contribution to the rhs in second line in \eqref{deco key} does not affect the gauge fixing condition. This is similar to the situation discussed around \eqref{gauge fixing cond}.

Because of the pair $(\hat{\Omega},J)$ induces a K\"ahler-type metric structure on $\overline{\mathcal{A}}$, the path integral measure over $\overline{\mathcal{A}}$ can, formally, be written in the symplectic form\footnote{If $(X,\hat{\Omega})$ is a symplectic manifold of dimension $2n$ with symplectic/K\"ahler form $\hat{\Omega}$, the Liouville volume form is $\hat{\Omega}^{n}/n!$. The latter can be represented by $\text{exp}\, ( \hat{\Omega})$, where we implicitly pick out from the series expansion of the exponential the term which is of top degree on $X$. This is formally extended to the infinite dimensional case via expressions of the form \eqref{coset measure}. }
\begin{equation}
\mathcal{D}\mathbb{A}|_{\overline{\mathcal{A}}}=\text{exp}\;i \hat{\Omega},\text{ \qquad \qquad }
\hat{\Omega}=-\frac{1}{2}\dint\nolimits_{\mathbb{M}}\omega_{\zeta} \wedge
\kappa \wedge \text{Tr}\big(  \hat{\delta }\mathbb{A}\wedge \hat{\delta }\mathbb{A}\big), \label{coset measure}
\end{equation}
where we have used \eqref{key shift} to replace $\Omega$ by $\omega_{\zeta}$ in \eqref{pre-symplectic}. The integrand in $\hat{\Omega}$ being an equivariant 4-form, imply that the quantity $i\hat{\Omega}$ is real-valued. 

Finally, we comment on the idea that motivated the introduction of the pre-symplectic form \eqref{pre-symplectic}, which is an important object in our approach to the 4d CS theory. Consider the embeddings $C \hookrightarrow \text{M}\hookrightarrow \mathbb{M}$ and their corresponding spaces of connections $\mathcal{A}_{C}\subset \mathcal{A}_{\text{M}}\subset \mathcal{A}_{\mathbb{M}}$ formed by real-valued connections. The latter denoted simply by $\mathcal{A}$. On $\mathcal{A}_{C}$ we consider the Atiyah-Bott symplectic form \cite{Atiyah-Bott}
\begin{equation}
\hat{\Omega}_{\text{AB}}:=-\frac{1}{2}\dint\nolimits_{C}\text{Tr}\big(  \hat{\delta }\mathbb{A}\wedge \hat{\delta }\mathbb{A}\big).
\end{equation}
This symplectic form can be embedded into $\mathcal{A}_{\text{M}}$ in the form of the Beasley-Witten pre-symplectic form \cite{NA loc CS}, defined by
\begin{equation}
\hat{\Omega}_{\text{BW}}:=-\frac{1}{2}\dint\nolimits_{\text{M}}
\alpha_{r} \wedge \text{Tr}\big(  \hat{\delta }\mathbb{A}\wedge \hat{\delta }\mathbb{A}\big). \label{BW}
\end{equation} 
Notice that \eqref{BW} is only symplectic when restricted to the quotient space formed by the space of connections $\mathcal{A}_{\text{M}}$ modulo connections of the form $s\alpha_{r}$. By the same token, we can embed \eqref{BW} into $\mathcal{A}$ by introducing
\begin{equation}
\hat{\Omega}_{r}:=-\frac{1}{2}\dint\nolimits_{\mathbb{M}}d\tau \wedge
\alpha_{r} \wedge \text{Tr}\big(  \hat{\delta }\mathbb{A}\wedge \hat{\delta }\mathbb{A}\big).
\end{equation}
This 2-form is symplectic only when restricted to the quotient space formed by the space of connections $\mathcal{A}$ modulo connections of the form $sd\tau+s'\alpha_{r}$. If we promote the real differential forms $(d\tau,\alpha_{r})$ to the equivariant ones $(\Omega, \kappa)$ presented in \eqref{solutions omega, kappa} and consider $\mathbb{A}$ as a complex equivariant connection, we end up with \eqref{pre-symplectic} or \eqref{coset measure}, which is symplectic only when restricted to $\overline{\mathcal{A}}$. Indeed, if we decompose $\mathbb{A}$ as in \eqref{dec A} and insert it in $\hat{\Omega}$ in \eqref{coset measure}, we get that $\hat{\Omega}\big|_{\overline{\mathcal{A}}}\neq 0$ and $\hat{\Omega}\big|_{\mathcal{S}}= 0$. 

\subsubsection{Measure on $L\mathcal{O}$} \label{6.1.3}

The construction of a Liouville measure on a complex Lie group (co)-adjoint orbit has its peculiarities, when compared to its real counterpart, as we shall now see. 

Consider the smallest number of adjoint orbits insertions allowed by the reality conditions considered in $\S \eqref{3}$, which is $N=2$ or equivalently $k=1$, as determined by \eqref{defect data'}, giving
\begin{equation}
\big(U_{1},U_{2}=\hat{\tau}(U_{1})  \big). \label{2 orbits}
\end{equation}
The latter, when applied to the expressions \eqref{orbit symp form}, \eqref{orbit symp form'}, gives 
\begin{equation}
i\hat{\Omega}'=-l\oint\nolimits_{S^{1}}\kappa \wedge \text{Tr}\big[\lambda \theta \wedge \theta+ \hat{\tau}(\lambda) \hat{\tau}(\theta) \wedge \hat{\tau}(\theta)  \big]=-2l\oint\nolimits_{S^{1}}\kappa \wedge \text{Re}\big[ \text{Tr}(  \lambda\, \theta \wedge \theta )  \big], \label{Real CASF}
\end{equation}
where we have set $\lambda_{1}=\lambda$ and defined $\theta:=f^{-1}\hat{\delta}f \in \Omega_{LG}^{1}$.

K\"ahler structures on (co)-adjoint orbits are well-understood and their construction heavily relies on the fact that the underlying orbit Lie groups are real and compact, with associated K\"ahler metrics that are positive definite, see for instance $\S 8$ in \cite{Besse}. This is an issue for our discussion as the Lie groups in our approach to the 4d CS theory are, from the outset, inherently complex and hence often non-compact. Thus, it is not clear how an expression such as \eqref{Real CASF} fits within what is known about usual K\"ahler structures on (co)-adjoint orbits. In order to make some progress in that direction, a generalization of the construction of \cite{Besse} to (co)-adjoint orbits of complex Lie groups is required. The outcome is that the associated K\"ahler metrics for such type of orbits are not necessarily positive definite, turning these orbits into pseudo-K\"ahler manifolds, see \eqref{signature'} for an example.  

The main goal for the remaining of this section is to interpret the expression \eqref{Real CASF}. In order to do this, we bring some facts reproduced from $\S 3$ of \cite{Wagner} almost verbatim, related to the structure of (co)-adjoint orbits of complex Lie groups. The reader is directed to that reference for proofs and further details concerning the technicalities\footnote{References in \cite{Wagner} mentioned along the text are also useful.}. Fortunately, the general treatment of \cite{Wagner} on this topic fits nicely into our discussion about eq. \eqref{Real CASF} but, for the sake of simplicity, we will mainly focus on the specific situation of (co)-adjoint orbits associated to complex Lie groups that are reductive and the reason for this is that in this way we get a straightforward generalization to the discussion of \cite{Wilson NA loc}, covering (co)-adjoint orbits of real compact Lie groups. In what follows, we will display upper/lower indexes $\mathbb{R}$ and $\mathbb{C}$ in order to facilitate the tracking of real and complex quantities as we proceed with the introduction of the two complex structures that coexist on the (co)-adjoint orbit of a complex Lie group. 

Consider a real Lie algebra $\mathfrak{g}^{\mathbb{R}}$ with Lie group $G^{\mathbb{R}}$ and endow $\mathfrak{g}^{\mathbb{R}}$ with a symmetric, non-degenerate and ad-invariant bilinear form $\text{Tr}_{\mathbb{R}}(\cdot, \cdot):\mathfrak{g}^{\mathbb{R}} \times\mathfrak{g}^{\mathbb{R}} \rightarrow \mathbb{R}$. Denote by $\mathcal{O}_{\mathbb{R}}\subset \mathfrak{g}^{\mathbb{R}}$, the orbit associated to the adjoint action of $G^{\mathbb{R}}$ on $\mathfrak{g}^{\mathbb{R}}$.

Now we introduce a complex structure $J$ on $\mathcal{O}_{\mathbb{R}}$. An adjoint orbit only admits a canonical complex structure $J_{w}$ if the orbit contains a skew-symmetric element $w\in \mathfrak{g}^{\mathbb{R}}$ and this means that its associated adjoint map $\text{ad}_{w}:=[w,\cdot \,]\in \text{End}\big(\mathfrak{g}^{\mathbb{R}}\big)$ satisfies the following two conditions:
\begin{itemize}
\item[(i)] Its complexification $\text{ad}_{w}:=[w,\cdot \,]\in \text{End}\big(\mathfrak{g}^{\mathbb{C}}\big)$ is diagonalizable. The action of $\text{ad}_{w}$, with $w \in \mathfrak{g}^{\mathbb{R}}$ is extended to $\mathfrak{g}^{\mathbb{C}}$ by $\mathbb{C}$-linearity.  

\item[(ii)]  The non-vanishing eigenvalues of $\text{ad}_{w}=[w,\cdot \,]\in \text{End}\big(\mathfrak{g}^{\mathbb{C}}\big)$ are purely imaginary.
\end{itemize}

Skew-symmetry of $\text{ad}_{w}$ follows from $\text{Tr}_{\mathbb{R}}\big(  \text{ad}_{w} (x)\, y \big)=-\text{Tr}_{\mathbb{R}}\big(  x\, \text{ad}_{w}(y) \big)$, for $x,y,w \in \mathfrak{g}^{\mathbb{R}}$, consequence of the ad-invariance of the bilinear form on $\mathfrak{g}^{\mathbb{R}}$, while the existence of purely imaginary eigenvalues is a consequence of requiring that $\text{Tr}_{\mathbb{R}}$ is positive definite, e.g. when $G^{\mathbb{R}}$ is a compact Lie group. Some immediate consequences of (i) and (ii), are the following:
\begin{itemize}
\item The Lie algebra $\mathfrak{g}^{\mathbb{R}}$ admits the splitting $\mathfrak{g}^{\mathbb{R}}=\text{ker}_{\text{ad}_{w}}\oplus \text{im}_{\text{ad}_{w}}$, for every skew-symmetric element $w\in \mathfrak{g}^{\mathbb{R}}$.

\item If $w \in \mathfrak{g}^{\mathbb{R}}$ is skew-symmetric, the non-vanishing eigenvalues of $\text{ad}_{w}$ come in pairs $(i\mu,-i\mu)$ with $\mu >0$. This follows from $\text{ad}_{w}(\overline{v})=\overline{\text{ad}_{w}(v)}$, for $v \in \mathfrak{g}^{\mathbb{C}}$, when $\text{ad}_{w}v=i\mu  v$. The bar denoting complex conjugation.

\item Every element in the adjoint orbit of an skew-symmetric element $w$ is also skew-symmetric. This can be seen from the identity
\begin{equation}
\text{ad}_{\text{Ad}_{f}w}=\text{Ad}_{f}\circ \text{ad}_{w}\circ \text{Ad}_{f^{-1}},\text{\qquad \qquad} \forall f\in G^{\mathbb{R}},
\end{equation}
where $\text{Ad}_{f}(u)=fuf^{-1}$, $u\in \mathfrak{g}^{\mathbb{R}}$. If an adjoint orbit contains a skew-symmetric element, then entire orbit is called skew-symmetric.
\end{itemize}

Skew-symmetric elements are important because the subspace $V^{\mathbb{R}}:=\text{im}_{\text{ad}_{w}}\subset \mathfrak{g}^{\mathbb{R}}$ is naturally equipped with a canonical complex structure defined by
\begin{equation}
J_{w}v:=\frac{1}{\mu}\text{ad}_{w}(v), \text{\qquad \qquad} \forall v\in E_{\mu}, \label{O complex structure}
\end{equation}
where $E_{\mu}:=V^{\mathbb{R}} \cap \big(E_{i\mu}\oplus E_{-i\mu}  \big)$ and $E_{\pm i \mu}\subset \mathfrak{g}^{\mathbb{C}}$ are the eigenspaces of $\text{ad}_{w}$ with eigenvalues $\pm i\mu$, $\mu >0$, respectively. Decompose
\begin{equation}
V^{\mathbb{R}}:= \bigoplus_{\mu >0}E_{\mu}.\label{V}
\end{equation} 
Any $v\in E_{\mu}$ can be written as $v=u+\overline{u}$, where $u\in E_{i \mu}$ and as a consequence, $J_{w}^{2}v=-v$. As usual, this allows to introduce an (anti)-holomorphic decomposition for $V^{\mathbb{C}}$, given by
\begin{equation}
V^{\mathbb{C}}=V^{(1,0)}\oplus V^{(0,1)}, \label{splitting}
\end{equation} 
where
\begin{equation}
V^{(1,0)}=\bigoplus_{\mu >0}E_{i\mu}, \text{\qquad \qquad}V^{(0,1)}=\bigoplus_{\mu >0}E_{-i\mu}. \label{splitting'}
\end{equation}
Elements in $V^{(1,0)}$ and $V^{(0,1)}$ have, respectively, eigenvalues $\pm i$ when acted with $J_{w}$.

The construction of $J_{w}$ on the vector space $\text{im}_{\text{ad}_{w}}$ is local. In order to equip any skew-symmetric adjoint orbit $\mathcal{O}_{\mathbb{R}}$ passing through $w$ with a full-fledged complex structure, it is necessary to identify $T_{w}\mathcal{O}_{\mathbb{R}}$ with $\text{im}_{\text{ad}_{w}}\subset \mathfrak{g}^{\mathbb{R}}$ for every point $w \in \mathcal{O}_{\mathbb{R}}$, to show that $J_{w}$ is $G^{\mathbb{R}}$-invariant and furthermore, that it is integrable. Fortunately, this is possible as any skew-symmetric adjoint orbit $\mathcal{O}_{\mathbb{R}}\subset \mathfrak{g}^{\mathbb{R}}$ carries a canonical $G^{\mathbb{R}}$-invariant integrable complex structure $J$. See lemma 3.2.18 in $\S 3$ of \cite{Wagner} and paragraphs above it for the proof. 

Denote by $G_{w}^{\mathbb{R}}$ the stabilizer of $w$ in $G^{\mathbb{R}}$. Adjoint and coadjoint orbits are identified via $\text{Tr}_{\mathbb{R}}$ and in addition, the coset space $G^{\mathbb{R}}/G_{w}^{\mathbb{R}}$, where $f\sim fh$, $f\in G^{\mathbb{R}}$, $h\in G_{w}^{\mathbb{R}}$ and the adjoint orbit $\mathcal{O}_{\mathbb{R}}$ passing through $w$, are equivalent due to the map\footnote{See also \cite{Wagner} for further technical details about this statement.}
\begin{equation}
fG_{w}^{\mathbb{R}}\longrightarrow fw f^{-1}\label{map}.
\end{equation}
Because of \eqref{map}, the left action of $G^{\mathbb{R}}$ on $G^{\mathbb{R}}/G_{w}^{\mathbb{R}}$ descends to a transitive action of $G^{\mathbb{R}}$ on $\mathcal{O}_{\mathbb{R}}$. Furthermore, the space $\mathfrak{g}^{\mathbb{R}}\ominus \mathfrak{g}_{w}^{\mathbb{R}}$ is geometrically identified with the tangent space to $\mathcal{O}_{\mathbb{R}}\cong G^{\mathbb{R}}/G_{w}^{\mathbb{R}}$ at the identity coset and the real 2-form $\hat{\omega}\in \Omega_{G^{\mathbb{R}}}^{2}$, defined by
\begin{equation}
\hat{\omega}_{w}:=\text{Tr}_{\mathbb{R}}\big(w\, \theta\wedge \theta   \big), \label{Kahler nu}
\end{equation}
where\footnote{By a slight abuse of notation, we have denoted the canonical left-invariant 1-form $\theta$ on $\Omega_{G^{\mathbb{R}}}^{1}$ and $\Omega_{LG}^{1}$ above in \eqref{Real CASF}, by the same symbol.} $\theta:=f^{-1}\hat{\delta}f \in \Omega_{G^{\mathbb{R}}}^{1}$, is invariant under the respective left and right actions of $G^{\mathbb{R}}$ and $G_{w}^{\mathbb{R}}$ on $f$ and defines a closed and non-degenerate symplectic form on $\mathcal{O}_{\mathbb{R}}$. Fundamental vector fields in $T_{w}\mathcal{O}_{\mathbb{R}}$ associated to $u\in \mathfrak{g}^{\mathbb{R}}$ are of the form $X_{u}(w)=-\text{ad}_{w}(u)$ and the contraction of $\hat{\omega}_{w}$ against two vector fields is given by the expression
\begin{equation}
\hat{\omega}_{w}\big(X_{u}(w), X_{u}(w)  \big)=\text{Tr}_{\mathbb{R}}\big(w[u,v]  \big), \label{contracted}
\end{equation} 
for any $w \in \mathcal{O}_{\mathbb{R}}$ and $u,v \in \mathfrak{g}^{\mathbb{R}}$. We can restrict to $u,v \in V^{\mathbb{R}}$, for obvious reasons.

The pair $(\hat{\omega},J)$ is compatible, in the sense that $\hat{\omega}(J \cdot,J \cdot)=\hat{\omega}$, and the metric on $\mathcal{O}_{\mathbb{R}}$ defined by
\begin{equation}
g(\cdot, \cdot):=\hat{\omega}(\cdot,J \cdot) \label{Her metric}
\end{equation}
is K\"ahler, being positive definite if $\text{Tr}_{\mathbb{R}}$, when restricted to $\text{im}_{\text{ad}_{w}} \subset \mathfrak{g}^{\mathbb{R}}$, is positive definite. This can be seen from the explicit result
\begin{equation}
g_{w}(X_{u}(w), X_{v}(w))=\hat{\omega}_{w}\big(X_{u}(w), J_{w}X_{v}(w)  \big)=-\text{Tr}_{\mathbb{R}}\big(X_{u}(w)J_{w}v  \big),
\end{equation} 
where we have used $J_{w}X_{v}(w)=X_{J_{w}v}(w)$. If we further restrict to $v\in E_{\mu}$, for generic $\mu >0$, we find that
\begin{equation}
g_{w}(X_{u}(w), X_{v}(w))=\frac{1}{\mu}\text{Tr}_{\mathbb{R}}\big(  X_{u}(w) X_{v}(w) \big). \label{R adjoint inner}
\end{equation}
However, by the symmetry of $g(\cdot \, , \cdot)$ we conclude that \eqref{V} is an orthogonal decomposition. Thus, a non-zero result requires that $u\in E_{\mu}$ and finally we get that, $\mu g(\cdot \, , \cdot)$ and $\text{Tr}_{\mathbb{R}}(\cdot \, , \cdot)$ coincide. Hence, the triple $(\mathcal{O}_{\mathbb{R}}, \hat{\omega},J)$ defines a K\"ahler manifold. See theorem 3.2.24 and the definition 3.2.23 in $\S 3$ of \cite{Wagner}, for further details.

As an example, take $G^{\mathbb{R}}$ compact with positive definite $\text{Tr}_{\mathbb{R}}$ and set $w=\lambda \in \mathfrak{t}^{\mathbb{R}}$, Cf. $\S 4$ of \cite{Wilson NA loc}. Identify $\text{im}_{\text{ad}_{w}}=\mathfrak{g}^{\mathbb{R}}\ominus \mathfrak{g}_{\lambda}^{\mathbb{R}}$. The splitting \eqref{splitting} can, in practice, be obtained from any decomposition of the root system $\mathfrak{R}$ of $G^{\mathbb{R}}$ into positive and negative subsets $\mathfrak{R}_{\pm}$, so that $\mathfrak{R}=\mathfrak{R}_{+}\cup \mathfrak{R}_{-}$. As usual, $\alpha> 0$ denotes a positive root $\alpha \in \mathfrak{R}_{+}$ and $\alpha< 0$ a negative root $\alpha \in \mathfrak{R}_{-}$. Choosing an element $\lambda \in \mathfrak{t}^{\mathbb{R}}$ that is regular\footnote{For simplicity, we will not consider here elements $\lambda$ that are irregular.} and in the positive Weyl chamber $C_{+}\subset \mathfrak{t}^{\mathbb{R}}$, defined by
\begin{equation}
C_{+}=\bigl\{\xi \in \mathfrak{t}^{\mathbb{R}}\, \big|\, \langle \alpha, \xi  \rangle \geq 0,\, \forall \alpha >0   \bigr\},
\end{equation} 
where $\langle \cdot, \cdot \rangle$ denotes the pairing between $\mathfrak{g}^{\mathbb{R}}$ and its dual, it follows that $\mathfrak{R}_{+}$ corresponds to the subset of $\mathfrak{R}$ satisfying the condition $\langle \alpha, \lambda  \rangle >0$. In addition, we get from the regularity of $\lambda$, that $\mathfrak{g}_{\lambda}^{\mathbb{R}}=\mathfrak{t}^{\mathbb{R}}$.

Given the decomposition $\mathfrak{R}=\mathfrak{R}_{+}\cup \mathfrak{R}_{-}$, there is an associated decomposition
\begin{equation}
V^{\mathbb{C}}=V^{+}\oplus V^{-}
\end{equation}  
into positive and negative root-spaces
\begin{equation}
V^{+}=\bigoplus_{\alpha >0}\mathfrak{g}_{\alpha}^{\mathbb{C}}, \text{\qquad \qquad}V^{-}=\bigoplus_{\alpha >0}\mathfrak{g}_{-\alpha}^{\mathbb{C}},
\end{equation}
where $\mathfrak{g}_{\alpha}^{\mathbb{C}}$ is the 1d root-space defined by
\begin{equation}
[\xi, x_{\alpha}]=+i\langle \alpha, \xi   \rangle x_{\alpha},\text{\qquad \qquad}x_{\alpha} \in \mathfrak{g}^{\mathbb{C}}_{\alpha}, \label{key lie bracket}
\end{equation} 
for any element $\xi \in \mathfrak{t}^{\mathbb{R}}$. It is then natural, Cf. \eqref{splitting'}, to identify $E_{i\mu}=\mathfrak{g}_{\alpha}^{\mathbb{C}}$ with $\mu =\langle \alpha, \lambda  \rangle >0$ and to declare that
\begin{equation}
V^{(1,0)}=V^{+}, \text{\qquad \qquad}V^{(0,1)}=V^{-}.
\end{equation}
In summary, the complex structure $J$ acting on $V^{\mathbb{C}}$ and defined by the skew-symmetric element $\lambda$, is naturally induced by the root-space decomposition of $\mathfrak{g}^{\mathbb{R}}$. 

Consider now the case in which $G$ is a reductive complex Lie group with Lie algebra $\mathfrak{g}$, i.e. $G$ is a complex Lie group whose real form $G^{\mathbb{R}}\subset G$, fixed by $\hat{\tau}$, is a compact Lie group with Lie algebra $\mathfrak{g}^{\mathbb{R}}$. By definition, $G$ comes equipped with a canonical complex structure $I$. Denote by $\mathcal{O}\subset \mathfrak{g}$, the orbit associated to the adjoint action of $G$ in $\mathfrak{g}$.

In order to understand how the orbits $\mathcal{O}$ and $\mathcal{O}_{\mathbb{R}}\subset \mathcal{O}$ and their corresponding complex structures $I$ and $J$ are intertwined, we bring to the discussion the following two useful results:
\begin{itemize}
\item[{(1)}] The symmetric, non-degenerate and ad-invariant (under $\mathfrak{g}^{\mathbb{R}}$) bilinear form $\text{Tr}_{\mathbb{R}}(\cdot, \cdot):\mathfrak{g}^{\mathbb{R}} \times\mathfrak{g}^{\mathbb{R}} \rightarrow \mathbb{R}$, extends to a unique symmetric, non-degenerate and ad-invariant (under $\mathfrak{g}$) bilinear form $\text{Tr}(\cdot \, , \cdot):\mathfrak{g} \times \mathfrak{g} \rightarrow \mathbb{C}$ on $\mathfrak{g}$.

\item[(2)] The complexification of a skew-symmetric adjoint orbit $\mathcal{O}_{\mathbb{R}}$ is again skew-symmetric. Equivalently, the complexification of diagonalizable linear maps with purely imaginary eigenvalues are also diagonalizable with purely imaginary eigenvalues. In particular, the action of $\text{ad}_{w}$, with $w \in \mathfrak{g}^{\mathbb{R}}$ is extended to $\mathfrak{g}$ by $\mathbb{C}$-linearity. 
\end{itemize} 

Proofs of these statements are provided by the propositions 3.3.8 and 3.3.10 in $\S 3$ of \cite{Wagner}. 

Let us briefly verify how this works. It is convenient to express the complex Lie algebra in the form $\mathfrak{g}=\mathfrak{g}^{\mathbb{R}}\oplus I_{e}\mathfrak{g}^{\mathbb{R}}$, where $I_{e}$ is the canonical $G$-invariant and integrable complex structure $I$ of $G$ evaluated at the group identity. The latter satisfy the relations 
\begin{equation}
I_{e}[u,v]=[I_{e}u,v]=[u,I_{e}v],\text{\qquad \qquad}\forall u,v \in \mathfrak{g}, \label{I def}
\end{equation}
implying that $I_{e}$ commutes with $\text{ad}_{u}$ for any $u\in \mathfrak{g}$. 

For the first result we have, explicitly, that $\text{Tr}(\cdot \, ,\cdot)$ is given by
\begin{equation}
\text{Tr}(uv):=\text{Tr}'_{\mathbb{R}}(uv)-i\text{Tr}'_{\mathbb{R}}\big(I_{e}(u)v\big),\label{inner diff}
\end{equation}
where $u,v \in \mathfrak{g}$ are of the form $u=u_{1}+I_{e}u_{2}$ and $v=v_{1}+I_{e}v_{2}$, for $u_{1}, u_{2}, v_{1}, v_{2} \in \mathfrak{g}^{\mathbb{R}}$ and $\text{Tr}'_{\mathbb{R}}(\cdot \, , \cdot):\mathfrak{g} \times \mathfrak{g}\rightarrow \mathbb{R}$ is defined by the formula 
\begin{equation}
\text{Tr}'_{\mathbb{R}}(uv):=\text{Tr}_{\mathbb{R}}(u_{1}v_{1})-\text{Tr}_{\mathbb{R}}(u_{2}v_{2}).\label{real diff}
\end{equation}
Clearly, $\text{Tr}'_{\mathbb{R}}$ restricts to $\text{Tr}_{\mathbb{R}}$ on $\mathfrak{g}^{\mathbb{R}}$, satisfies $\text{Tr}'_{\mathbb{R}}( I_{e}\cdot \,, I_{e}\cdot )=-\text{Tr}'_{\mathbb{R}}(\cdot \, , \cdot)$ and turns the decomposition $\mathfrak{g}=\mathfrak{g}^{\mathbb{R}}\oplus I_{e}\mathfrak{g}^{\mathbb{R}}$ orthogonal under it. Furthermore, $\text{Tr}$ is non-degenerate because $\text{Tr}_{\mathbb{R}}$ is non-degenerate and it is invariant because of $\text{Tr}_{\mathbb{R}}'$ is ad-invariant under the action of $\mathfrak{g}$, consequence of \eqref{I def}, the ad-invariance of $\text{Tr}_{\mathbb{R}}$ under the action of $\mathfrak{g}^{\mathbb{R}}$ and the fact that $I_{e}^{2}=-1$. Notice that the second contribution to the rhs of \eqref{inner diff}, takes the form
\begin{equation}
-\text{Tr}'_{\mathbb{R}}\big(I_{e}(u)v\big)=\text{Tr}_{\mathbb{R}}(u_{1}v_{2})+\text{Tr}_{\mathbb{R}}(u_{2}v_{1}).\label{complex diff}
\end{equation}
Not surprisingly, the expressions \eqref{real diff} and \eqref{complex diff} coincide, respectively, with the ones obtained in \eqref{ReIm parts} from the usual representation $\mathfrak{g}=\mathfrak{g}^{\mathbb{R}}\oplus i\mathfrak{g}^{\mathbb{R}}$, where now $u=u_{1}+iu_{2}$ and $v=v_{1}+iv_{2}$. Also notice that the signature of $\text{Tr}'_{\mathbb{R}}(uv)$ is $(n,n)$, where $n:=\text{dim}_{\mathbb{R}}\mathfrak{g}^{\mathbb{R}}$. 

For the second result, consider the spaces\footnote{It can be defined more generally in terms of \eqref{splitting'}.}
\begin{equation}
\tilde{E}_{\mu}:=E_{\mu}\oplus I_{e}E_{\mu},\text{\qquad \qquad}\tilde{E}_{\pm i\mu}:=E_{\pm i\mu}\oplus I_{e}E_{\pm i\mu},
\end{equation}
generalizing the vector spaces $E_{\mu}$ and $E_{\pm i \mu}$ introduced above around equation \eqref{O complex structure}. Notice that
\begin{equation}
\text{ad}_{w}v_{\pm}=\pm i\mu v_{\pm}\longrightarrow \text{ad}_{w}\big(I_{e}v_{\pm}\big)=I_{e}\big(\text{ad}_{w}v_{\pm}\big)=\pm i\mu \big(I_{e}v_{\pm}\big),
\end{equation}
for $v_{\pm}\in E_{\pm i \mu}$. Thus, $v_{\pm}$ and $I_{e}v_{\pm}$ are eigenvectors of $\text{ad}_{w}$ with the same eigenvalues. 

The complex Lie algebra $\mathfrak{g}$ then admits a well-defined scalar product $\text{Tr}(\cdot \, , \cdot)$ and a skew-symmetric adjoint orbit $\mathcal{O}\in \mathfrak{g}$. In this case, \eqref{Kahler nu} and \eqref{contracted} extend, respectively, to
\begin{equation}
\hat{\omega}_{w}=\text{Tr}'_{\mathbb{R}}\big(w\, \theta\wedge \theta   \big),\text{\qquad \qquad}\hat{\omega}_{w}\big(X_{u}(w), X_{u}(w)  \big)=\text{Tr}'_{\mathbb{R}}\big(w[u,v]  \big), \label{Com omega}
\end{equation} 
where $\theta=f^{-1}\hat{\delta}\theta \in \Omega_{G}^{1}$ and $u,v \in \mathfrak{g}$. The norm of any vector $X_{u}(w)\in T_{w}\mathcal{O}$ with $u \in \tilde{E}_{\mu}$, is obtained from the expression \eqref{R adjoint inner}, after setting $v=u$ and replacing $\text{Tr}_{\mathbb{R}}$ by $\text{Tr}'_{\mathbb{R}}$. Thus, norms are not positive definite due to the signature of \eqref{real diff}.

By restricting $I_{e}$ to $\text{im}_{\text{ad}_{w}}\cong T_{w}\mathcal{O}$, an almost complex structure is obtained, which turns out to be integrable too. In addition, the pair $(\hat{w}, I_{e})$ is anti-compatible, in the sense that $\hat{\omega}(I_{e}\cdot, I_{e}\cdot)=-\hat{\omega}(\cdot,\cdot)$, and $J$ as well as the 2-form defined by
\begin{equation}
\hat{\mathbf{\omega}}:=\hat{\omega}-i\hat{\omega}(I_{e}\cdot,\cdot   ),
\end{equation}
are $I_{e}$-holomorphic. These statements comprises the proof of theorem 3.3.4 in $\S 3$ in \cite{Wagner}. Thus, the 4-tuple $(\mathcal{O},\hat{\omega}, J, I_{e})$ defines a holomorphic K\"ahler manifold, see definition 3.3.1, theorem 3.3.11 and corollary 3.3.14 in $\S 3$ of \cite{Wagner} as well for further details. In a nutshell, holomorphic K\"ahler manifolds are symplectic manifolds endowed with two mutually commuting integrable and $G$-invariant complex structures $(J,I_{e})$ that are, respectively, compatible and anticompatible with $\hat{\omega}$. This is a vast topic, but at the present level of analysis, the discussion just presented suffices in giving a sensible interpretation to the expression \eqref{Real CASF}. 

After identifying $\text{Re}\big[ \text{Tr}(  \cdot \,  , \cdot ) \big]=\text{Tr}_{\mathbb{R}}'(\cdot \, , \cdot)$ and setting $w=\lambda\in \mathfrak{t}^{\mathbb{R}}$, we find that \eqref{Real CASF}, takes the form
\begin{equation}
i\hat{\Omega}'=-2l\oint\nolimits_{S^{1}}\kappa \wedge \hat{\omega}_{\lambda}, \label{2 orbits'}
\end{equation} 
where we have extended $\hat{\omega}_{\lambda}$ point-wise into $S^{1}$, wedged it against $\kappa$ and integrated the result. By restricting to $\mathfrak{g}^{\mathbb{R}} $, i.e. to the first copy in $\mathfrak{g}=\mathfrak{g}^{\mathbb{R}}\oplus i \mathfrak{g}^{\mathbb{R}}$ we recover, up to some constant factors, the symplectic form first introduced in $\S 4.3$ of \cite{Wilson NA loc}. Thus, \eqref{2 orbits} characterizes a single orbit instead of two, as shown by \eqref{2 orbits'}.    

The loopspace $L\mathcal{O}_{\lambda}$ then carries a metric induced by the metric $g$ on the complex orbit $\mathcal{O}_{\lambda}$ and this allows to formally write the path integral measure
\begin{equation}
\mathcal{D}U=\mathcal{D}U_{1}\times...\times \mathcal{D}U_{N} \label{LO e DU}
\end{equation}
on $L\mathcal{O}=L\mathcal{O}_{\lambda_{1}}\times ... \times L\mathcal{O}_{\lambda_{N}}$, in terms of the symplectic expression
\begin{equation}
\mathcal{D}U= \text{exp}\;i \hat{\Omega}',\label{sym U measure}
\end{equation}
where ($k=N/2$)
\begin{equation}
i\hat{\Omega}'=-2l\sum_{j=1}^{k}\left(\oint\nolimits_{S^{1}}\kappa \wedge \hat{\omega}_{\lambda_{j}}\right),\text{\qquad \qquad}\lambda_{j} \in \mathfrak{t}^{\mathbb{R}}. \label{facto3}
\end{equation}

\section{Localization formula}\label{7}

In order to implement the quantum analogue of the classical round-trip diagram introduced at the end of section $\S \eqref{2.1.3}$,
the relevant object to be considered now is the path integral for the extended 4d CS theory action with $N$ co-adjoint orbit defect insertions \eqref{ext action N}, i.e.
\begin{equation}
\mathcal{Z}_{\text{ext}}=\int\nolimits_{\mathcal{A}\times L\mathcal{O} \times \mathcal{S}_{\kappa}} \mathcal{D}\mathbb{A}\mathcal{D}U \mathcal{D}\Phi  \; \text{exp}\left[\frac{i}{\hbar} S\big(\mathbb{A}-\kappa \Phi, U \big)_{\text{reg}}\right].\label{Zext}
\end{equation}
Right above, we have used \eqref{LO e DU} and the fact that $\Phi$, is in the orbit of $\mathcal{S}_{\kappa}$ (i.e. $\kappa$-shifts) and is a fixed point of $\mathcal{S}_{\omega_{\zeta}}$ (i.e. $\omega_{\zeta}$-shifts), in order to define the integration domain for the measure $\mathcal{D}\Phi$. To implement the duality manipulations at the level of path integrals, it is convenient to write the action \eqref{ext action N} in the form
\begin{equation}
S\big(\mathbb{A}-\kappa \Phi, U \big)_{\text{reg}}=S(\mathbb{A},U)_{\text{con}} +ic \dint_{\mathbb{M}}\gamma_{\text{top}}\text{Tr}\Big[ \big(\Phi-\tilde{\Phi}_{\text{on}}   \big)^{2}  \Big],
\end{equation} 
where we have used the second equation in \eqref{two term} to eliminate the boundary term contribution.

Now, we perform the same field theory maneuvers of \cite{NA loc CS} to \eqref{Zext}. On the one hand, using the $\kappa$-shift symmetry to fix $\Phi=0$, with unit path integral measure Jacobian, we get
\begin{equation}
\mathcal{Z}_{\text{reg}}=\left( \int\nolimits_{\mathcal{S}_{\kappa}} \mathcal{D}\Phi   \right) \int\nolimits_{\mathcal{A}\times L\mathcal{O}} \mathcal{D}\mathbb{A}\mathcal{D}U \; \text{exp}\left[\frac{i}{\hbar} S\big(\mathbb{A}, U\big)_{\text{reg}}\right]. \label{reg path}
\end{equation}
On the other hand, after integrating out the quadratic field $\Phi$, we obtain 
\begin{equation}
\mathcal{Z}_{\text{con}}=\mathcal{N}'  \int\nolimits_{\overline{\mathcal{A}}\times L\mathcal{O}}\mathcal{D}\mathbb{A}\mathcal{D}U \; \text{exp}\left[\frac{i}{\hbar} S(\mathbb{A},U)_{\text{con}}  \right ], \label{contact path}
\end{equation} 
where
\begin{equation}
\mathcal{N}'=\bigg( \int\nolimits_{\mathcal{S}_{\kappa}} \mathcal{D}\Phi \; \text{exp}\left[-\frac{c}{\hbar} \int\nolimits_{\mathbb{M}}\gamma_{\text{top}} \text{Tr}\left( \Phi ^{2}\right) \right] \bigg) \left( \int\nolimits_{\mathcal{S}} \mathcal{D}\mathbb{A}  \right).
\end{equation}
Using the result \eqref{orto decom} for the orthogonal decomposition, we get that $\mathcal{D}\mathbb{A}=\mathcal{D}\mathbb{A}\big|_{\overline{\mathcal{A}}}\times \mathcal{D}\mathbb{A}\big|_{\mathcal{S}}$. Also, using \eqref{coset measure} and \eqref{sym U measure} to deal, respectively, with the path integral measures $\mathcal{D}\mathbb{A}\big|_{\overline{\mathcal{A}}}$ and $\mathcal{D}U$ and \eqref{Generalized 4CS action} to rewrite the contact 4d CS theory action with defects in the quadratic form, we get, after equating \eqref{reg path} and \eqref{contact path}, the desired localization formula
\begin{equation}
\int\nolimits_{\mathcal{A}\times L\mathcal{O}} \mathcal{D}\mathbb{A}\mathcal{D}U \; \text{exp}\left[\frac{i}{\hbar} S\big(\mathbb{A}, U\big)_{\text{reg}}\right]= \mathcal{N}\int\nolimits_{\overline{\mathcal{A}}\times L\mathcal{O}}\text{exp} \left[ \hat{\mathbf{\Omega}}-\frac{1}{2\epsilon}\big(\mathbf{\mu},\mathbf{\mu}\big)  \right ], \label{main result}
\end{equation}
where $\epsilon:=\hbar/2c$ and
\begin{equation}
\mathcal{N}=\mathcal{N}'/\left( \int\nolimits_{\mathcal{S}_{\kappa}} \mathcal{D}\Phi   \right)\text{\qquad \qquad} \mathbf{\hat{\Omega}}:=i\big(\hat{\Omega}+\hat{\Omega}'   \big).
\end{equation}
 
Finally, let us suppress in \eqref{main result}, for simplicity, all the defects. In the limit $\zeta \rightarrow 0$, we showed that 
\begin{equation}
S(\mathbb{A})_{\text{reg}}\rightarrow S(\mathbb{A})_{4d\text{-CS}}=ic\dint\nolimits_{\mathbb{M}}\omega \wedge CS(\mathbb{A}). 
\end{equation}
Thus, at the path integral level 
\begin{equation}
\int\nolimits_{\mathcal{A}}\, \mathcal{D}\mathbb{A}\, \text{exp}\left[\frac{i}{\hbar} S(\mathbb{A})_{4d\text{-CS}}\right]=\underset{\zeta \rightarrow 0}{\text{lim}}\left( \mathcal{N}\int\nolimits_{\overline{\mathcal{A}}}\text{exp} \left[ i\hat{\Omega}-\frac{1}{2\epsilon}\big(\mu,\mu\big)  \right ]  \right), \label{Main Result}
\end{equation}
suggesting the following updated duality diagram
\begin{equation*}
\begin{array}{ccc}
\mathcal{Z}_{\text{ext}}=\small{\text{eq.\,}} \eqref{Zext} & \qquad {\overset{\Phi \text{-integration}}{\xrightarrow{\hspace*{2.5cm}}  } }& \qquad  \boxed{\mathcal{Z}_{\text{con}}=\small\text{eq.\,} \eqref{contact path}}
\\ 
&  &  \\ 
\; \;  \Bigg\downarrow
\begin{array}{c}
 \scriptstyle{ \Phi=0}  
\end{array}
&  & \quad \quad \, \Bigg\downarrow \, 

\! \! \! \! \! \! \! \!  \! \! \! \! \! \!   \! \!  \! \!  

\begin{array}{c}
 \quad   \quad  \quad  \! \footnotesize\text{step I: }\, \scriptstyle{P^{\perp}(\mathbb{A})=0} \\ 
\qquad \! \!  \!   \footnotesize\text{step II: }\, \scriptstyle{\zeta \rightarrow 0}%
\end{array}
 \\ 
&  &  \\ 
\boxed{\mathcal{Z}_{\text{reg}}=\small\text{eq.\,}\eqref{reg path}}
& \qquad  {\overset{\zeta \rightarrow 0 }{\xrightarrow{\hspace*{2.5cm}} }} & \qquad  \mathcal{Z}_{\text{2d-IFT}}%
\end{array}
\end{equation*}
By the existing equivalence between the two boxed path integrals, we are compelled to conjecture the existence of a relation between the path integral for certain 2d IFT's of the non-ultralocal type and an equivariant localization principle.



\section{Concluding remarks}\label{8}

By refining the results of \cite{Yo}, we have constructed a regularized 4d CS theory that is compatible with the properties of the usual 4d CS theory and the dualization procedure necessary to deduce a localization formula. Some features of this construction are of a different nature when compared to the known results. For instance, in the 3d CS theory case on M \cite{NA loc CS, Wilson NA loc}, the underlying Lie groups are real and compact, while in the 4d CS theory formulation on $\mathbb{R}\times \text{M}$ (or in the usual CY formulation on $\Sigma \times C$ as well), they are all by definition, complex. This introduces a fundamental difference in the localization formula, as it is constructed now on a symplectic manifold that is also pseudo-K\"ahler. Nevertheless, at least by judiciously applying the dualization recipe to the regularized theory in a `marginal' way, a formal localization integral reveals itself. Thus, if one is to make some progress based on this approach, a finer understanding of the localization mechanism on this kind of generalized non-compact symplectic manifolds is necessary. 

It is important to recall that for our approach to achieve full gauge invariance, it remains to be shown that the WZ term contribution,
\begin{equation}
\dint\nolimits_{\mathbb{R}\times \text{M}}\omega\wedge \chi(g),
\end{equation} 
Cf. \eqref{WZ term} vanishes. This can be done, in principle, by adapting the techniques of \cite{Homotopical} applied to $\mathbb{R}^{2}\times \mathbb{CP}^{1}$, to the more general case of $\mathbb{R}\times \text{M}$, with M defined as in \eqref{n bundle}. We expect to address this problem in the future, at least for $C=\mathbb{CP}^{1}$, including as well, the regularization procedure introduced in \cite{Homotopical} that is necessary to handle non-simple poles.

In the original CY 4d CS theory formulation, ultralocal and non-ultralocal IFT's are associated with twist forms without and with zeroes and order and disorder defects, respectively \cite{CY}. In relation to the regularized 4d CS theory, there is no apparent obstruction\footnote{In relation to the absence of zeroes.} for describing ultralocal IFT's as well and the choices made here, i.e. the non-ultralocal integrable $\sigma$-models of $\S \eqref{4}$, were simply a matter of author's personal choice. Although there is an important detail related to the surface defects that couple to the bulk theory, in the case of ultralocal IFT's. We comment on this right below. 

There are some interesting scenarios, not considered in this work, that we plan to address elsewhere in the future. Let us list some of them that are more obvious at this stage:

$\bullet$ \textit{Ultralocal IFT's} \cite{CY}. They have a well-behaved perturbative description, due to the absence of zeroes in $\omega_{C}$ and this would provide a good opportunity for testing and exploring the localization approach under controlled conditions, hopefully producing results that are comparable against known data. However, it is not clear yet how to generate the dof at the order defects, as they should emerge naturally from the symplectic approach in the quadratic term $(\mu,\mu)$. Perhaps, an extension of \eqref{pre-symplectic} is necessary. 

$\bullet$ \textit{Integrable lattice models} \cite{C1,C2,CWY1,CWY2}. Coadjoint orbits defects based on compact real Lie groups are an essential ingredient for realizing these kind of models within the 4d CS theory due to their relation to Wilson loops. In our case, it is not clear yet how to approach them from the perspective of localization and this is because of the coadjoint orbits of complex Lie groups, as the ones considered in $\S \eqref{6.1.3}$, are less understood. We elaborate a bit more on this below. Nevertheless, there is no apparent impediment for, at least experimenting, with their real counterparts. 

$\bullet$ \textit{Non-ultralocal IFT's on (semi)-symmetric spaces} \cite{CY,GS-CS,PS-CS}. Including their deformations \cite{YB-super,lambda-super} based on \cite{eta-def bos,derivation,eta-def fer,lambda-bos,lambda-fer,PS lambda} as well, for example. A hint of how to modify $C$ for the case of symmetric spaces was given in \cite{dihedral,exchange}. It is equally important to perform explicit calculations, for the integrable models of the PCM type considered here, in order to gain a deeper understanding of the formula \eqref{main result} itself. These cases are already under scrutiny.

Now, we make some comments about the presence of Wilson loops and coadjoint orbits in the 4d CS theory and a possible relation to the IFT's monodromy matrix.

Consider a particular transport matrix associated to two points $(\sigma_{1},\sigma_{2})$ belonging to an arbitrary circle fiber $S^{1}$ on top of a generic point $p$ on the base manifold $C$, defined by the expression
\begin{equation}
T(\sigma_{2},\sigma_{1})=P\text{exp}\left(-\int_{\sigma_{1}} ^{\sigma_{2}}d\sigma A_{\sigma}(\sigma)  \right),
\end{equation}
where we have used the local coordinates $(\tau,\sigma,z)$ on $\mathbb{M}$. This object depends explicitly on the coordinates $(\tau, z)$, but for the sake of simplicity, we have omitted them. Under an arbitrary variation $\delta$ and a formal gauge transformations \eqref{general gauge transf}, it obeys
\begin{equation}
\begin{aligned}
\delta T(\sigma_{2},\sigma_{1})&=-\int_{\sigma_{1}}^{\sigma_{2}}d\sigma T(\sigma_{2},\sigma)\delta A_{\sigma}(\sigma)T(\sigma,\sigma_{1}), \\
^{g}T(\sigma_{2},\sigma_{1})&=g^{-1}(\sigma_{2})T(\sigma_{2},\sigma_{1})g(\sigma_{1}).\label{T relations}
\end{aligned}
\end{equation}

In the gauge fixing situation considered in $\S \eqref{4}$ that also involves the $\zeta\rightarrow 0$ limit, when the connection $\mathbb{A}$ is gauge rotated into the 2d flat Lax connection $\mathscr{L}$ defined on $\Sigma$, Cf. \eqref{rotation} and \eqref{L eom}, the first equation in \eqref{T relations} reduces, for $\delta=\partial_{\tau}$, to\footnote{Use $\partial_{\sigma_{1}}T(\sigma_{2},\sigma_{1})=T(\sigma_{2},\sigma_{1})A_{\sigma}(\sigma_{1})$ and $\partial_{\sigma_{2}}T(\sigma_{2},\sigma_{1})=-A_{\sigma}(\sigma_{2})T(\sigma_{2},\sigma_{1})$.}
\begin{equation}
\partial_{\tau}T(\sigma_{2},\sigma_{1})=T(\sigma_{2},\sigma_{1})\mathscr{L}_{\tau}(\sigma_{1})-\mathscr{L}_{\tau}(\sigma_{2})T(\sigma_{2},\sigma_{1}).\label{time flow T}
\end{equation}
It then follows that if we impose periodic boundary conditions upon the field content of the theory, the gauge invariant Wilson loop along $S^{1}$ defined by
\begin{equation}
W_{R}=\text{Tr}_{R}P\text{exp}\left(-\oint_{S^{1}}\mathbb{A}  \right), \label{Wilson}
\end{equation}
where $R$ is an irreducible representation of the Lie group $G$ defining the formal gauge group $\mathcal{G}$, can be seen as the precursor of the usual monodromy matrix associated to the 2d IFT that is obtained after ending the whole process of gauge fixing, namely 
\begin{equation}
M_{R}=\text{Tr}_{R}P\text{exp}\left(-\oint_{S^{1}}\mathscr{L}  \right). \label{monodromy}
\end{equation}
Conservation of the latter under time flow follows directly from \eqref{time flow T}, while the equivalence $W_{R}=M_{R}$ follows as a consequence of the second relation in \eqref{T relations}, when we take $g=\hat{g}^{-1}$.

Recall the relation between coadjoint orbits of the 1d CS type and Wilson loops \cite{Wilson NA loc}, see also \cite{Jones}. From \cite{Wilson NA loc}, we bring the result
\begin{equation}
\text{Tr}_{R}P\text{exp}\left(-\oint_{S^{1}}\mathbb{A}  \right)=\dint_{L\mathcal{O}_{\lambda}}\mathcal{D}U\, \text{exp}\Big[i\, \textbf{cs}_{\lambda} \big( U,\mathbb{A}\big|_{S^{1}} \big)   \Big], \label{Wilson-Coadjoint}
\end{equation}
where
\begin{equation}
\textbf{cs}_{\lambda} \big( U,\mathbb{A}\big|_{S^{1}} \big)=\oint_{S^{1}}\text{Tr}\big(\lambda f^{-1}d_{\mathbb{A}}f\big)
\end{equation}
is, up to a multiplicative constant, formally equal to the 1d CS theory defect action introduced above in \eqref{1d CS}. In \cite{Wilson NA loc}, when M is a Seifert manifold and $G$ is a compact, connected, simply-connected and simple real Lie group, the equivalence \eqref{Wilson-Coadjoint}, which is based on the orbit method \cite{orbit method}, is essential for modeling in the symplectic language, the expectation values of Wilson loops in the path integral formulation of the 3d CS theory on M, while extending the localization formula of \cite{NA loc CS} to that general situation. Unfortunately, a similar equivalence valid for arbitrary\footnote{Non-trivial connected complex Lie groups are essentially non-compact, as compact connected complex Lie groups are Abelian.} complex Lie groups is, to the author's knowledge, not known\footnote{The orbit method has a limited applicability for non-compact complex or, non-compact semisimple Lie groups, where a direct universal one-to-one correspondence between all unitary irreducible representations and coadjoint orbits does not hold as occurs in other cases. The success of the method strongly depends on the underlying Lie group structure, e.g. if it is compact, nilpotent, solvable, etc \cite{orbit method}. }, but as we have seen, modifying the regularized 4d CS theory with (co)-adjoint orbit defects insertions is still consistent with the 4d CS theory symplectic approach to localization. It would be interesting to study if there is a relation of that form, at least for reductive\footnote{As they have compact real forms.} complex Lie groups as considered in $\S \eqref{6.1.3}$, because if such a relation holds, assuming \eqref{Wilson-Coadjoint} in that case and considering the partially integrated path integral for the regularized action with $N$ adjoint orbit defects \eqref{reg action N}, we would obtain something of the form
\begin{equation}
\int_{L\mathcal{O}}\mathcal{D}U\, \text{exp}\left[\frac{i}{\hbar} S\big(\mathbb{A}, U \big)_{\text{reg}}\right]=\prod_{j=1}^{N}W_{R_{j}}(z_{j})\, \text{exp}\left[\frac{i}{\hbar} S\big(\mathbb{A} \big)_{\text{reg}}\right], \label{partial path int}
\end{equation}
where we have defined $l:=\hbar/2c$ in order to apply the result \eqref{Wilson-Coadjoint} on the rhs. This would allow, after gauge fixing, to include the expectation values of monodromy matrices inserted at the points $z_{j}$, as given by \eqref{monodromy}, in the would-be 2d IFT path integral formulation. 

For the reasons just explained, we have found prudent to limit the discussion of complex coadjoint orbits to their symplectic properties without trying to related them, for example\footnote{The author thanks the referee for pointing these references.}, to the coadjoint orbits used to relate lattice models and ultralocal IFT's in the thermodynamic limit, as in \cite{Ashwinkumar} or their possible identification with impurities, as in \cite{Gaiotto}. 

We finish with some comments on some less obvious complementary approaches that might be useful in exploring new relations between the 4d CS theory and other field theories:

$\bullet$ \textit{Caloron correspondence} \cite{Mickler}. In this paper the 3d CS theory on a manifold M of the form \eqref{n bundle}, is considered. The Lie groups involved being real and compact though. The approach in this work is based on the Caloron correspondence and provides a new formulation to the Beasley-Witten localization formula \cite{NA loc CS}. It is shown that the 3d CS theory on M is equivalent to a new 2d TQFT on the base space $C$, called Caloron BF theory. The advantage of this approach is that it is naturally defined on non-trivial principal $G$-bundles over M. Recall that in our case, triviality was imposed from the outset, Cf. footnote 24 in  $\S \eqref{5}$. This would allow to generalize our construction to non-trivial $G$-bundles and, perhaps, to construct new IFT's on the pole defect surface $\mathfrak{p}\times \Sigma$, as well. It would be equally interesting to discover if there is always an equivalent 2d Caloron type TQFT on $C$, associated to any 2d IFT on $\mathfrak{p}\times \Sigma$.

$\bullet$ \textit{Abelianisation} \cite{Abelianization}. In this paper 3d CS theory on a manifold M of the form \eqref{n bundle} and with a complex gauge group is considered. It is presented a complete non-perturbative evaluation of the path integral (the partition function and certain expectation values of Wilson loops). The approach used is based on the method of Abelianisation rather than on the method of NA localization and has the advantage of making explicit calculations more tractable. The method basically relies on fiber integration and results on an equivalent 2d field theory on $C$ that, under some circumstances, is of the BF type. For compact real groups, $q$-deformed YM theory is obtained \cite{real Abeli}. They also show that in certain cases, the path integral can be seen to factorize neatly into holomorphic and anti-holomorphic parts. It would be interesting to see if this technique applies to the original/regularized 4d CS theory as well and under what conditions the partition function factorizes. It would also be equally interesting to identify, if any, the resulting lower dimensional theory produced in this approach that is, in principle, associated to any of the 2d IFT's on $\mathfrak{p}\times \Sigma$.

All this suggest a quantum duality,
\begin{equation}
\{\text{2d IFT on $ \mathfrak{p}\times \Sigma$}\}\quad \overset{ ?}{\xleftrightarrow{\hspace*{1.5cm}}  }\quad \{\text{2d gauge theory on $C$}\}.
\end{equation}
The intuition is that, somehow the contact 4d CS theory sources this duality. We expect to address this problem in the future, but this seem to be a quite challenging problem.

$\bullet$ \textit{D-brane description} \cite{non perturbative 4d CS}. The authors show how integrable lattice models described by the usual 4d CS theory can be realized via a stack of D4-branes ending on an NS5-brane in type IIA string theory, with D0-
branes on the D4-brane worldvolume sourcing a meromorphic RR 1-form, and fundamental strings
forming the lattice. This description provides a nonperturbative integration cycle for the usual 4d CS theory. By applying T- and S-duality, they also showed how the R-matrix, the Yang-Baxter equation and the Yangian can be categorified, that is, obtained via the Hilbert space of a 6d gauge theory. It would be interesting to see if there is a variant of such a 6d theory formulation that is associated to the regularized 4d CS theory \eqref{reg reg CS} or not. This would shed some light into the higher dimensional origin of the regularized 4d CS theory.

Some of these problems are currently under investigation \cite{progress}.

\section*{Acknowledgements}

The author is indebted to M. Ashwinkumar, M. Blau, R. Mickler and B. Vicedo for correspondence, questions and suggestions. The author also thanks the organizers of the conference `\textit{Integrability, Dualities and Deformations, IDD 24}' held at Swansea, UK, and the organizers of the conference `\textit{IV Theoretical Physics Meeting
GIFT – UNIFEI}', held at Itajub\'a, Brazil in 2024 for the opportunity of presenting part of the results of this paper prior to publication. Finally, the author thanks the referee for very appropriate suggestions that helped to improve the  manuscript's clarity. 
\center{\textcolor{cyan}{$\therefore$}}


\end{document}